\documentclass[pra,aps,showpacs,groupedaddress,superscriptaddress,twocolumn,toc=flat]{revtex4-1}

\usepackage[colorlinks=true,linkcolor=blue,urlcolor=blue,citecolor=blue]{hyperref}

\usepackage[utf8x]{inputenc}
\usepackage{color}
\usepackage{bbm} 

\usepackage{amsfonts,amsmath,amssymb,stmaryrd}

\usepackage{graphicx}
\usepackage{subfigure}  
\usepackage{bbm} 
\usepackage{hyperref}
\usepackage{epsfig}
\usepackage{mathrsfs}
\usepackage{verbatim}
\usepackage{centernot}
\usepackage{ulem}
\usepackage{nicefrac}

\renewcommand{\l}{\left(}
\renewcommand{\r}{\right)}

\newcommand{\bra}[1]{\langle#1|}
\newcommand{\ket}[1]{|#1\rangle}

\renewcommand{\ij}{{\langle i, j \rangle}}
\renewcommand{\H}{\hat{\mathcal{H}}}

\renewcommand{\c}{\hat{c}}
\renewcommand{\a}{\hat{a}}
\newcommand{\s}{\hat{s}}
\newcommand{\cd}{\hat{c}^\dagger}

\newcommand{\ad}{\hat{a}^\dagger}
\newcommand{\sd}{\hat{s}^\dagger}
\newcommand{\bd}{\hat{b}^\dagger}

\renewcommand{\b}{\hat{b}}
\newcommand{\hd}{\hat{h}^\dagger}
\newcommand{\h}{\hat{h}}

\newcommand{\hc}{\text{h.c.}}

\usepackage{array}

\usepackage{cancel,ifthen}
\newcommand{\cmnt}[2][NoInPuT]{\ifthenelse{\equal{#1}{NoInPuT}}{}{{\color{red}\sout{#1}}} {\color{blue} #2}}

\usepackage{bm}	
\renewcommand{\vec}[1]{\bm{#1}}

\begin{document}
\normalem	

\title{Parton theory of magnetic polarons: Mesonic resonances and signatures in dynamics}

\author{F. Grusdt}
\affiliation{Department of Physics, Harvard University, Cambridge, Massachusetts 02138, USA}

\author{M. K\'anasz-Nagy}
\affiliation{Department of Physics, Harvard University, Cambridge, Massachusetts 02138, USA}

\author{A. Bohrdt}
\affiliation{Department of Physics and Institute for Advanced Study, Technical University of Munich, 85748 Garching, Germany}
\affiliation{Department of Physics, Harvard University, Cambridge, Massachusetts 02138, USA}

\author{C. S. Chiu}
\affiliation{Department of Physics, Harvard University, Cambridge, Massachusetts 02138, USA}

\author{G. Ji}
\affiliation{Department of Physics, Harvard University, Cambridge, Massachusetts 02138, USA}

\author{M. Greiner}
\affiliation{Department of Physics, Harvard University, Cambridge, Massachusetts 02138, USA}

\author{D. Greif}
\affiliation{Department of Physics, Harvard University, Cambridge, Massachusetts 02138, USA}

\author{E. Demler}
\affiliation{Department of Physics, Harvard University, Cambridge, Massachusetts 02138, USA}


\date{\today}

\begin{abstract}
When a mobile hole is moving in an anti-ferromagnet it distorts the surrounding N\'eel order and forms a magnetic polaron. Such interplay between hole motion and anti-ferromagnetism is believed to be at the heart of high-temperature superconductivity in cuprates. In this article we study a single hole described by the $t-J_z$ model with Ising interactions between the spins in two dimensions. This situation can be experimentally realized in quantum gas microscopes with Mott insulators of Rydberg-dressed bosons or fermions, or using polar molecules. We work at strong couplings, where hole hopping is much larger than couplings between the spins. In this regime we find strong theoretical evidence that magnetic polarons can be understood as bound states of two partons, a spinon and a holon carrying spin and charge quantum numbers respectively. Starting from first principles, we introduce a microscopic parton description which is benchmarked by comparison with results from advanced numerical simulations. Using this parton theory, we predict a series of excited states that are invisible in the spectral function and correspond to rotational excitations of the spinon-holon pair. This is reminiscent of mesonic resonances observed in high-energy physics, which can be understood as rotating quark antiquark pairs carrying orbital angular momentum. Moreover, we apply the strong coupling parton theory to study far-from equilibrium dynamics of magnetic polarons observable in current experiments with ultracold atoms. Our work supports earlier ideas that partons in a confining phase of matter represent a useful paradigm in condensed-matter physics and in the context of high-temperature superconductivity in particular. While direct observations of spinons and holons in real space are impossible in traditional solid-state experiments, quantum gas microscopes provide a new experimental toolbox. We show that, using this platform, direct observations of partons in and out-of equilibrium are now possible. Extensions of our approach to the $t-J$ model are also discussed. Our predictions in this case are relevant to current experiments with quantum gas microscopes for ultracold atoms. 
\end{abstract}

\maketitle

\begin{figure}[t!]
\centering
\epsfig{file=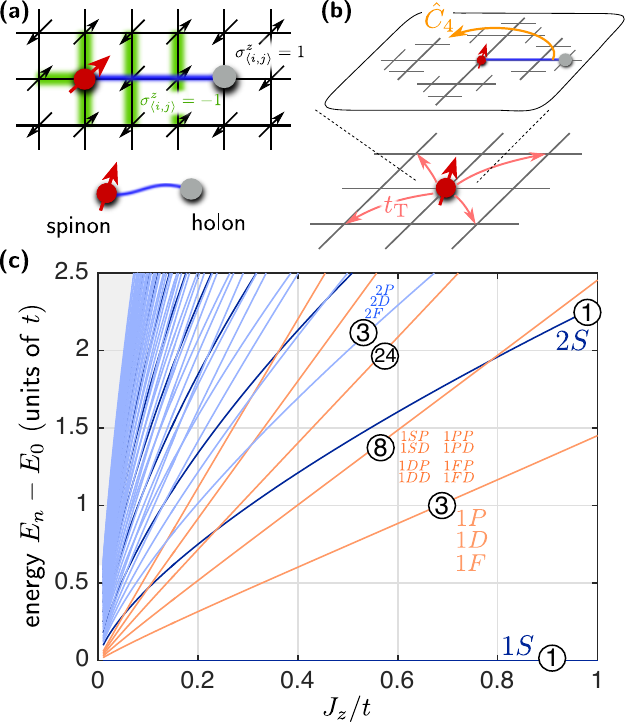, width=0.47\textwidth}
\caption{We consider the dynamics of a hole propagating in a N\'eel-ordered spin environment. (a) The hole creates a distortion of the spin state (green). This distortion carries a fractional spin $S^z=1/2$. We introduce two types of partons, spinons and holons, describing the spin and charge quantum numbers respectively. They are confined by a string of displaced spins connecting them (blue), similar to mesons which can be understood as bound states of confined quark antiquark pairs. (b) In the strong coupling limit the slow dynamics of the spinon can be decoupled from the fast dynamics of the holon. This leads to coherent motion of the spinon on the original square lattice, which is independent of the holon motion. The latter takes place on a fractal Bethe lattice which is co-moving with the spinon. The holon states on the Bethe lattice obey a discrete $\hat{C}_4$ symmetry that corresponds to rotations of the string configuration around the spinon position. (c) As a result, discrete rotational and vibrational excited states of the magnetic polaron can be formed. They are characterized by quantum numbers $n m_4 ...$, where $e^{i \pi m_4/2}$ denotes the eigenvalue of $\hat{C}_4$ and $n$ labels vibrational excitations. In the figure we used labels $S,P,D,F$ for $m_4=0,1,2,3$ and indicated the multiplicity of the degenerate states by numbers in circles. Different colors correspond to the different types of ro-vibrational excitations. For the calculation we used $S=1/2$ and linear string theory (LST, described later in the text) with strings of length up to $\ell_{\rm max} = 100$.}
\label{fig:Overview}
\end{figure}

\section{Introduction}
\label{sec:Intro}

Understanding the dynamics of charge carriers in strongly correlated materials constitutes an important prerequisite for formulating an effective theory of high-temperature superconductivity. It is generally assumed that the Fermi Hubbard model provides an accurate microscopic starting point for a theoretical description of cuprates \cite{Emery1987,Dagotto1994,Lee2006}. At strong couplings, this model can be mapped to the $t-J$ model, which describes the motion of holes (hopping $t$) inside a strongly correlated bath of spins with strong anti-ferromagnetic (AFM) Heisenberg couplings (strength $J$). To grasp the essence of the complicated $t-J$ Hamiltonian, theorists have also studied closely related variants, most prominently the $t-J_z$ model for which Heisenberg couplings are replaced with Ising interactions ($J_z$) between the spins \cite{Dagotto1994,Chernyshev1999}. 

While the microscopic $t-J$ and $t-J_z$ models are easy to formulate, understanding the properties of their ground states is extremely challenging. As a result, most theoretical studies so far have relied on large-scale numerical calculations \cite{LeBlanc2015} or effective field theories which often cannot capture microscopic details \cite{Anderson1987,Kivelson1987,Zhang1988,Shraiman1988,Auerbach1991,Ribeiro2006,Punk2015PNASS,Sachdev2016}. Even the problem of a single hole propagating in a state with N\'eel order \cite{Nagaoka1966,Bulaevskii1968,Brinkman1970,SchmittRink1988,Trugman1988,Kane1989,Sachdev1989,Shraiman1988,Dagotto1990,Elser1990,Trugman1990,Boninsegni1991,Auerbach1991,Liu1992,Boninsegni1992,Chernyshev1999,White2001,Mishchenko2001}, see Fig.~\ref{fig:Overview} (a), is so difficult in general, that heavy numerical methods are required for its solution. This is true in particular at strong couplings $t \gg J,J_z$, where the tunneling rate $t$ of the hole exceeds the couplings $J,J_z$ between the spins. The strong coupling regime is also relevant for high-temperature cuprate superconductors for which typically $t/J \approx 3$ \cite{Lee2006}. While several theoretical approaches have been developed, which are reliable in the weak-to-intermediate coupling regime $t \lesssim J,J_z$, to date there exist only a few theories describing the strong coupling limit \cite{Kane1989} and simple variational wavefunctions in this regime are rare. Even calculations of qualitative ground state properties of a hole in an anti-ferromagnet, such as the renormalized dispersion relation, require advanced theoretical techniques. These include effective model Hamiltonians \cite{SchmittRink1988,Kane1989}, fully self-consistent Green's function methods \cite{Liu1992}, non-trivial variational wave functions \cite{Sachdev1989,Trugman1990,Boninsegni1992} or sophisticated numerical methods such as Monte-Carlo \cite{Boninsegni1992,Mishchenko2001} and DMRG \cite{White2001,Zhu2014} calculations. The difficulties in understanding the single-hole problem add to the challenges faced by theorists trying to unravel the mechanisms of high-$T_c$ superconductivity.

Here we study the problem of a single hole moving in an anti-ferromagnet from a different perspective, focusing on the $t-J_z$ model for simplicity. In contrast to most earlier works, we consider the strong coupling regime, $t \gg J_z$. Starting from first principles, we derive a microscopic parton theory of magnetic polarons. This approach not only provides new conceptual insights to the physics of magnetic polarons, but it also enables semi-analytical derivations of their properties. We benchmark our calculations by comparison to the most advanced numerical simulations known in literature. Notably, our approach is not limited to low energies but provides an approximate description of the entire energy spectrum. This allows us, for example, to calculate magnetic polaron dynamics far from equilibrium. Note that in the extreme limit when $J_z=0$, Nagaoka has shown that the ground state of this model has long-range ferromagnetic order \cite{Nagaoka1966}. We will work in a regime where this effect does not yet play a role, see Ref.~\cite{White2001} for a discussion.

\subsection{Partons and the $t-J_z$ model}
\label{subsecPartons}
Partons have been introduced in high energy physics to describe hadrons \cite{Feynman1988}. Arguably, the most well known example of partons is provided by quarks. In quantum chromodynamics (QCD), the quark model elegantly explains mesons (baryons) as composite objects consisting of two (three) valence quarks. On the other hand, individual quarks have never been observed in nature, and this has been attributed to the strong confining force between a pair of quarks mediated by gauge fields \cite{Wilson1974}. Even though there is little doubt that quarks are truly confined and can never be separated at large distances, a strict mathematical proof is still lacking, and the quark confinement problem is still attracting considerable attention in high-energy physics, see for example Ref.~\cite{Greensite2003}. 

To understand how the physics of holes moving in a spin environment with strong AFM correlations is connected to the quark confinement problem, consider removing a spin from a two-dimensional N\'eel state. When the hole moves around, it distorts the order of the surrounding spins. In the strong coupling regime, $t \gg J, J_z$, these spins have little time to react and the hole can distort a large number of AFM bonds. Assuming for the moment that the hole motion is restricted to a straight line, as illustrated in Fig.~\ref{fig:Overview} (a), we notice that a string of displaced spins is formed. At one end, we find a domain wall of two aligned spins, and the hole is located on the opposite end. By analyzing their quantum numbers, we note that the domain wall corresponds to a spinon -- it carries half a spin and no charge -- whereas the hole becomes a holon -- it carries charge but no spin. The spinon and holon are the partons of our model. Because a longer string costs proportionally more energy, the spinon can never be separated from the holon. This is reminiscent of quark confinement. 

Partons also play a role in various phenomena of condensed matter physics. A prominent example is the fractional quantum Hall effect \cite{Tsui1982,Stormer1999}, where electrons form a strongly correlated liquid with elementary excitations (the partons) which carry a quantized fraction of the electron's charge \cite{Laughlin1983,Picciotto1997,Saminadayar1997}. This situation is very different from the case of magnetic polarons which we consider here, because the fractional quasiparticles of the quantum Hall effect are to a good approximation non-interacting and can be easily separated. Similar fractionalization has also been observed in one-dimensional spin chains \cite{Giamarchi2003,Kim1996,Segovia1999,Kim2006,Kruis2004,Kruis2004a,Hilker2017}, where holes decay into pairs of independent holons and spinons as a direct manifestation of spin-charge separation. Unlike in the situation described by Fig.~\ref{fig:Overview} (a), forming a string costs no energy in one dimension and spinons and holons are deconfined in this case. 

Confined phases of partons are less common in condensed matter physics. It has first been pointed out by B\'eran et al.~\cite{Beran1996} that this is indeed a plausible scenario in the context of high-temperature superconductivity, and the $t-J$ model in particular. In Ref.~\cite{Beran1996} theoretical calculations of the dispersion relation and the optical conductivity of magnetic polarons were analyzed, and it was concluded that their observations can be well explained by a parton theory of confined spinons and holons. A microscopic description of those partons has not yet been provided, although several models with confined spinon-holon pairs have been studied \cite{Ribeiro2006,Punk2012,Punk2015PNASS}. 

The most prominent feature of partons that has previously been discussed in the context of magnetic polarons is the existence of a set of resonances in the single-hole spectral function \cite{Shraiman1988a,Dagotto1990,Liu1991,Liu1992,Beran1996,Mishchenko2001,Manousakis2007} which can be measured by angle-resolved photoemission spectroscopy (ARPES), see e.g. Ref.~\cite{Damascelli2003} for a discussion. Such long-lived states in the spectrum can be understood as vibrational excitations of the string created by the motion of a hole in a N\'eel state \cite{Bulaevskii1968,Brinkman1970,Trugman1988,Shraiman1988a,Manousakis2007}. In the parton theory they correspond to vibrational excitations of the spinon-holon pair \cite{Beran1996}, where in a semi-classical picture the string length is oscillating in time. 

In this paper we present additional evidence for the existence of confined partons in the two-dimensional $t-J_z$ model at strong coupling. Using the microscopic parton theory, we show that besides the known vibrational states an even larger number of rotational excitations of magnetic polarons exist. This leads to a complete analogy with mesons in high-energy physics which we discuss next (in \ref{subsecRotationalExcitations}). The rotational excitations of magnetic polarons have not been discussed before, partly because they are invisible in traditional ARPES spectra. Quantum gas microscopy \cite{Bakr2009,Sherson2010} represents a new paradigm for studying the $t-J_z$ model, and we discuss below (in \ref{subsecQuantGasMic}) how it enables not only measurements of rotational excitations, but also direct observations of the constituent partons in current experiments with ultracold atoms.

\subsection{Rotational excitations of parton pairs}
\label{subsecRotationalExcitations}
Mesons can be understood as bound states of two quarks and thus are most closely related to the magnetic polarons studied in this paper. The success of the quark model in QCD goes far beyond an explanation of the simplest mesons, including for example pions ($\bm \pi$) and kaons ($\bm K$). Collider experiments that have been carried out over many decades have identified an ever growing zoo of particles. Within the quark model, many of the observed heavier mesons can be understood as excited states of the fundamental mesons. Aside from the total spin $s$, heavier mesons can be characterized by the orbital angular momentum $\ell$ of the quark antiquark pair \cite{Micu1969} as well as the principle quantum number $n$ describing their vibrational excitations. In Table \ref{tabMesons} we show a selected set of excited mesons, together with the quantum numbers of the involved quark-antiquark pair. Starting from the fundamental pion (kaon) state $\bm \pi$ ($\bm K$), many rotational states with $\ell = 1,2,3$ ($P,D,F$) can be constructed \cite{Micu1969,Amsler2008} which have been observed experimentally \cite{Olive2014}. By changing $n$ to two, the excited states $\bm \pi(1300)$ and $\bm K(1460)$ can be constructed. Due to the deep theoretical understanding of quarks, all these mesons are considered as composites instead of new fundamental particles.

\begin{table}[t!]
\centering
\begin{tabular}{ p{1.6cm} | p{2cm} | p{1.6cm}}
  $n^{2s+1} \ell_J $ & $u \overline{d}$ & $u \overline{s}$ \\
  \hline
  $1^1 S_0$ & $\bm \pi$ & $\bm K$   \\
  $1^1 P_1$ &  $\bm b_1(1235)$ & $\bm K_{1B}$   \\
    $1^1 D_2$ & $\bm \pi_2(1670)$ & $\bm K_{2}(1770)$   \\
  $1^3 F_4$ & $\bm a_4(2040)$ & $\bm K^*_{4}(2045)$   \\
  $2^1 S_0$ &  $\bm \pi(1300)$ & $\bm K(1460)$ 
\end{tabular}
\caption{Examples of meson resonances corresponding to rotational ($\ell$) and vibrational ($n$) excitations of the quark antiquark pair ($q \overline{q}$). This list is incomplete and the data was taken from Ref.~\cite{Amsler2008}. The numbers in brackets denote the mass of the excited meson state in units of $MeV / c^2$.}
\label{tabMesons}
\end{table}

Similarly, rotationally and vibrationally excited states of magnetic polarons can be constructed in the $t-J_z$ model. They can be classified by the angular momentum (rotational) and radial (vibrational) quantum numbers $\ell$ and $n$ of the spinon-holon pair, as well as the spin $\sigma$ of the spinon. An important difference to mesons is that we consider a lattice model where the usual angular momentum is not conserved. However, there still exist discrete rotational symmetries which can be defined in the spinon reference frame. 

To understand this, let us consider the Hilbert space $\mathscr{H}_{\rm p}$ of the parton theory introduced in this paper. As illustrated in Fig.~\ref{fig:Overview} (b), it can be described as a direct product of the space of spinon positions on the square lattice $\mathscr{H}_{\rm s}$ and the space of string configurations $\mathscr{H}_{\Sigma}$ emerging from the spinon,
\begin{equation}
\mathscr{H}_{\rm p} = \mathscr{H}_{\rm s} \otimes \mathscr{H}_{\Sigma}.
\label{eqHpartontJz}
\end{equation}
The latter is equivalent to the Hilbert space of a single particle hopping on a fractal Bethe lattice with four bonds emerging from each site. We recognize a discrete four-fold rotational symmetry $\hat{C}_4$ of the parton theory, where the string configurations are cyclically permuted around the spinon position, see Fig.~\ref{fig:Overview} (b). Therefore we can construct excited magnetic polaron states with eigenvalues $e^{i \pi m_4 /2}$ of $\hat{C}_4$ with $m_4=1,2,3~(P, D, F)$, which are analogous to the rotational excitations of mesons. In the ground state, $m_4=0$. Similarly, there exist three-fold permutation symmetries $\hat{P}_3$ in the parton theory corresponding to cyclic permutations of the string configuration around sites one lattice constant away from the spinon. This leads to a  second quantum number $m_3=0,1,2$ required to classify all eigenstates. 

In Fig.~\ref{fig:Overview} (c) we calculate the excitation energies of rotational and vibrational states in the magnetic polaron spectrum. We applied the linear string approximation, where self-interactions of the string connecting spinon and holon are neglected. It will be shown in the main part of this paper, that this description is justified for the low lying excited states of magnetic polarons in the $t-J_z$ model. In analogy with the meson resonances listed in Table \ref{tabMesons}, we have labeled the eigenstates in Fig.~\ref{fig:Overview} (c) by their ro-vibrational quantum numbers. Similar to the pion, the magnetic polaron corresponds to the ground state $1 S$. The lowest excited states are given by $1P, 1D, 1F$. In contrast to their high-energy analogues \cite{Greensite2003} these states are degenerate which will be shown to be due to lattice effects. Depending on the ratio of $J_z/t$, the next higher states correspond to a vibrational excitation ($2S$), or a second rotational excitation (states $1 m_4 m_3$ for $m_3=P, D$ and $m_4=S,P,D,F$).

The parton theory of magnetic polarons provides an approximate description over a wide range of energies at low doping. This makes it an excellent starting point for studying the transition to the pseudogap phase observed in cuprates at higher hole concentration \cite{Damascelli2003,Shen2005,Lee2006}, because the effective Hilbert space is not truncated to describe a putative low-energy sector of the theory. We discuss extensions of the parton theory to finite doping and beyond the $t-J_z$ model in a forthcoming work.

\subsection{Quantum gas microscopy of the $t-J_z$ model}
\label{subsecQuantGasMic}

Experimental studies of individual holes are challenging in traditional solid state systems. Ultracold atoms in optical lattices provide a promising alternative platform for realizing this scenario and investigating microscopic properties of individual holes in a state with N\'eel order. The toolbox of atomic physics offers unprecedented coherent control over individual particles. In addition, many powerful methods have been developed to probe these systems, including the ability to measure correlation functions \cite{Hart2015,Parsons2016,Cheuk2016,Boll2016}, non-local string order parameters \cite{Hilker2017} and spin-charge correlations \cite{Boll2016} on a single-site level. Moreover bosonic \cite{Bakr2009,Sherson2010} and fermionic \cite{Parsons2015,Cheuk2015,Omran2015,Haller2015,Edge2015,Greif2016,Brown2017} quantum gas microscopes offer the ability to realize arbitrary shapes of the optical potential down to length scales of a single site as set by the optical wavelength \cite{Weitenberg2011,Zupancic2016}. 

These capabilities have recently led to the first realization of an anti-ferromagnet with N\'eel order across a finite system of ultracold fermions with $SU(2)$ symmetry at temperatures below the spin-exchange energy scale $J$ \cite{Mazurenko2017}. In another experiment, canted anti-ferromagnetic states have been realized at finite magnetization \cite{Brown2017} and the closely related attractive Fermi Hubbard model has been investigated \cite{Mitra2017}. In systems of this type individual holes can be readily realized in a controlled setting and studied experimentally. 

\subsubsection{Implementation of the $t-J_z$ model}
\label{subsubsecImplmttJz}
The $t-J_z$ model was long considered as a mere toy model, closely mimicking some of the essential features known from more accurate model Hamiltonians relevant in the study of high-temperature superconductivity. By using ultracold atoms one can go beyond this paradigm, and realize the $t-J_z$ model experimentally.

One approach suggested in Ref.~\cite{Gorshkov2011tJ} is to use ultracold polar molecules which introduce anisotropic and long-range dipole-dipole couplings between their internal spin states. The flexibility to adjust the anisotropy also allows to realize Ising couplings as required for the $t-J_z$ model. When the molecules are placed in an optical lattice, the ratio $t/J_z$ between tunneling and spin-spin couplings can be tuned over a wide range.

\begin{figure}[t!]
\centering
\epsfig{file=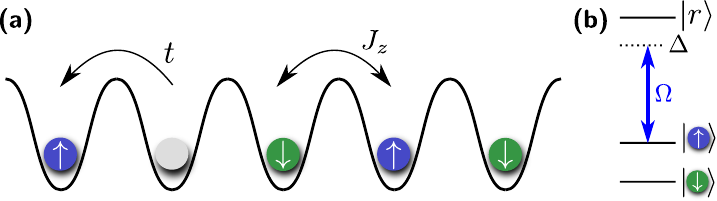, width=0.47\textwidth}
\caption{The $t-J_z$ model can be implemented for ultracold atoms with two internal (pseudo-) spin states in a quantum gas microscope by starting from a Mott insulator (a). Ising interactions can be realized by including Rydberg dressing for one of the two spin states, as demonstrated experimentally in Refs.~\cite{Zeiher2016,Zeiher2017}. Mobile holes can be doped into the system by removing atoms from the Mott insulator. Their statistics is determined by the underlying particles forming the Mott state. In principle the method works in arbitrary dimensions, although many Rydberg states have anisotropic interactions which can break the lattice symmetries.}
\label{figtJzRydberg}
\end{figure}

Other methods to implement Ising couplings between spins involve trapped ions \cite{Porras2004,Britton2012} or arrays of Rydberg atoms \cite{Labuhn2016,Zeiher2016,Zeiher2017}. We now discuss the second option in more detail, because it allows a direct implementation of the $t-J_z$ model in a quantum gas microscope with single-particle and single-site resolution. 

As illustrated in Fig.~\ref{figtJzRydberg} (a), one can start from a Mott insulating state of fermions or bosons with two internal (pseudo-) spin states. Strong Ising interactions between the spins can be realized by Rydberg dressing only one of the two states, see Fig.~\ref{figtJzRydberg} (b). This situation can be described by the effective Hamiltonian \cite{Henkel2010,Pupillo2010,Zeiher2016}
\begin{equation}
\H_{\rm Ising} = \frac{1}{2} \sum_{i,j} J_z^{i,j} \hat{S}^z_i \hat{S}^z_j.
\label{eqHIsing}
\end{equation}
Here, we have assumed a large homogeneous system and ignored an additional energy shift that only depends on the total, conserved magnetization in this case. As demonstrated in Ref.~\cite{Zeiher2016}, the couplings $J_z^{i,j}$ in Eq.~\eqref{eqHIsing} decay quickly with the distance $r_{i,j}$ between two spins, $J_z^{i,j} =U_0 / \l 1 + (r_{i,j}/R_c)^6 \r$ where $ U_0 = \hbar \Omega^4 / (8 |\Delta|^3)$. Here $\Omega$ is the Rabi frequency of the Rydberg dressing laser with detuning $\Delta$ from resonance, and the critical distance $R_c$ below which Rydberg blockade plays a role \cite{Saffman2010} is determined by the Rydberg-Rydberg interaction potential, see Refs.~\cite{Henkel2010,Pupillo2010,Zeiher2016} for details. 

By realizing sufficiently small $R_c/a \approx 1$, where $a$ denotes the lattice constant, a situation can be obtained where nearest neighbor AFM Ising couplings $J_z$ are dominant. By doping the Mott insulator with holes, this allows to implement an effective $t-J_z$ Hamiltonian with tunable coupling strengths. 

The statistics of the holes in the resulting $t-J_z$ model are determined by the statistics of the underlying particles forming the Mott insulator. For the study of a single magnetic polaron in this paper, quantum statistics play no role. At finite doping magnetic polarons start to interact and their statistics become important. Studying the effects of quantum statistics on the resulting many-body states is an interesting future direction.

\subsubsection{Direct signatures of strings and partons}
Quantum gas microscopes provide new capabilities for the direct detection of the partons constituting magnetic polarons, as well as the string of displaced spins connecting them. The possibility to perform measurements of the instantaneous quantum mechanical wavefunction directly in real space allows one to detect non-local order parameters \cite{Endres2011,Hilker2017} and is ideally suited to unravel the physics underlying magnetic polarons. Now we provide a brief summary of the most important signatures of the parton theory which can be directly accessed in quantum gas microscopes and will be discussed in this paper. The following considerations apply to a regime of temperatures $T < J_z$ where the local anti-ferromagnetic correlations are close to their zero-temperature values, although most of the phenomenology is expected to be qualitatively similar at higher temperatures \cite{Nagy2017PRB}.

\emph{Ro-vibrational excitations.--}
As mentioned in Sec.~\ref{subsecPartons}, the ro-vibrational excitations of magnetic polarons provide direct signatures for the parton nature of magnetic polarons. Their energies can be directly measured: Vibrational states are visible in ARPES spectra, which can also be performed in a quantum gas microscope \cite{Bohrdt2017spec}. Later we also discuss alternative spectroscopic methods based on magnetic polaron dynamics in a weakly driven system, which enable direct measurements of the rotational resonances. 

\emph{Direct imaging of strings and partons.--}
In a quantum gas microscope, the instantaneous spin configuration around the hole can be directly imaged \cite{Boll2016}. Up to loop configurations, this allows one to directly observe spinons, holons and strings and extract the full counting statistics of the string length for example. We show in this paper that this method works extremely accurately in the case of the $t-J_z$ model.

\emph{Scaling relations at strong couplings.--}
When $t \gg J_z$, the motion of the holon relative to the spinon can be described by an effective one-dimensional Schr\"odinger equation with a linear confining potential at low energies. This leads to an emergent scaling symmetry which allows to relate solutions at different ratios $J_z/t$ by a simple re-scaling of lengths \cite{Bulaevskii1968}: $x \to \lambda^{1/3} x$ when $J_z \to \lambda J_z$. In ultracold atom setups the ratio $t/J_z$ can be controlled and a wide parameter range can be simulated. For doping with a single hole this allows to observe the emergent scaling symmetry by showing a data collapse after re-scaling all lengths as described above. The scaling symmetry also applies for the expectation values of potential and kinetic energies in the ground state. In the strong coupling regime, $t \gg J_z$, to leading order both depend linearly on $t^{1/3} J_z^{2/3}$ \cite{Bulaevskii1968}. By simultaneously imaging spin and hole configurations \cite{Boll2016}, the potential energy can be directly measured using a quantum gas microscope and the linear dependence on $(J_z/t)^{2/3}$ can be checked.

\subsubsection{Far-from-equilibrium experiments}
Ultracold atoms allow a study of far-from equilibrium dynamics of magnetic polarons \cite{Zhang1991,Mierzejewski2011,Kogoj2014,Lenarcic2014,DalConte2015,Carlstrom2016PRL,Nagy2017PRB}. For example, a hole can be pinned in a N\'eel state and suddenly released. This creates a highly excited state with kinetic energy of the order $t \gg J_z$, which is quickly transferred to spin excitations \cite{Mierzejewski2011,Eckstein2014,Kogoj2014}. Such dynamics can be directly observed in a quantum gas microscope. It has been suggested that this mechanism is responsible for the fast energy transfer observed in pump-probe experiments on cuprates \cite{DalConte2015}, but the coupling to phonons in solids complicates a direct comparison between theory and experiment. 

As a second example, external fields (i.e. forces acting on the hole) can be applied and the resulting transport of a hole through the N\'eel state can be studied \cite{Mierzejewski2011}. As will be shown, this allows direct measurements of the mesonic excited states of the magnetic polaron, analogous to the case of polarons in a Bose-Einstein condensate \cite{Bruderer2010,Grusdt2014BO}.

\subsection{Magnetic polaron dynamics}
\label{subsec:DynOverview}
To study dynamics of magnetic polarons in this paper, we use the strong coupling parton theory to derive an effective Hamiltonian for the spinon and holon which describes the dynamics of a hole in the AFM environment. By convoluting the probability densities for the holon and the spinon, we obtain the density distribution of the hole, which can be directly measured experimentally. Even though the spinon dynamics are slow compared to the holon motion at strong coupling, they determine the hole distribution at long times because the holon is bound to the spinon. We consider different non-equilibrium situations which can all be realized in current experiments with ultracold atoms. 

\begin{figure}[t!]
\centering
\epsfig{file=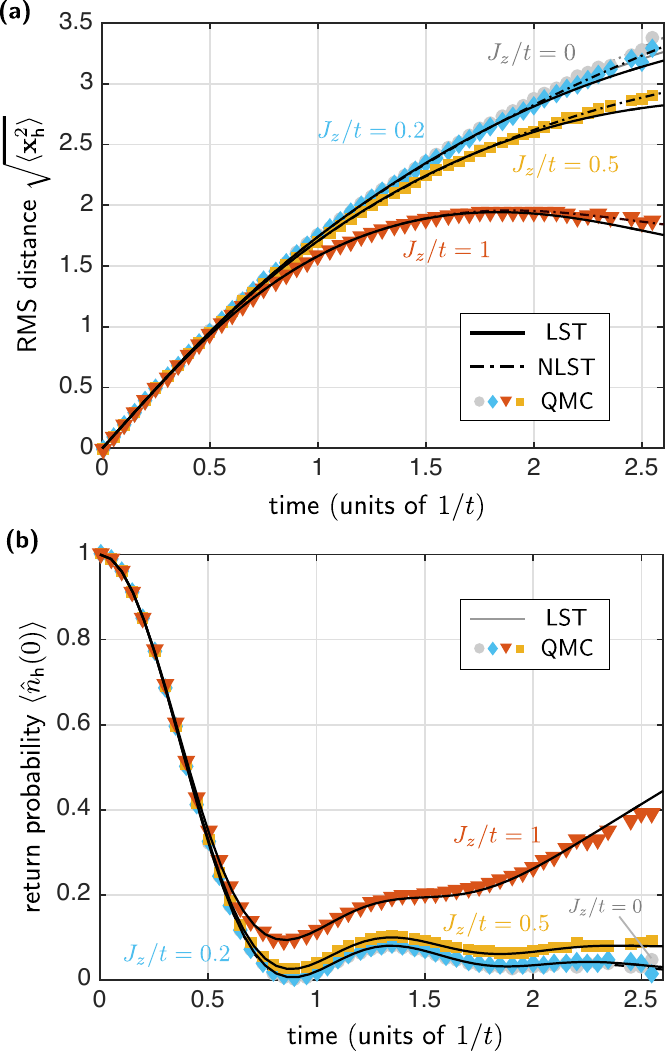, width=0.45\textwidth} $~$
\caption{We benchmark the strong coupling parton theory of magnetic polarons by comparing to numerically exact time-dependent quantum Monte Carlo (QMC) calculations. A quantum quench in the $t-J_z$ model with spin $S=1/2$ is considered, where initially a single hole is created in a N\'eel state on a square lattice by removing the spin on the central site. We calculate the root mean square distance, see Eq.~\eqref{eq:defRMS}, of the hole from the origin (a) and the return probability of the hole (b). For all values of $J_z/t$ and for times accessible by the time-dependent Monte Carlo approach we obtain excellent agreement with calculations based on the linear string theory (LST) described in this paper. Non-linear string theory (NLST) predicts only small corrections and agrees with the QMC results slightly better at intermediate times. For details on the QMC method, see Appendix \ref{apdx:TdepQMC}. We compare results from NLST and LST at longer times in Appendix \ref{apdx:NLSTvsLST}.}
\label{fig:QMC}
\end{figure}

\emph{Benchmark.--}
We benchmark our parton theory by comparing to time-dependent quantum Monte Carlo calculations \cite{Carlstrom2016PRL,Nagy2017PRB} of the hole dynamics in the two-dimensional $t-J_z$ model. To this end we study the far-from equilibrium dynamics of a hole which is initialized in the system by removing the central spin from the N\'eel state. A brief summary of the quantum Monte Carlo method used for solving this problem can be found in Appendix \ref{apdx:TdepQMC}.

In Fig.~\ref{fig:QMC} we show our results for the return probability $\langle \hat{n}_{\rm h}(0) \rangle$ and the extent of the hole wave function $\sqrt{\langle  \hat{\vec{x}}_{\rm h}^2 \rangle}$, for times accessible with our quantum Monte Carlo method. Here $\hat{n}_{\rm h}(\vec{x})$ is the density operator of the hole at site $\vec{x}$, and $\hat{\vec{x}}_{\rm h}$ is the position operator of the hole in first quantization. For all considered values of $J_z$ we obtain excellent agreement of the strong coupling parton theory with numerically exact Monte Carlo results. 

\emph{Pre-spin-charge separation.--}
For the largest considered value of $J_z=t$ we observe a pronounced slow-down of the hole expansion in Fig.~\ref{fig:QMC}. This is due to the restoring force mediated by the string which connects spinon and holon. However, the expansion does not stop completely \cite{Trugman1988,Chernyshev1999}. Instead it becomes dominated by slow spinon dynamics at longer times, as we show by an explicit calculation in Sec.~\ref{subsec:FarFromEqul} and Fig.~\ref{fig:HolonRelease}.

For smaller values of $J_z/t$, the hole expansion slows down at later times before it becomes dominated by spinon dynamics. In the strong coupling regime, $t \gg J_z$, we obtain a large separation of spinon and holon time scales. This can be understood as a pre-cursor of spin-charge separation: although the holon is bound to the spinon, it explores its Hilbert space defined by the Bethe lattice independently of the spinon dynamics. As illustrated in Fig.~\ref{fig:Overview} (b) the entire holon Hilbert space is co-moving with the spinon. This is a direct indicator for the parton nature of magnetic polarons.

At short-to-intermediate times the separation of spinon and holon energies gives rise to universal holon dynamics. Indeed, the expansion observed in Fig.~\ref{fig:QMC} at strong coupling $t \gg J_z$ is similar to the case of hole propagation in a spin environment at infinite temperature \cite{Carlstrom2016PRL}. In that case an approximate mapping to the holon motion on the Bethe lattice is possible too \cite{Nagy2017PRB}. 

\emph{Coherent spinon dynamics.--}
We consider a situation starting from a spinon-holon pair in its ro-vibrational ground state. In contrast to the far-from equilibrium dynamics discussed above, the holon is initially distributed over the Bethe lattice in this case. We still start from a state where the spinon is localized in the center of the system. In this case there exist no holon dynamics on the Bethe lattice within the strong coupling approximation, and a measurement of the hole distribution allows to directly observe coherent spinon dynamics. 

We also present an adiabatic preparation scheme for the initial state described above, where the magnetic polaron is in its ro-vibrational ground state. The scheme can be implemented in experiments with ultracold atoms. The general strategy is to first localize the hole on a given lattice site by a strong pinning potential. By slowly lowering the strength of this potential, the ro-vibrational ground state can be prepared with large fidelity, as we demonstrate in Fig.~\ref{fig:MagPolPrep}. Details are discussed in Sec.~\ref{subsec:SpinonDynamics}. 

\emph{Spectroscopy of ro-vibrational excitations.--}
To test the strong coupling parton theory experimentally, we suggest measuring the energies of rotational and vibrational eigenstates of the spinon-holon bound state directly. In Sec.~\ref{subsec:BreathingOscillations} we demonstrate that rotational states can be excited by applying a weak force to the system. As before, we start from the ro-vibrational ground state of the magnetic polaron. The force induces oscillations of the density distribution of the hole, which can be directly measured in a quantum gas microscope. We demonstrate that the frequency of such oscillations is given very accurately by the energy of the first excited state, which has a non-trivial rotational quantum number. This is another indicator for the parton nature of magnetic polarons.

Vibrational excitations can be directly observed in the spectral function \cite{Mishchenko2001}. In Sec.~\ref{subsec:Spectroscopy} we briefly explain its properties in the strong coupling regime. Possible measurements with ultracold atoms are also discussed.

\subsection{Outline}
\label{subsecOrganization}
This paper is organized as follows. In Sec.~\ref{sec:SpinonHolonModel} we introduce the microscopic parton theory describing holes in an anti-ferromagnet in the strong coupling regime, starting from first principles. We solve the effective Hamiltonian in Sec.~\ref{secStringExcitations} and derive the rotational and vibrational excited states of magnetic polarons. Direct signatures in the string-length distribution and its measurement in a quantum gas microscope are also discussed. Sec.~\ref{sec:TrugmanDisp} is devoted to a discussion of the effective spinon dispersion relation. We derive and benchmark a semi-analytical tight-binding approach to describe the effects of Trugman loops, the fundamental processes underlying spinon dynamics in the $t-J_z$ model. In Sec.~\ref{sec:DynProp} we apply the strong coupling parton theory to solve different problems involving magnetic polaron dynamics, which can be realized in experiments with ultracold atoms. Extensions of the parton theory to the $t-J$ model are discussed in Sec.~\ref{sectJoutlook}. We close with a summary and by giving an outlook in Sec.~\ref{sec:Summary}.

\section{Microscopic parton theory of magnetic polarons}
\label{sec:SpinonHolonModel}
In this section we introduce the strong coupling parton theory of holes in the $t-J_z$ model, which builds upon earlier work on the string picture of magnetic polarons \cite{Bulaevskii1968,Brinkman1970,Trugman1988,Manousakis2007}. After introducing the model in \ref{subsec:Model} and explaining our formalism in \ref{subsec:LargeSholehopping} we derive the parton construction in Sec.~\ref{subsec:SpinonHolon}.

\subsection{The $t-J_z$ model}
\label{subsec:Model}
For a single hole, $\sum_{j,\sigma} \cd_{j,\sigma} \c_{j,\sigma} = N-1$, where $N$ is the number of lattice sites, the $t-J_z$ Hamiltonian can be written as
\begin{multline}
\H_{t-J_z} = \H_J + \H_t = \sum_\ij  J_z \hat{S}^z_i \hat{S}^z_j +\\
+ \hat{\mathcal{P}} \biggl[ - t  \sum_{\ij,\sigma} \l \cd_{i,\sigma} \c_{j,\sigma} + \hc \r \biggr] \hat{\mathcal{P}}.
\label{eq:model}
\end{multline}
Here $\cd_{j,\sigma}$ creates a boson or a fermion with spin $\sigma = \uparrow, \downarrow$ on site $j$ and $\hat{\mathcal{P}}$ projects onto the subspace without double occupancies. $\sum_\ij$ denotes a sum over all bonds $\ij$ between neighboring sites $i$ and $j$, where every bond is counted once. The spin operators are defined by $\hat{S}^z_j = \sum_{\sigma,\tau} \cd_{j,\sigma} \sigma^z_{\sigma,\tau} \c_{j,\tau} /2$. 

The second line of Eq.~\eqref{eq:model} describes the hopping of the hole with amplitude $t$ and the first line corresponds to Ising interactions between the spins. See e.g. Ref.~\cite{Chernyshev1999} and references therein for previous studies of the $t-J_z$ model. Experimental implementations of the $t-J_z$ Hamiltonian were discussed in Sec.~\ref{subsubsecImplmttJz}. From now on we consider the case when $\cd_{j,\sigma}$ describes a fermion for concreteness, but as long as a single hole is considered the physics is identical if bosons were chosen.

\subsubsection{Schwinger-boson representation and constraint}
\label{subsubsec:SchwingerBosonRep}
For our discussion of the $t-J_z$ model \eqref{eq:model} we find it convenient to choose a parameterization in terms of Schwinger-bosons $\b_{j \sigma}$ and spinless fermionic holon operators $\h_i$ satisfying the following constraint, 
\begin{equation}
\sum_\sigma \bd_{j \sigma} \b_{j \sigma} = 2 S \bigl( 1 - \hd_j \h_j \bigr)  \qquad \forall j.
\label{eq:constraintTJ}
\end{equation}
In the original Hamiltonian from Eq.~\eqref{eq:model}, the length of the spins is $S=1/2$ and Eq.~\eqref{eq:constraintTJ} is equivalent to the condition of no double occupancy of lattice sites. More generally, for $S \geq 1/2$, Eq.~\eqref{eq:constraintTJ} ensures that a given lattice site is either occupied by a single holon and no Schwinger-bosons or no holon but exactly $2S$ Schwinger-bosons.

From now on we will consider more general models with arbitrary integer or half-integer values of $S$. This approach is similar to the usual $1/S$-expansion of the $t-J$ and related Hamiltonians, see e.g. Ref.~\cite{Auerbach1998}, except that in the latter a different constraint is used: $\sum_\sigma \bd_{j \sigma} \b_{j \sigma} + \hd_j \h_j = 2 S$. For $S=1/2$ the two constraints are identical, and we discuss in Appendix \ref{apdxQuantFlucGenS} how they differ for larger $S > 1/2$. In both cases, the spin operators are
\begin{flalign}
\hat{S}^z_j = \frac{1}{2} \l \bd_{j \uparrow} \b_{j \uparrow} - \bd_{j \downarrow} \b_{j \downarrow} \r, \quad \\
\hat{S}_j^+ = \bd_{j \uparrow} \b_{j \downarrow}, \qquad \hat{S}_j^- = \bd_{j \downarrow} \b_{j \uparrow}.
\end{flalign}

In terms of holons and Schwinger-bosons, the second term in the $t-J_z$ Hamiltonian becomes
\begin{equation}
\H_t = t \sum_\ij \l  \hd_i \h_j \hat{\mathcal{F}}^\dagger_{ij}(S)  + \hc \r,
\label{eq:HtSB}
\end{equation}
where $\hat{\mathcal{F}}^\dagger_{ij}(S)$ involves only Schwinger-bosons and will be explained shortly. Note that here, in contrast to Eq.~\eqref{eq:model}, the hopping rate $t$ comes with a positive sign because holon creation corresponds to fermion annihilation, $\cd_j \c_i  \to  \h_j \hd_i = - \hd_i \h_j$.

The term $\hat{\mathcal{F}}^\dagger_{ij}(S)$ describes the re-ordering of the spins during the hopping of the holon. This is required by the Schwinger-boson constraint in Eq.~\eqref{eq:constraintTJ}. In particular, $\hat{\mathcal{F}}_{ij}(S)$ describes how the spin state on site $j$ is moved to site $i$ while the hole is hopping from $i$ to $j$. In this work, we are interested in the case $S=1/2$, where $\hat{\mathcal{F}}_{ij}(S)$ becomes \cite{Auerbach1998}
\begin{equation}
\hat{\mathcal{F}}_{ij}(1/2) = \bd_{i \uparrow} \b_{j \uparrow} +  \bd_{i \downarrow} \b_{j \downarrow},
\label{eq:Fij12}
\end{equation}
see Fig.~\ref{fig:largeSholeHopping} (a). Because the constraint Eq.~\eqref{eq:constraintTJ} is fulfilled, the projectors $\hat{\mathcal{P}}$ from Eq.~\eqref{eq:model} can be dropped in the Schwinger-boson representation.

\subsection{Generalized $1/S$ expansion}
\label{subsec:LargeSholehopping}
In this section we introduce a generalization of our model to large spins, $S > 1/2$. Technically we only achieve a re-formulation of the original Hamiltonian, and in the case of the $t-J_z$ model no real progress is made. However, as discussed in more detail in Appendix \ref{apdxQuantFlucGenS}, the formalism developed in this section can be straightforwardly generalized to include quantum fluctuations. In such cases, the leading order generalized $1/S$ expansion represents a $t-J_z$ model again, and the parton theory which we develop in Sec.~\ref{subsec:SpinonHolon} can be applied here, too. This establishes our parton construction as a valuable starting point for analyzing a larger class of models. In a forthcoming work we apply the generalized $1/S$ expansion introduced below to situations including quantum fluctuations in the spin environment described by a general XXZ Hamiltonian \cite{Huse1988,Tamaribuchi1991}. 

\begin{figure}[t!]
\centering
\epsfig{file=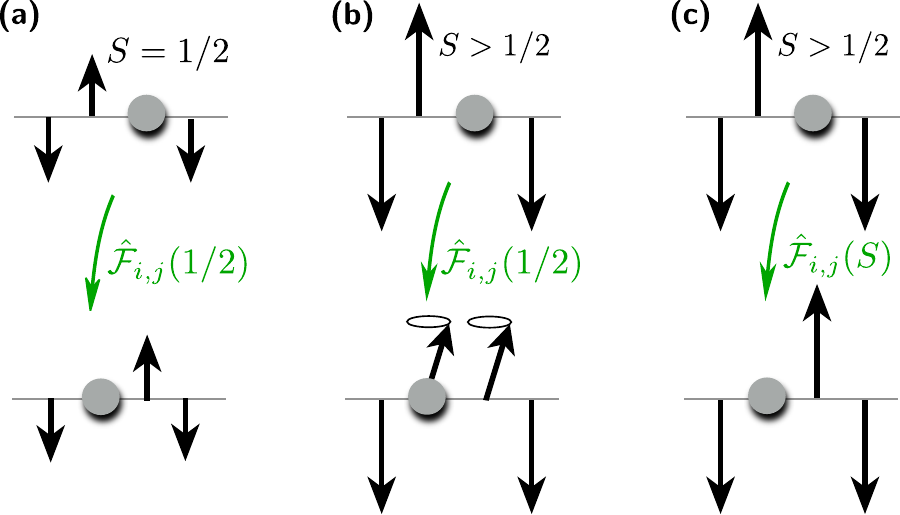, width=0.41\textwidth} $\quad$
\caption{Illustration of hole hopping in a large-$S$ spin environment along one direction. (a) For $S=1/2$ the hole distorts the N\'eel order of the AFM spin environment. (b) For large $S>1/2$, but using the spin-$1/2$ hole hopping operator $\hat{\mathcal{F}}_{ij}(1/2)$, the N\'eel order parameter is not distorted but spin excitations (magnons) are created on top of the AFM by the movement of the hole. (c) For large $S$ the hole hopping described by $\hat{\mathcal{F}}_{ij}(S)$ exchanges the spin states on neighboring sites and fully captures the distortion of the N\'eel order around the hole.}
\label{fig:largeSholeHopping}
\end{figure}

In the conventional $1/S$ expansion, the holon motion is described by the operator $\hat{\mathcal{F}}_{ij}(S) \equiv \hat{\mathcal{F}}_{ij}(1/2)$ from Eq.~\eqref{eq:Fij12} and $S$ only enters in the corresponding Schwinger-boson constraint, $\sum_\sigma \bd_{j \sigma} \b_{j \sigma} + \hd_j \h_j = 2 S$. As we explain in detail in Appendix \ref{subsubsecShortComingConventionalS}, this approach cannot capture strong distortions of the local N\'eel order parameter, or the local staggered magnetization, defined by
\begin{equation}
\hat{\Omega}_{j} = (-1)^j \hat{S}^z_j.
\label{eqDefOmgaj}
\end{equation}
Here $(-1)^j$ denotes the sublattice parity, which is $+1$ ($-1$) for $j$ from the A (B) sublattice. Within the conventional extension of the $t-J_z$ model to large values of $S$, the motion of the holon from site $i$ to $j$ is accompanied by changes of the spins $S^z_{i}$ and $S^z_{j}$ by $\pm 1/2$, see Fig.~\ref{fig:largeSholeHopping} (b). As a result, the sign of the local N\'eel order parameter $\Omega_j$ cannot change when $S \gg 1/2$ is large, unless the holon performs multiple loops. 

To avoid these problems of the conventional $1/S$ expansion, and to ensure that the generalized Schwinger-boson constraint Eq.~\eqref{eq:constraintTJ} is satisfied by $\H_t$ in Eq.~\eqref{eq:HtSB}, we replace $\hat{\mathcal{F}}_{ij}(1/2)$ from Eq.~\eqref{eq:Fij12} by a new term $\hat{\mathcal{F}}_{ij}(S)$. As we explain in detail in Appendix \ref{subsubsec:LargeSholonHopping}, this \emph{generalized holon-hopping} operator $\hat{\mathcal{F}}_{ij}(S)$ describes a transfer of the entire spin state from site $j$ to $i$, see also Fig.~\ref{fig:largeSholeHopping} (c).

\subsubsection{Formalism: Ising variables and the distortion field}
\label{subsubsecFormalismDistortion}
Now we introduce some additional formalism which is useful for the formulation of the microscopic parton theory. Our discussion is kept general and applies to arbitrary values of the spin length $S$. 

\emph{Zero doping.--}
To describe the orientation of the local spin of length $S$, we introduce an Ising variable $\tilde{\tau}^z_j$ on the sites of the square lattice,
\begin{equation}
\tilde{\tau}^z_j = \begin{cases}
+1, \quad \ket{S}_j \\
-1, \quad \ket{-S}_j
\end{cases}.
\end{equation}
For $S=1/2$ the Ising variable $\tilde{\tau}^z_j = 2 \hat{S}^z_j$ is identical to the local magnetization. For $S>1/2$ the situation is different because $\tilde{\tau}^z_j = \pm 1$ can still only take two possible values, whereas $\hat{S}^z_j = -S,-S+1,...,S$. 

The classical N\'eel state with AFM ordering along the $z$-direction corresponds to a configuration where $\tilde{\tau}^z_j=+1$ on the $A$-sublattice, and $\tilde{\tau}^z_j=-1$ on the $B$-sublattice. To take the different signs into account, we define another Ising variable $\hat{\tau}^z_j$ describing the staggered magnetization,
\begin{equation}
\hat{\tau}^z_j = \begin{cases}
\tilde{\tau}^z_j, \quad j \in A \\
-\tilde{\tau}^z_j, \quad j \in B
\end{cases}.
\end{equation}
The N\'eel state corresponds to the configuration \cite{Note1} 
\begin{equation}
\hat{\tau}^z_j \equiv 1 \qquad \text{for all} ~ j.
\end{equation} 

\emph{Doping.--}
Now we consider a systems with one hole. Our general goal in this section is to construct a complete set of one-hole basis states. This can be done starting from the classical N\'eel state by first removing a spin of length $S$ on site $j$ and next allowing for distortions of the surrounding spins. In this process, we assign the value of $\hat{\tau}_j^z=1$ to the lattice site where the hole was created. Note that this value is associated with a sublattice index of the hole and it reflects the spin $\sigma$ which was initially removed when creating the hole.

\emph{Distortions.--}
The holon motion, described by $\H_t$, introduces distortions into the classical N\'eel state. Using the staggered Ising variable $\hat{\tau}_j^z$ they correspond to sites with $\hat{\tau}_j^z=-1$. We will now show that the $t-J_z$ Hamiltonian can be expressed entirely in terms of the product defined on links,
\begin{equation}
\hat{\sigma}^z_{\ij} = \hat{\tau}^z_j \hat{\tau}^z_i,
\label{eqDefSigmaij}
\end{equation}
which will be referred to as the \emph{distortion field}. On bonds with $\hat{\sigma}^z_{\ij}=1$ (respectively $\hat{\sigma}^z_{\ij}=-1$) the spins are anti-aligned (aligned), see Fig.~\ref{fig:Overview} (a) for an illustration. 

\emph{Effective Hamiltonian.--}
The term $\H_J$ in the $t-J_z$ Hamiltonian Eq.~\eqref{eq:model} can be re-formulated as
\begin{equation}
\H_J = -2 N J_z S^2 + J_z S^2 \sum_\ij \l 1 - \hat{\sigma}^z_\ij \r.
\label{eqHJdistortion}
\end{equation}
The first term corresponds to the ground state energy of the undistorted N\'eel state, where $N$ is the number of lattice sites. The second term describes the energy cost of creating distortions. In Appendix \ref{apdxQuantFlucGenS} we explain how quantum fluctuations can be included within the generalized $1/S$ expansion.

\begin{figure}[t!]
\centering
\epsfig{file=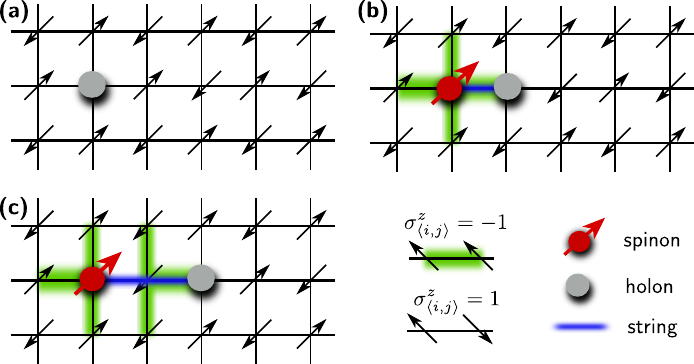, width=0.46\textwidth} $\quad$
\caption{When a hole is moving in a spin state with AFM order, it distorts the N\'eel order (green) and creates a string of displaced spins (blue), (a)-(c). In the formalism used for the generalized $1/S$ expansion this is described by the coupling of the holon to the distortion field $\sigma^z_\ij$ on the bonds. The end of the string (red) can be associated with a parton, the spinon, which carries the spin $S$ of the magnetic polaron. Its charge is carried by another parton, the holon (gray).}
\label{fig:holeHopping}
\end{figure}

The term $\H_t$ in the $t-J_z$ Hamiltonian Eq.~\eqref{eq:model} can also be formulated in terms of the distortion field. Consider the motion of the holon from site $i$ to $j$. This corresponds to a movement of the spin on site $j$ to site $i$, which changes the distortion field $\hat{\sigma}^z_{l}$ on links $l=\langle r,k \rangle$ including sites $r=i$ and $r=j$. Such changes depend on the original orientation of the involved spins on sites $k$ and can be described by the operators $\hat{\sigma}^x_{l}$. We obtain the expression 
\begin{multline}
\H_t =  t \sum_\ij \hd_j \h_i   \biggl[ \prod_{\langle k,i \rangle} \hat{\sigma}^x_{\langle i,k \rangle} \frac{1}{2} \l 1 + \hat{\sigma}^z_\ij \r \\
+ \prod_{\langle k,j \rangle} \hat{\sigma}^x_{\langle j,k \rangle} \frac{1}{2} \l 1 - \hat{\sigma}^z_\ij \r  \biggr] + \hc.
\label{eq:Ht0}
\end{multline}
Because $\hat{\sigma}^x_\ij$ and $\hat{\sigma}^z_\ij$ are not commuting, the distortion field $\hat{\sigma}^z_\ij$ begins to fluctuate in the presence of the mobile hole with $t \neq 0$. The effect of the hole hopping term \eqref{eq:Ht0} is illustrated in Fig.~\ref{fig:holeHopping}. 

By combining Eqs.~\eqref{eqHJdistortion} and \eqref{eq:Ht0} we obtain an alternative formulation of the single-hole $t-J_z$ model at $S=1/2$. Even when $S>1/2$, the leading order result in the generalized $1/S$ expansion is a similar effective $S=1/2$ Hamiltonian, formulated in terms of the same distortion field. This method goes beyond the conventional $1/S$ expansion, where the distortion field $\hat{\sigma}^z_{\ij} \equiv 1$ is kept fixed on all bonds and does not fluctuate. To describe the distortion of the N\'eel state introduced by the holon motion in the conventional $1/S$ expansion, one has to resort to magnon fluctuations on top of the undisturbed N\'eel state, which only represent sub-leading corrections in the generalized $1/S$ formalism. This property of the generalized $1/S$ expansion makes it much more amenable for an analytical description of the strong coupling regime where $t \gg J$ and the N\'eel state can be substantially distorted even by a single hole.

\subsection{The spinon-holon picture and string theory}
\label{subsec:SpinonHolon}
So far we have formulated the $t-J_z$ model using two fields, whose interplay determines the physics of magnetic polarons: The holon operator $\h_j$ and the distortion field $\hat{\vec{\sigma}}_\ij$ on the bonds. By introducing magnons $\a_j$, more general models with quantum fluctuations can also be considered, but such terms are absent in the $t-J_z$ case. Our goal in this section is to replace the distortion field $\hat{\vec{\sigma}}_\ij$ by a simpler description of the magnetic polaron, which is achieved by introducing partons.

\subsubsection{Spinons and holons}
The Hamiltonian \eqref{eq:Ht0} describing the motion of the holon in the distorted N\'eel state determined by $\hat{\vec{\sigma}}_\ij$, is highly non-linear. To gain further insights, we study more closely how the distortion field $\hat{\sigma}_\ij^z$ is modified by the holon motion. In particular we will argue that it carries a well-defined spin quantum number.

\emph{Quantum numbers.--}
Let us start from the classical N\'eel state and create a hole by removing the spin on the central site of the lattice. This changes the total charge $Q$ and spin $S^z$ of the system by $\Delta Q = -1$ and $\Delta S^z=\pm S$, where the sign of $\Delta S^z$ depends on the sublattice index of the central site. When the hole is moving, both $S^z$ and $Q$ are conserved and we conclude that the magnetic polaron (mp) carries spin $S_{\rm mp}^z=\pm S$ and charge $Q_{\rm mp}=-1$. 

There exists no true spin-charge separation for a single hole in the 2D N\'eel state \cite{Bulaevskii1968,Trugman1988,Mishchenko2001}, i.e. the spin degree of freedom of the magnetic polaron cannot completely separate from the charge. We can understand this for the case $S=1/2$, where the magnetic polaron carries \emph{fractional} spin $S^z_{\rm mp} = \pm 1/2$. Because the elementary spin-wave excitations of the 2D anti-ferromagnet carry spin $S^z=\pm 1$, see e.g. Ref.~\cite{Auerbach1998}, they cannot change the fractional part $S^z_{\rm mp} {\rm mod}~1$ of the magnetic polaron's spin, which is therefore bound to the charge. This is in contrast to the 1D case, where fractional spinon excitations exist in the spin chain even at zero doping \cite{Giamarchi2003} and the hole separates into independent spinon and holon quasiparticles \cite{Kim1996,Segovia1999,Kim2006,Kruis2004,Kruis2004a,Hilker2017}. 

\emph{Partons.--}
Now we show that the magnetic polaron in a 2D N\'eel state can be understood as a bound state of two partons, the holon $\h_j$ carrying charge and a spinon $\s_i$ carrying spin. In the strong coupling regime, $t \gg J_z$, we predict a mesoscopic precursor of spin-charge separation: While the spinon and holon are always bound to each other, their separation can become rather large compared to the lattice constant, and they can be observed as two separate objects. Their bound state can be described efficiently by starting from two partons with an attractive interaction between them, similar to quarks forming a meson. When $J_z=0$ spinon-holon pairs can be completely separated \cite{Carlstrom2016PRL,Nagy2017PRB}. 

In contrast to the usual slave-fermion (or slave-boson) approach \cite{Wen2004}, we will not define the holon and spinon by breaking up the original fermions $\c_{j,\sigma}$ on site $j$. Instead we notice that the spin quantum number $S^z_{\rm mp}$ of the magnetic polaron is carried by the distortion field $\hat{\sigma}_\ij^z$. The latter determines the distribution of the spin on the square lattice. We have already introduced the spin-less holon operators $\h_j$ in Eq.~\eqref{eq:constraintTJ} by using a Schwinger-boson representation of the $t-J_z$ model.

\begin{figure}[b!]
\centering
\epsfig{file=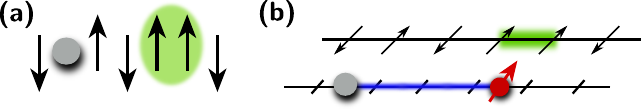, width=0.4\textwidth}
\caption{String-theory in the one-dimensional Ising model. (a) The spinon corresponds to a domain wall across which the direction of the N\'eel order changes sign. (b) The string-theory becomes exact in 1D. This picture is equivalent to the squeezed-space description of spin-charge separation \cite{Kruis2004}, where the holon occupies the bonds between neighboring sites of a spin chain, see also Ref.~\cite{Bohrdt2017spec}.}
\label{fig:1DStringTheory}
\end{figure}

\begin{figure*}[t!]
\centering
\epsfig{file=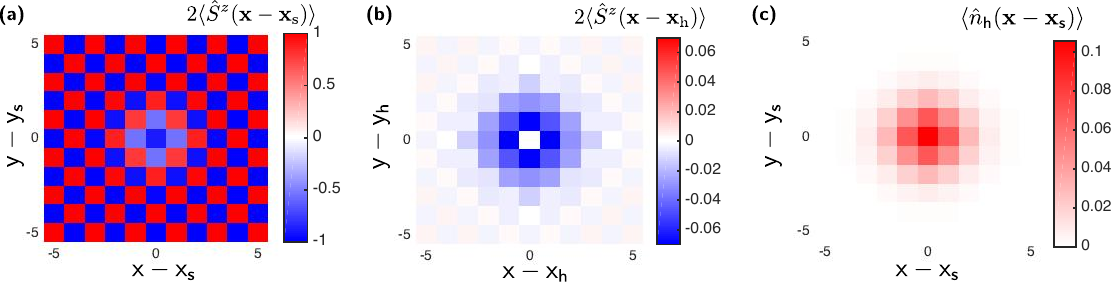, width=0.95\textwidth}
\caption{Partons forming the magnetic polaron: The spinon is defined as the end of the string of distorted spins created by the motion of the spin-less holon. Their position operators are denoted by $\mathbf{x}_{\rm s}$ and $\mathbf{x}_{\rm h}$, respectively. (a) The spinon corresponds to a localized magnetic moment, as can be seen from the magnetization $\langle \hat{S}^z(\vec{x}-\vec{x}_{\rm s}) \rangle$ calculated in the frame co-moving with the spinon. (b) In the frame co-moving with the holon, in contrast, the magnetization $\langle \hat{S}^z(\vec{x}-\vec{x}_{\rm h}) \rangle$ is extended over an area a few lattice sites wide. (c) The charge $\langle \hat{n}_{\rm h}(\vec{x} - \vec{x}_{\rm s}) \rangle$ calculated in the spinon frame is extended over a similar area around the spinon. We have performed calculations for the non-linear string theory of the $t-J_z$ model as described in the text, at $J_z=0.1 t$, $S=1/2$ and for a maximum string length $\ell_{\rm max} = 10$.}
\label{fig:SlaveParticles}
\end{figure*}

\emph{Spin and charge distribution.--}
To understand how well the spin $S^z_{\rm mp} = \pm 1/2$ of the magnetic polaron is localized on the square lattice, we study the motion of the holon described by Eq.~\eqref{eq:Ht0}. When the hole is moving it leaves behind a string of displaced spins \cite{Bulaevskii1968,Trugman1988}, see Fig.~\ref{fig:holeHopping}. At the end of the string which is not attached to the hole, we identify a site $i$ from which three excited bonds ($\hat{\sigma}^z_\ij=-1$) emerge, unless the hole returns to the origin. This corresponds to a surplus of spin on this site relative to the original N\'eel state, and we identify it with the location $\vec{x}_{\rm s}$ of the spinon. The spin $\sigma$ of the spinon is opposite for the two different sublattices. 

Our definition of spinons can be considered as a direct generalization of domain walls in the one-dimensional Ising model, see Fig.~\ref{fig:1DStringTheory} (a). In both cases, the fractional spin carried by the spinon is not strictly localized on one lattice site but extends over a small region around the assigned spinon position. To demonstrate this explicitly, we calculate the average magnetization $2 \langle \hat{S}^z(\vec{x} - \vec{x}_{\rm s}) \rangle$ in the spinon frame in Fig.~\ref{fig:SlaveParticles} (a). We use the $t-J_z$ model in the strong coupling regime with $t = 10 J_z$. We observe that the checkerboard pattern of the N\'eel order parameter is completely retained, except for a spin-flip in the center. This shows that the spin of the magnetic polaron $S^z_{\rm mp}$ is localized around the spinon position, i.e. at the end of the string defined by the holon trajectory. 

This result should be contrasted to the magnetization calculated in the holon frame, $2 \langle \hat{S}^z(\vec{x} - \vec{x}_{\rm h}) \rangle$ where $\vec{x}_{\rm h}$ is the holon position, shown in Fig.~\ref{fig:SlaveParticles} (b). In that case, the spin of the magnetic polaron is distributed over a wide area around the holon. The AFM checkerboard structure is almost completely suppressed because it is favorable for the holon to delocalize equally over both sublattices at this large value of $t/J_z = 10$. Similarly, the charge distribution $\langle \hat{n}_{\rm h}(\vec{x}-\vec{x}_{\rm s}) \rangle$ covers an extended area around the spinon, see Fig.~\ref{fig:SlaveParticles} (c).

\emph{Formal definition of spinons.--}
Formally we add an additional label $\sd_{i,\sigma} \ket{0}$ to the quantum states. Here $i$ denotes the site of the spinon as defined above, and the spin index $\sigma$ depends on the sublattice index of site $i$ and will be suppressed in the following. When holon trajectories are included which are not straight but return to the origin, this label is not always unique, see Ref.~\cite{Trugman1988} or Sec.~\ref{subsec:TrugmanLoops}. Thus, by adding the new spinon label to the wave function, we obtain an over-complete basis. We will deal with this issue later in Sec.~\ref{sec:TrugmanDisp} and argue that the use of the over-complete basis is a useful approach. 

We also note that the spinon label basically denotes the site where we initialize the hole and let it move around to construct the basis of the model. In a previous work by Manousakis \cite{Manousakis2007}, this site has been referred to as the ``birth" site without drawing a connection to the magnetization (the spin) of the magnetic polaron localized around this site, as shown in Fig.~\ref{fig:SlaveParticles} (a).

\subsubsection{String theory: an over-complete basis}
\label{subsubsec:StringTheory}
Our goal in this section is to describe the distortion field $\hat{\vec{\sigma}}_\ij$ by a conceptually simpler \emph{string} on the square lattice. This idea goes back to the works by Bulaevskii et al.~\cite{Bulaevskii1968} and Brinkman and Rice \cite{Brinkman1970}, as well as works by Trugman \cite{Trugman1988} and more recently by Manousakis \cite{Manousakis2007}. In Refs.~\cite{Trugman1988,Manousakis2007} a set of variational states was introduced, based on the intuition that the holon leaves behind a string of displaced spins, see Fig.~\ref{fig:holeHopping}.

\emph{Replacing the basis.--}
Within the approximations so far, the orthogonal basis states are labeled by the value of the distortion field $\hat{\sigma}^z_\ij$ on all bonds and the spinon and holon positions, 
\begin{equation}
 \sd_j \ket{0} ~\hd_i \ket{0} ~ \ket{ \{ \sigma^z_\ij \}_\ij }.
\label{eq:BasisStatesDistortionField}
\end{equation}
The string description can be obtained by replacing this basis by a closely related, but conceptually simpler set of basis states. 

When the holon propagates in the N\'eel state, starting from the spinon position, it modifies the distortion field $\hat{\sigma}^z_\ij$ differently depending on the trajectory $\Sigma$ it takes. Here we use the convention that \emph{trajectories} $\Sigma$ are defined only up to self-retracing components, in contrast to \emph{paths} which contain the complete information where the holon went. The holon motion thus creates a memory of its trajectory in the spin environment. 

Given a trajectory $\Sigma$ and the spinon position, we can easily determine the corresponding distortion field $\hat{\sigma}^z_\ij(\Sigma,\vec{x}_{\rm s})$. In the following we will assume that for all relevant quantum states, the opposite is also true. Namely, that given the distortion of the N\'eel state $\hat{\sigma}^z_\ij$, we can reconstruct the trajectory $\Sigma$ defined up to self-retracing components, as well as the spinon position. We will show that this is an excellent approximation. Using a quantum gas microscope this one-to-one correspondence can be used for accurate measurements of holon trajectories $\Sigma$ in the N\'eel state by imaging instantaneous spin and hole configurations. We analyze the efficiency of this mapping in detail in Sec.~\ref{subsecSpinReconstruction}.

In some cases our assumption is strictly correct, for example in the one-dimensional Ising model. In that case the spinon corresponds to a domain wall in the anti-ferromagnet, see Fig.~\ref{fig:1DStringTheory} (a). When its location is known, as well as the distance of the holon from the spinon (i.e. the trajectory $\Sigma$), the spin configuration $\hat{\sigma}_\ij^z$ can be reconstructed, see Fig.~\ref{fig:1DStringTheory} (b). A second example, which is experimentally relevant for ultracold atoms, involves a model where the hole can only propagate along one dimension inside a fully two-dimensional spin system \cite{GrusdtMixedDim2017}.  

For the fully two-dimensional magnetic polaron problem, there exist sets of different trajectories $\Sigma$ which give rise to the same spin configuration $\hat{\sigma}_\ij^z$. Trugman has shown \cite{Trugman1988} that the leading-order cases correspond to situations where the holon performs two steps less than two complete loops around an enclosing area, see Fig.~\ref{fig:TrugmanLoops} (a). Choosing a single plaquette, this requires a minimum of six hops of the holon before two states cannot be distinguished by the corresponding holon trajectory anymore. 

By performing six steps, a large number of states can be reached in principle: In the first step, starting from the spinon, there are four possible directions which the holon can choose, followed by three possibilities for each of the next five steps. This makes a total of $4 \times 3^5 = 972$ states over which the holon tends to delocalize in order to minimize its kinetic energy. In contrast, there are only eight distinct Trugman loops involving six steps which lead to spin configurations that cannot be uniquely assigned to a simple holon trajectory.

In the following we will use an over-complete set of basis states, labeled by the spinon position and the holon trajectory $\Sigma$,
\begin{equation}
 \sd_j \ket{0} ~ \hd_i \ket{0} ~\ket{\Sigma}.
 \label{eqBasisSpinonHolon}
\end{equation}
$\Sigma$ will be referred to as the \emph{string} which connects the spinon and the holon. We emphasize that the string $\Sigma$ is always defined only up to self-retracing components; i.e. two paths $p_{1,2}$ taken by the holon correspond to the same string $\Sigma$ if $p_1$ can be obtained from $p_2$ by eliminating self-retracing components. The distortion field $\hat{\sigma}^z_\ij(\Sigma)$ is uniquely determined by the string configuration and no longer appears as a label of the basis states. Note that two inequivalent states in the over-complete basis can be identified with the same physical state if their holon positions as well as the corresponding distortion fields $\hat{\sigma}^z_\ij(\Sigma)$ coincide.

Geometrically, the over-complete space \cite{Trugman1988} of all strings $\Sigma$ starting from one given spinon position corresponds to the fractal Bethe lattice. The latter is identical to the tree defined by all possible holon trajectories without self-retracing components. In Fig.~\ref{fig:Overview} (b) this correspondence is illustrated for a simple trajectory taken by the holon. When $r=r(\Sigma)$ denotes the site on the Bethe lattice defined with the spinon in its origin and $\hd(r)$ creates the holon in this state, we can formally write the basis states as
\begin{equation}
 \sd_j \ket{0} ~ \hd_i \ket{0} ~\ket{\Sigma} =  \sd_j \ket{0} ~ \hd(r) \ket{0}.
  \label{eq:BasisStringTheory}
\end{equation}

\emph{Linear string theory.--} 
Next we derive the effective Hamiltonian of the system using the new basis states \eqref{eq:BasisStringTheory}. From Eq.~\eqref{eq:Ht0} we obtain an effective hopping term of the holon on the Bethe lattice,
\begin{equation}
\H_t = t \sum_{\langle r,s \rangle \in {\rm BL}} \hd(r) \h(s) + \hc,
\label{eq:HtBethe}
\end{equation}
where $\langle r,s \rangle \in {\rm BL}$ denotes neighboring sites on the Bethe lattice. This reflects the fact that the system keeps a memory of the holon trajectory.

The spin Hamiltonian Eq.~\eqref{eq:HheisenbergFluc} without magnons can be analyzed by first considering straight strings. Their energy increases linearly with their length $\ell$ with a coefficient $4 J_z S^2 \ell$. To obtain the correct energy of the distorted state, we have to include the zero-point energies $4 J_z S^2$ of the holon and $2 J_z S^2$ of the spinon, both measured relative to the energy $E_0^{\rm cl} = -2 N J_z S^2$ of the classical N\'eel state. Because the state with zero string length $\ell=0$ has energy $4 J_z S^2$, which is $2 J_z S^2$ smaller than the sum of holon and spinon zero-point energies, we obtain a point-like spinon-holon attraction. 

The resulting Hamiltonian reads
\begin{multline}
\H_{\rm LST} =\H_t + 4 J_z S^2 \sum_{r \in {\rm BL}} \left[ 1 + \ell(r) \right] \hd(r) \h(r)\\
+ 2 J_z S^2 \sum_j \sd_j \s_j - 2 J_z S^2 \hd(0) \h(0).
\label{eq:HLST}
\end{multline}
Here $\ell(r)$ denotes the length of the string defined by site $r$ on the Bethe lattice and $\hd(0)$ creates a string with length zero. Because we neglect self-interactions of the string which can arise for configurations where the string is not a straight line, we refer to Eq.~\eqref{eq:HLST} as \emph{linear string theory} (LST). Note that we have written Eq.~\eqref{eq:HLST} in second quantization for convenience. It should be noted however, that the spinon and the holon can only exist together and a state with only one of them is not a well defined physical state. Later we will include additional terms describing spinon dynamics.

\emph{Non-linear string theory.--}
When the length $\ell$ of the string is sufficiently large, it can start to interact with itself. For example, when a string winds around a loop or crosses its own path, the energy of the resulting state becomes smaller than the value $4 J_z S^2 \ell$ used in LST. We can easily extend the effective Hamiltonian from Eq.~\eqref{eq:HLST} by taking into account self-interactions of the string. If $\H_J$ denotes the potential energy of the spin configuration, determined from Eq.~\eqref{eq:HheisenbergFluc} by using $\hat{\sigma}^z_\ij(\Sigma)$, we can formally write:
\begin{equation}
\H_{\rm NLST} =\H_t + \sideset{}{'} \sum_\Sigma \H_J(\sigma^z_\ij(\Sigma))  ~\ket{\Sigma} \bra{\Sigma}.
\label{eq:HNLST}
\end{equation}

The sum in Eq.~\eqref{eq:HNLST}, denoted with a prime, has to be performed over all string configurations $\Sigma$ which do not include Trugman loops. This is required to avoid getting a highly degenerate ground state manifold, because Trugman loop configurations correspond to strings with zero potential energy. Such states are parametrized by different spinon positions in our over-complete basis from Eq.~\eqref{eq:BasisStringTheory}. As will be discussed in detail in Sec.~\ref{sec:TrugmanDisp}, Trugman loops give rise to spinon dynamics, i.e. they induce changes of the spinon position. Their kinetic energy lifts the large degeneracy in the ground state of the potential energy operator. 

If, on the other hand, we remove the spinon label from the basis states in Eq.~\eqref{eq:BasisStringTheory} and exclude all string configurations with loops leading to double-counting of physical states in the basis, Eq.~\eqref{eq:HNLST} corresponds to an exact representation of the single-hole $t-J_z$ model. By removing only the shortest Trugman loops and states with zero potential energy while allowing for spinon dynamics, one obtains a good truncated basis for solving the $t-J_z$ model.

Using exact numerical diagonalization, the spectrum of $\H_{\rm NLST}$ can be easily obtained. Because of the potential energy cost of creating long strings, the holon and the spinon are always bound, see for example Fig.~\ref{fig:SlaveParticles} (b) and (c). In Sec.~\ref{secStringExcitations} we will discuss their excitation spectrum and the resulting different magnetic polaron states.

\subsubsection{Parton confinement and relation to lattice gauge theory}
\label{subsubsecConfinementAndLGT}
Finally we comment on the definition of partons in our work and in the context of lattice gauge theories. In the latter case, one usually defines a gauge field on the links of the lattice which couples to the charges carried by the partons. Hence the partons interact via the gauge field, and the question whether they are confined or not becomes a question about the gauge fluctuations \cite{Wilson1974,Kogut1979}. 

In our parton construction so far, we have not specified the gauge field, and partons interact via the string connecting them. Because the string is defined by displaced spins, it can be directly measured, see also Sec.~\ref{subsecSpinReconstruction}, and thus represents a gauge-invariant quantity. An interesting question, which we devote to future research however, is whether a lattice gauge theory can be constructed which has a gauge-invariant field strength corresponding to the string. This would allow to establish even more direct analogies between partons in high-energy physics and holes in the $t-J$ or $t-J_z$ model.

\subsection{Strong coupling wavefunction}
\label{subsec:StrongCouplingWvfct}
Because of the single-occupancy constraint enforcing either one spin or one hole per lattice site, see Eq.~\eqref{eq:constraintTJ}, the spin and charge sectors are strongly correlated in the original $t-J_z$ Hamiltonian. Even when a separation of timescales exists, as provided by the condition $t \gg J_z$, no strong-coupling expansion is known for conventional approaches developed to describe magnetic polarons, e.g. for the usual $1/S$ expansion \cite{SchmittRink1988}. This is in contrast to conventional polaron problems with density-density interactions, where strong coupling approximations can provide important analytical insights \cite{Landau1946,Landau1948,Feynman1955,Devreese2013}. 

In the effective parton theory, the holon motion can be described by a single particle hopping on the fractal Bethe lattice. This already builds strong correlations between the holon and the surrounding spins into the formalism. Because the characteristic spinon and holon time scales are given by $1/J_z$ and $1/t$ respectively, the magnetic polaron can be described within the Born-Oppenheimer approximation at strong couplings. This corresponds to using an ansatz wavefunction of the form
\begin{equation}
\ket{\psi_{\rm mag.pol.}} = \ket{\psi_{\rm spinon}} \otimes \ket{\psi_{\rm holon}}.
\label{eqStrongCplgWvfct}
\end{equation}
We can solve the fast holon dynamics for a static spinon, and derive an effective low-energy Hamiltonian for the spinon dressed by the holon afterwards. We will make use of this strong-coupling approach throughout the following sections.

\section{String excitations}
\label{secStringExcitations}
New insights about the magnetic polaron can be obtained from the simplified LST Hamiltonian in Eq.~\eqref{eq:HLST} by making use of its symmetries. Because the potential energy grows linearly with the distance between holon and spinon, they are strongly bound, i.e. the spinon and holon form a confined pair. The bound state can be calculated easily by mapping the LST to an effective one-dimensional problem, see Ref.~\cite{Bulaevskii1968}. After providing a brief review of this mapping, we generalize it to calculate the full excitation spectrum of magnetic polarons including rotational states. We check the validity of the effective LST by comparing our results to numerical calculations using NLST. 

\begin{figure}[b!]
\centering
\epsfig{file=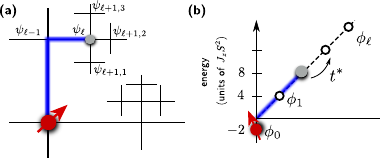, width=0.45\textwidth}
\caption{Symmetric holon states on the Bethe lattice. (a) For rotationally invariant holon eigenstates on the Bethe lattice, the wave function $\psi_{\ell,s}$ does not depend on the angular variable $s$. (b) Such states can be mapped to eigenstates $\phi_\ell$ on a semi-infinite one-dimensional lattice, with renormalized hopping strength $t^*$ and a potential linear in $\ell$.}
\label{fig:SymmetricHolonStates}
\end{figure}

\subsection{Mapping LST to one dimension: a brief review}
\label{subsec:LSTmapping1D}
The Schr\"odinger equation for the holon moving between the sites of the Bethe lattice can be written in compact form as \cite{Note2}
\begin{equation}
 t \sum_{s} \psi_{\ell+1,s} + t \psi_{\ell-1} +  V_\ell \psi_\ell =  E \psi_\ell.
 \label{eq:BetheLatticeSEq}
\end{equation}
Here the linear string potential is given by $V_\ell =  2 J_z S^2 \l 2 \ell - \delta_{\ell,0} \r$, $E$ denotes the energy, and $z=4$ is the coordination number of the square lattice. In general the wave function $\psi(\Sigma)$ depends on the index $\Sigma \in {\rm BL}$ corresponding to a site on the Bethe lattice, or equivalently a string $\Sigma$. A useful parameterization of $\Sigma \in {\rm BL}$ is provided by specifying the length $\ell$ of the string as well as $\ell$ angular coordinates $\vec{s} = s_1,...,s_\ell$ with values $s_1=1...z$ and $s_j = 1 ... z-1$ for $j>1$. This formalism is used in Eq.~\eqref{eq:BetheLatticeSEq} and illustrated in Fig.~\ref{fig:SymmetricHolonStates} (a). In Eq.~\eqref{eq:BetheLatticeSEq} only the dependence on $s=s_\ell$ is shown explicitly. Because we started from the LST, the potential $V_\ell$ is independent of $\vec{s}$. The normalization condition is given by 
\begin{equation}
\sum_{\Sigma \in {\rm BL}} |\psi(\Sigma)|^2 = \sum_{\ell, \vec{s}} |\psi_{\ell,\vec{s}}|^2 = 1
\end{equation}
where the sum includes all sites of the Bethe lattice.

The simplest symmetric wave functions $\psi(\Sigma)$ only depend on $\ell$ and are independent of $\vec{s}$. We will first consider this case, which realizes the rotational ground state of the magnetic polaron. It is useful to re-parametrize the wave function $\psi_\ell$ by writing
\begin{flalign}
\psi_{\ell,s} &= \underbrace{\frac{(-1)^\ell}{2} (z-1)^{(1-\ell)/2}}_{ = \lambda_\ell} ~ \phi_\ell, \qquad \ell \geq 1, \label{eq:reParametrize}\\
 \psi_0 &= \phi_0.
\end{flalign}
The normalization for the new wave function $\phi_\ell$ is given by the usual condition, $\sum_{\ell=0}^\infty  |\phi_\ell |^2 = 1$, corresponding to a single particle in a semi-infinite one-dimensional system with lattice sites labeled by $\ell$.

The Schr\"odinger equation \eqref{eq:BetheLatticeSEq} for the 1D holon wave function $\phi_\ell$ becomes \cite{Bulaevskii1968},
\begin{flalign}
- t^* \frac{2}{\sqrt{z-1}} \phi_{1}  + V_0 \phi_0 &= E  \phi_0, 
\label{eq:SchroedingerPhiEll} \\
- t^* \l \phi_{2} + \frac{2}{\sqrt{z-1}} \phi_{0} \r + V_1 \phi_1 &= E  \phi_1, \\
- t^* \l \phi_{\ell+1} + \phi_{\ell-1} \r + V_\ell \phi_\ell &= E  \phi_\ell, \qquad \ell \geq 2. \label{eq:SchroedingerPhiEll1}
\end{flalign}
Away from the origin $\ell=0$, the effective hopping constant $t^*$ in the 1D model is given by \cite{Bulaevskii1968,Brinkman1970}
\begin{equation}
t^* = t \sqrt{z-1}.
\end{equation}
The tunneling rate between $\ell=0$ and $1$, on the other hand, is given by $2 t^*/\sqrt{z-1} = 2 t$. 
 
Before we move on, we consider the continuum limit of the effective 1D model where $\phi_\ell \to \phi(x)$ and $x \geq 0$ becomes a continuous variable, see Ref.~\cite{Bulaevskii1968}. This is a valid description in the strong coupling limit, where $t \gg J_z$. For simplicity we will ignore deviations of $V_\ell$ from the purely linear form at $\ell =0$, as well as the renormalization of the tunneling $t^* \to 2 t$ from site $\ell=0$ to $1$. As a result one obtains the Schr\"odinger equation \cite{Bulaevskii1968}
\begin{equation}
\l - \frac{\partial_x^2}{2 m^*} + V(x) \r \phi(x) = E \phi(x),
\label{eq:ContiSchrEq}
\end{equation}
where the effective mass is $m^* = 1/2 t^*$, and the confining potential is given by $V(x) = -2 t^* + 4 J_z S^2 x$.

By simultaneous rescaling of lengths, $x \to \lambda^{1/3} x$, and the potential $J_z \to \lambda J_z$, one can show that the eigen-energies $E$ in the continuum limit are given by \cite{Bulaevskii1968,Kane1989,Shraiman1988a}
\begin{equation}
E_n(t/J) = - 2 t \sqrt{z-1} + t a_n (t/J)^{-2/3},
\label{eq:EnScaling}
\end{equation}
for some numerical coefficients $a_n$. It has been shown in Refs.~\cite{Bulaevskii1968,Kane1989} that they are related to the eigenvalues of an Airy equation.

The scaling of the magnetic polaron energy like $t^{1/3} J_z^{2/3}$ is considered a key indicator for the string picture. It has been confirmed in different numerical works for a wide range of couplings \cite{Dagotto1990,Martinez1991,Liu1991,Liu1992,Mishchenko2001}, both in the $t-J$ and the $t-J_z$ models. Diagrammatic Monte Carlo calculations by Mishchenko et al.~\cite{Mishchenko2001} have moreover confirmed for the $t-J$ model that the energy $-2 \sqrt{3} t$ is asymptotically approached when $J \to 0$. However, for extremely small $J/t$ on the order of $0.03$ it is expected \cite{White2001} that the ground state forms a ferromagnetic polaron \cite{Nagaoka1966} with ferromagnetic correlations developing inside a finite disc around the hole. In this regime Eq.~\eqref{eq:EnScaling} is no longer valid.

Using ultracold atoms in a quantum gas microscope the universal scaling of the polaron energy can be directly probed when $J_z/t$ is varied and for temperatures $T < J$. To this end the super-exchange energy $\langle \H_J \rangle$ can be directly measured by imaging the spins around the hole. Note that $\langle \H_J \rangle$ has the same universal scaling with $t^{1/3} J_z^{2/3}$ as the ground state energy at strong couplings.

The excited states of the effective 1D Schr\"odinger equation \eqref{eq:ContiSchrEq} correspond to vibrational resonances of the meson formed by the spinon-holon pair, labeled by the vibrational quantum number $n$. In a semi-classical picture, they can be understood as states where the string length is oscillating in time. Now we generalize the mapping to a 1D problem for rotationally excited states. 

\subsection{Rotational string excitations in LST}
\label{subsec:LSTexcitations}
Within LST the entire spectrum of the magnetic polaron can easily be derived by making use of the symmetries of the holon Hamiltonian on the Bethe lattice. Around the central site, where $\ell=0$, we obtain a $C_4$ symmetry. The $C_4$-rotation operator has eigenvalues $e^{i \pi m_4 / 2}$ with $m_4=0,1,2,3$ and the eigenfunctions depend on the first angular variable $s_1$ in the following way: $e^{i \pi m_4 s_1 / 2}$. So far we assumed that the wave function $\psi_\ell$ only depends on the length of the string $\ell$, which corresponds to an eigenvalue of $C_4$ which is $m_4 = 0$. 

In addition, every node of the Bethe lattice at $\ell > 0$ is associated with a $P_3$ permutation symmetry. The $P_3$-permutation operator has eigenvalues $e^{i 2 \pi m_3 / 3}$ with $m_3=0,1,2$ and the eigenfunctions depend on the $j$-th angular variable $s_j$, $j > 1$, in the following way: $e^{i 2 \pi m_3 s_j / 3}$. The symmetric wave function $\psi_\ell$ discussed in Sec.~\ref{subsec:LSTmapping1D} so far had $m_3=0$ for all nodes. 

\subsubsection{First rotationally excited states}
We begin by considering cases where all $m_3=0$ are trivial, but $m_4 \neq 0$ becomes non-trivial. The Schr\"odinger equation in the origin at $\ell=0$ now reads
\begin{equation}
t \sum_{s_1=1}^4 \psi_{1,s_1}^{(n,m_4)} + V_0 \psi_0^{(n,m_4)} = E^{(n,m_4)} \psi_0^{(n,m_4)},
\label{eqSchrEffOrgnm4}
\end{equation}
where we introduced labels $(n,m_4)$ denoting the vibrational and the first rotational quantum numbers. 

Because the dependence of $\psi_{1,s_1}^{(n,m_4)}$ on the first angular variable $s_1$ is determined by the value of $m_4$ as explained above, the first term in Eq.~\eqref{eqSchrEffOrgnm4} becomes
\begin{equation}
\sum_{s_1=1}^4  \psi_{1,s_1}^{(n,m_4)} \propto \sum_{s_1=1}^4 e^{i \frac{\pi}{2} m_4 s_1} \propto \delta_{m_4,0}.
\label{eqKronm4}
\end{equation}
Because of the Kronecker delta function $\delta_{m_4,0}$ on the right hand side, we see that for $m_4 \neq 0$ Eq.~\eqref{eqSchrEffOrgnm4} becomes $V_0 \psi_0^{(n,m_4)} = E^{(n,m_4)} \psi_0^{(n,m_4)}$. Unless $E^{(n,m_4)}=V_0$, this equation only has the solution $\psi_0^{(n,m_4)} = 0$. It is only possible to have $\psi_0^{(n,m_4)} \neq 0$ if $E^{(n,m_4)} = V_0$, which is not a solution of ${\rm det} ( \H - E^{(n,m_4)}) = 0$ in general however.

Thus the first rotationally excited states are three-fold degenerate ($m_4 =1,2,3$) and given by
\begin{flalign*}
\psi_0^{(n,m_4)} = 0, \\
\psi_{\ell, s_1}^{(n,m_4)} = e^{i \frac{\pi}{2} m_4 s_1} \lambda_{\ell} \phi_\ell^{(n,m_4)},  \qquad & s_1=0,1,2,3.
\end{flalign*}
Here $\lambda_{\ell}$ was defined in Eq.~\eqref{eq:reParametrize} and the radial part $\phi_\ell^{(n,m_4)}$ is the solution of the Schr\"odinger equation \eqref{eq:SchroedingerPhiEll} - \eqref{eq:SchroedingerPhiEll1} for the potential $V_\ell \to V_\ell^{C_4}$, where
\begin{equation}
V_\ell^{C_4} = \begin{cases}
+ \infty, \quad \ell = 0\\
V_\ell, \quad \ell > 0
\end{cases}.
\end{equation}
For $\ell=0$ we introduced a large centrifugal barrier, preventing the holon from occupying the same site as the spinon. This takes into account the effect of the Kronecker-delta function in Eq.~\eqref{eqKronm4}, without the need to explicitly deal with the rotational variable $s_1$ in the wavefunction. Note that the effective 1D Schr\"odinger equation is independent of $m_4$ when $m_4 \neq 0$, and the same is true for the resulting eigenenergies $E^{(n,m_4)}$.

\subsubsection{Higher rotationally excited states}
Higher rotationally excited states with non-trivial $P_3$ quantum numbers $m_3 \neq 0$ at some node can be determined in a similar way. Let us consider $m_3 \neq 0$ at a node corresponding to a string of length $\ell_{\rm P}$. The Schr\"odinger equation at this node reads
\begin{multline}
t \sum_{s_{\ell_{\rm P}+1}=1}^3  \psi_{\ell_{\rm P}+1,s_{\ell_{\rm P}+1}}^{(n,m_3)} + t \psi_{\ell_{\rm P}-1}^{(n,m_3)} + V_{\ell_{\rm P}} \psi_{\ell_{\rm P}}^{(n,m_3)} =\\
= E^{(n,m_3)} \psi_{\ell_{\rm P}}^{(n,m_3)}.
\end{multline}
Again the first term is only non-zero when $m_3 = 0$. This can be seen from the dependence of $\psi_{\ell_{\rm P}+1,s_{\ell_{\rm P}+1}}^{(n,m_3)} \propto e^{i \frac{2\pi}{3} m_3 s_{\ell_{\rm P}+1}}$ on the angular variable $s_{\ell_{\rm P}+1}$, which yields a Kronecker delta $\delta_{m_3,0}$ when summed over $s_{\ell_{\rm P}+1} = 1,2,3$. 

For $m_3 \neq 0$, we obtain \emph{two} sets of independent eigenequations. The first involves only strings of length $\ell \leq \ell_{\rm P}$. If it has a non-trivial solution with $\psi_{\ell_{\rm P}}^{(n,m_3)} \neq 0$ and energy $E^{(n,m_3)}$, there is a second eigenequation involving strings of length $\ell \geq \ell_{\rm P}$. In general the second equation cannot be satisfied, because the energy $E^{(n,m_3)}$ is already fixed. The trivial choice $\psi_{\ell}^{(n,m_3)} = 0$ for $\ell > \ell_{\rm P}$ does not represent a solution because there exists a non-vanishing coupling to $\psi_{\ell_{\rm P}}^{(n,m_3)} \neq 0$. 

Therefore the only general solution is trivial for $\ell \leq \ell_{\rm P}$ and non-trivial for longer strings,
\begin{flalign*}
\psi_\ell^{(n,m_3)} = 0,  \qquad & \ell \leq \ell_{\rm P},\\
\psi_{\ell, s}^{(n,m_3)} = e^{i 2 \pi m_3 s /3} \lambda_{\ell} \phi_\ell^{(n,m_3)},  \qquad &  \ell > \ell_{\rm P},
\end{flalign*}
for $m_3 \neq 0$ and with $s=0,1,2$. The radial part $\phi_\ell^{(n,m_3)}$ of the rotationally excited string is determined by the Schr\"odinger equation \eqref{eq:SchroedingerPhiEll} - \eqref{eq:SchroedingerPhiEll1} for the potential $V_\ell \to V_\ell^{P_3}$, where
\begin{equation}
V_\ell^{P_3}(\ell_{\rm P}) = \begin{cases}
+ \infty, \quad \ell \leq \ell_{\rm P}\\
V_\ell, \quad \ell > \ell_{\rm P}
\end{cases}.
\end{equation} 
In this case there exists an even more extended centrifugal barrier than for the $C_4$ rotational excitations. It excludes all string configurations of length $\ell \leq \ell_{\rm P}$. Note that the effective 1D Schr\"odinger equation is independent of $m_3$ when $m_3 \neq 0$, and the same is true for the resulting eigenenergies $E^{(n,m_3)}$.

Note that the rotationally excited states of the $P_3$ operator are highly degenerate. Because the wave function vanishes for all $\ell \leq \ell_{\rm P}$, there exist $4 \times 3^{\ell_{\rm P}-1}$ decoupled sectors on the Bethe lattice where $\psi^{(n,m_3)}(\Sigma) \neq 0$. Together with the two choices $m_3=1$ and $2$, the total degeneracy $D_3(\ell_{\rm P})$ becomes
\begin{equation}
D_3(\ell_{\rm P})= 8 \times 3^{\ell_{\rm P} - 1}.
\end{equation}

\begin{figure}[b!]
\centering
\epsfig{file=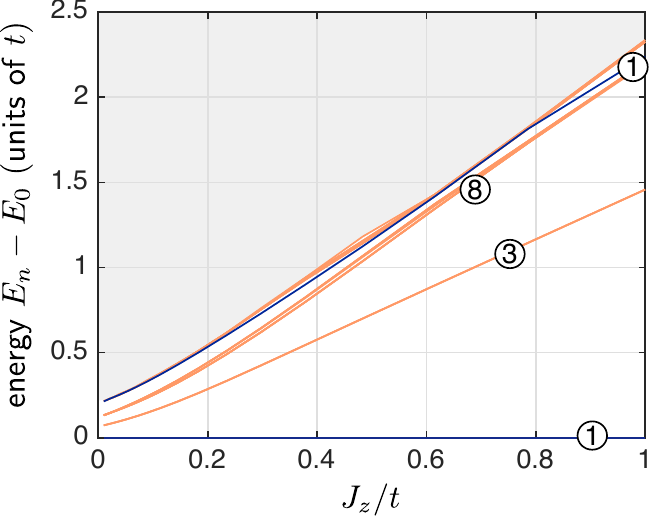, width=0.4\textwidth} $\quad$
\caption{Excitation spectrum of magnetic polarons in the strong coupling parton theory as a function of $J_z/t$. Note that the ground state energy was subtracted, which scales as $J_z^{2/3} t^{1/3}$. We used $S=1/2$ and NLST with strings of length up to $\ell_{\rm max} = 8$. A similar calculation is presented in Fig.~\ref{fig:Overview} using LST with $\ell_{\rm max}=100$. The degeneracies of the lowest excited manifolds of states are indicated in circles. Dark blue lines correspond to vibrationally excited states without rotational excitations, i.e. $m_3=m_4=0$. Orange lines correspond to purely rotationally excited states with at least one $m_3$ or $m_4$ non-vanishing. No calculations were performed for higher energies (shaded area). The finite gap predicted for small $J_z/t$ by NLST is a finite-size effect caused by the maximal string length $\ell_{\rm max}=8$ assumed in the calculations.}
\label{fig:StringExcitations}
\end{figure}

\subsection{Comparison to NLST and scaling laws}
\label{subsec:compNLST}
In Fig.~\ref{fig:Overview} (c) and Fig.~\ref{fig:StringExcitations} we show results for the eigenenergies from LST and NLST respectively. As indicated in the figures, we have confirmed for the lowest lying excited states that LST correctly predicts the degeneracies of the low-lying manifolds of states obtained from the more accurate NLST. While these degeneracies are exact for LST, the self-interactions of the string included in the NLST open small gaps between some of the states and lift the degeneracies. In general we find good qualitative agreement between LST and NLST. 

\subsubsection{Excitation energies}
For the energy of the first excited state with rotational quantum number $m_4>0$ and $m_3=0$, we find excellent quantitative agreement between LST and NLST. At small $J_z/t$ we note that NLST predicts a larger energy than LST which appears to saturate at a non-zero value when $J_z/t \to 0$. This is a finite-size effect caused by the restricted Hilbert space with maximum string length $\ell_{\rm max}=8$. Aside from this effect, we obtain the following scaling behavior
\begin{equation}
E_{\rm rot}(n=1)-E_0 ~ \propto~  J_z
\end{equation}
for all rotationally excited states without radial (i.e. vibrational) excitations ($n=1$).

In contrast, the excited states with vibrational excitations ($n>1$) show a scaling behavior
\begin{equation}
E_{\rm vib}(n>1)-E_0 ~ \propto~  J_z^{2/3} t^{1/3}
\label{eq:EvibScaling}
\end{equation}
as expected on general grounds from the effective one-dimensional Schr\"odinger equation, see Eq.~\eqref{eq:EnScaling}.

\subsubsection{String-length distribution}
\begin{figure}[t!]
\centering
\epsfig{file=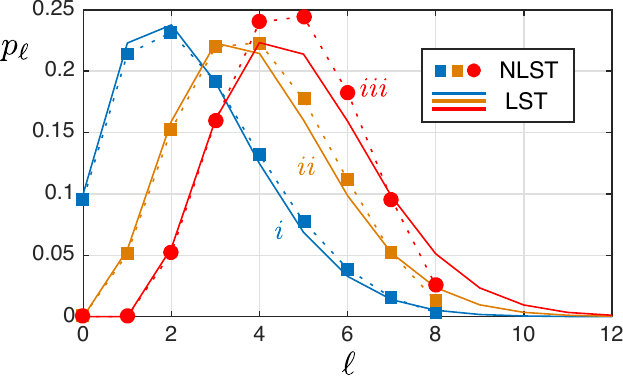, width=0.45\textwidth} $~$
\caption{String length distributions $p_\ell$ of the first excited string states at $J_z/t=0.1$: $i$ the ground state, $ii$ the first excited state with $\ell_{\rm P}=0$ and $m_4\neq0$, and $iii$ the fifth excited state with $\ell_{\rm P}=1$ and $m_3\neq0$. Calculations are performed for $S=1/2$ and a maximum string length of $\ell_{\rm max}=8$ ($\ell_{\rm max}=100$) was used for NLST (LST).}
\label{fig:StringExcitationsNell}
\end{figure}

In Fig.~\ref{fig:StringExcitationsNell} we calculate the distribution function $p_\ell$ of string lengths for $t/J_z=10$ well in the strong coupling regime. The comparison between results from NLST with a maximum string length $\ell_{\rm max} = 8$, and LST calculations with $\ell_{\rm max}=100$ shows excellent quantitative agreement for $p_\ell$. Only for the highest excited state considered, with the largest mean string length, we observe some discrepancies at large values of $\ell$ around the cut-off $\ell_{\rm max}$ used in the NLST.

We confirm a strong suppression of $p_\ell$ for $\ell \leq \ell_{\rm P}$ in the rotationally excited states due to the centrifugal barrier. We note that even within the NLST the $C_4$ rotational symmetry and the $P_3$ permutation symmetry at $\ell=1$ are strictly conserved. The good quantitative agreement for $p_\ell$ should be contrasted to the predicted energies, where larger deviations are observed between LST and NLST in the strong coupling regime.

\subsection{String reconstruction in a quantum gas microscope}
\label{subsecSpinReconstruction}
As explained in Sec.~\ref{subsecQuantGasMic} the spin configuration around the holon in a $t-J_z$ model can be directly accessed \cite{Boll2016}. Measurements of this general type are routinely performed in quantum gas microscopy. For example, they have been used to measure the full counting statistics of the staggered magnetization in a Heisenberg AFM, see Ref.~\cite{Mazurenko2017}, and non-local signatures of spin-charge separation, see Ref.~\cite{Hilker2017}. These capabilities should allow imaging of the string attached to the holon, and extract the full distribution function of the string length. This makes quantum gas microscopes ideally suited for direct observations of the different excited states of magnetic polarons. Now we will assume that both spin states and the density can be simultaneously imaged \cite{Boll2016}.

\begin{figure}[t!]
\centering
\epsfig{file=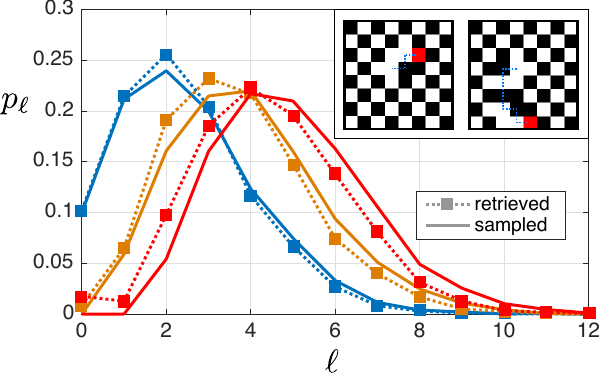, width=0.44\textwidth}
\caption{Retrieving strings in the $t-J_z$ model. We start from a perfect N\'eel state and create a spinon in the center. Within LST, the distribution of spin patterns is determined by the wavefunction of the holon on the Bethe lattice. The inset shows two snapshots for string lengths $\ell = 3$ and $\ell = 7$ where the holon position is marked red. We generated pictures by sampling the three different string length distributions plotted in the figure and choosing random directions of the string. By analyzing the happiness for every spin pattern as described in the text, we retrieve the string for every snapshot. We assume that both spin states and the density can be simultaneously measured. The distribution functions of the lengths of retrieved strings compare well with the sampled distributions, even for long strings.}
\label{fig:findingStrings}
\end{figure}

Knowledge of the spin configuration enables the determination of the distortion field $\hat{\sigma}^z_{\ij}$. In order to identify the string attached to a holon in a single shot, we now introduce a local measure for the distortion of the N\'eel state at a given site $i$,
\begin{equation}
\theta_i =\frac{1}{2} \sum_{\ij}{\vphantom{\sum}}' \l 1+ \hat{\sigma}^z_{\ij} \r.
\end{equation}
Here, $\sum_{\ij}'$ denotes a sum over all bonds $\ij$ to neighboring sites $j$, where both sites $i$ and $j$ are occupied by spins. Therefore, the \emph{happiness} $\theta_i$ assumes integer values between 0 and 4, where $0 ~(4)$ corresponds to aligned (anti-aligned) spins on all adjacent bonds. 

Since the AFM spin order is maintained along a string without loops, the distortion field $\hat{\sigma}^z_{\ij} =1$ is unity on bonds $\ij$ that belong to the string and do not include the holon position. Since spins on the string are displaced with respect to the surrounding AFM, it holds $\hat{\sigma}^z_{\ij} =- 1$ if site $i$ is occupied by a spin and belongs to the string and site $j$ is not part of the string. Therefore, the happiness $\theta_i$ on sites $i$ belonging to the string corresponds to the number of neighboring spins that are also part of the string. For sites $j$ outside of the string, $\theta_j = 4 - N_{i}^{\rm s}$, where $N_{i}^{\rm s}$ is the number of neighboring sites that belong to the string. By analyzing the happiness according to these rules, we can start from the holon position and reconstruct the attached string. 

The scheme described above allows to directly observe spinons and strings in realizations of the $t-J_z$ model with quantum gas microscopes. To mimic this situation, we start from a perfect N\'eel state and initiate a hole in the center of the system. For a given string length $\ell > 0 $, we first move it randomly to one out of its four neighboring sites. If $\ell >1$, we randomly choose one out of the three sites which the holon did not visit in the previous step. This step is repeated until a string of length $\ell$ is created. Thereby, loops are allowed in the string, but the hole cannot retrace its previous path. In Fig.~\ref{fig:findingStrings}, we sample strings according to a given string length distribution function $p_\ell$ and retrieve them from the resulting images by examining the values of $\theta_i$ as described above. Comparison of the sampled distribution, taken from LST, with the retrieved string length distribution shows that even long strings can be efficiently retrieved in the images. Corrections by loops only lead to a small over-counting (under-counting) of short (long) strings on the level of a few percent.

\begin{figure*}[t!]
\centering
\epsfig{file=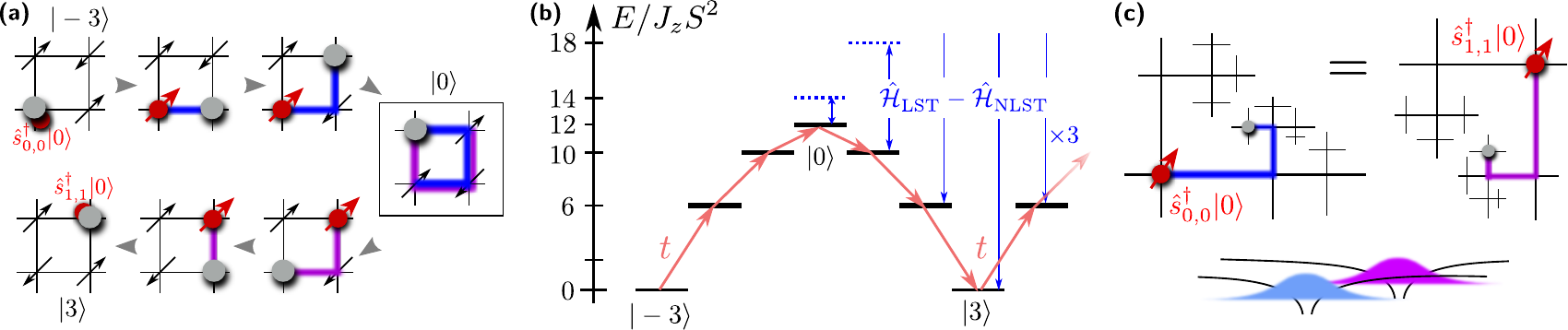, width=0.96\textwidth}
\caption{Trugman loops in the spinon-holon picture: (a) When the holon performs one-and-a-half loops around one plaquette (arrows), the spin configuration of the N\'eel state is restored \cite{Trugman1988}. This process can be associated with two different configurations of the spinon and the string in the over-complete Hilbert space of the string theory. In the first case the spinon is located at $(0,0)$ and the string is oriented counterclockwise (blue). In the second case the spinon is located at $(1,1)$ and the string is oriented clockwise (pink). Trugman loops correspond to a correlated pair tunneling of the spinon and the holon diagonally across the plaquette, for which the holon has to overcome the potential energy barrier shown in (b). At strong couplings, $t \gg J_z$, this barrier is shallow. The resulting effective spinon hopping element can be calculated by a tight-binding approach, where the difference in potential energies $\H_{\rm NLST} - \H_{\rm LST}$ between the full non-linear and linear string theory are treated perturbatively (c). This is similar to the tight-binding calculation of the hopping element of an electron between two atoms. Like the string theory the latter relies on using an over-complete basis, with one copy of the entire Hilbert space for each atom.}
\label{fig:TrugmanLoops}
\end{figure*}

\section{Spinon dispersion: tight-binding description of Trugman loops}
\label{sec:TrugmanDisp}
In the previous section we were concerned with the properties of the holon on the Bethe lattice. Motivated by the strong-coupling ansatz from Eq.~\eqref{eqStrongCplgWvfct} the spinon was treated as completely static so far, placed in the origin of the Bethe lattice. It has been pointed out by Trugman \cite{Trugman1988} that the magnetic polaron in the $t-J_z$ model can move freely through the entire lattice by performing closed loops around plaquettes of the square lattice which restore the N\'eel order, see Fig.~\ref{fig:TrugmanLoops} (a). By using the string theory in an over-complete Hilbert space we have not included the effects of such loops so far. 

Now we show that a conventional tight-binding calculation allows one to include the effects of Trugman loops in our formalism. They give rise to an additional term in the effective Hamiltonian describing spinon dynamics,
\begin{equation}
\H_{\rm T} = t_{\rm T} \sum_{\langle \ij \rangle_d} \sd_j \s_i + \hc.
\label{eq:Htrugman}
\end{equation}
Here $\langle \ij \rangle_d$ denotes a pair of two next-nearest neighbor sites $i$ and $j$ on opposite ends of the diagonal across a plaquette in the square lattice; every such bond is counted once. The new term is denoted by $\H_{\rm T}$ because it describes how Trugman loops contribute to the spinon dynamics. Below we will derive an analytic equation for calculating $t_{\rm T}$ and compare our predictions with exact numerical calculations of the spinon dispersion relation.

\subsection{Trugman loops in the spinon-holon picture}
\label{subsec:TrugmanLoops}
In the spinon-holon theory we use the over-complete basis introduced in Eq.~\eqref{eqBasisSpinonHolon}. To study corrections in our model introduced by this over-completeness, let us first consider only the LST approximation, see Eq.~\eqref{eq:HLST}, where the holon is localized around the spinon and the latter has no dynamics. In this case, the over-completeness of the basis leads to very small errors. 

To see this, note that the first physical state $\sigma_{\ij}^z$ which can be identified with two inequivalent basis states $\ket{\vec{x}_{\rm s},\Sigma}$, $\ket{\vec{x}_{\rm s}',\Sigma'}$ involves at least six string segments. Such states can be obtained from the so-called Trugman loops \cite{Trugman1988}: Starting with a holon at spinon position $\vec{x}_{\rm s}$ and performing one-and-a-half loops around a plaquette, one obtains a state $\ket{\vec{x}_{\rm s},\Sigma}$ with string length $\ell(\Sigma)=6$ which corresponds to the same physical state as the string length zero state $\ket{\vec{x}_{\rm s}', \Sigma'}$ with $\ell(\Sigma')=0$, where $\vec{x}_{\rm s}'$ is a diagonal next-nearest neighbor of $\vec{x}_{\rm s}$. This is illustrated in Fig.~\ref{fig:TrugmanLoops} (a), where after performing the Trugman loop the surrounding N\'eel state is not distorted.

When the confinement is tight, $J_z \gtrsim t$, the wavefunction $\phi_{\ell}$ decays exponentially with the string length, and physical states which are represented more than once in the over-complete basis are very weakly occupied. Even when $t \gg J_z$ -- but still assuming the LST Hamiltonian -- the fraction of physical states with multiple representations in the over-complete basis is small, see discussion above Eq.~\eqref{eqBasisSpinonHolon}. The reason is essentially that specific loop configurations need to be realized out of exponentially many possible string configurations. 

Now we consider the effect of non-linear corrections to the string energy, as described in Eq.~\eqref{eq:HNLST}. We can loosely distinguish between two types of corrections: (i) for strings without loops, attractive self-interactions between parallel string segments can lower the string energy, and (ii) for strings with Trugman loops, the potential energy can vanish completely. While the first effect (i) only leads to small quantitative corrections of the spinon-holon energy, the second effect leads to a large degeneracy within the over-complete basis and needs to be treated more carefully. 

When two states $\ket{\vec{x}_{\rm s},\Sigma}$ and $\ket{\vec{x}_{\rm s}',\Sigma'}$ in the over-complete basis have zero potential energy and their holon positions coincide, their distortion fields are identical, $\sigma^z_\ij(\Sigma) = \sigma^z_\ij(\Sigma')$ and they correspond to the same physical state. By this identification we notice that the effect of Trugman loops is to introduce spinon dynamics: consider starting from a string length zero state around a spinon at $\vec{x}_{\rm s}$. By holon hopping a final state in the over-complete basis can be reached which can be identified with another string length zero state but around a different spinon position $\vec{x}_{\rm s}'$. Our goal in this section is to start from degenerate eigenstates of the LST and describe quantitatively how the corrections by NLST, $\H_{\rm NLST} - \H_{\rm LST}$, introduce spinon dynamics. This can be achieved by a conventional tight-binding approach.

In the case when $t<J_z$ the Trugman loop process is strongly suppressed because it corresponds to a $6$th order effect in $t$ and the holon has to overcome an energy barrier of height $12 J_z S^2$, see Ref.~\cite{Trugman1988} and Fig.~\ref{fig:TrugmanLoops} (b). Although these perturbative arguments based on an expansion in $t/J_z$ no longer work when $t \gg J_z$, we will show that Trugman loop processes only lead to small corrections $t_{\rm T} \lesssim J_z$, a small fraction of $J_z$. The key advantage of using tight-binding theory is that the potential energy $\sim J_z$ rather than the holon hopping $\sim t$ is treated perturbatively.

\subsection{Tight-binding theory of Trugman loops}
\label{subsecTBofTrugmanLoops}
To explain how tight-binding theory allows us to take into account the over-completeness of the basis used in LST, see Sec.~\ref{subsubsec:StringTheory}, we draw an analogy with conventional tight-binding calculations for Bloch bands. We start by a brief review of the tight-binding approach and explain how, implicitly, use is being made of an over-complete basis. See e.g. Ref.~\cite{Ashcroft1976} for an extended discussion of conventional tight-binding calculations.

\subsubsection{Tight-binding theory in a periodic potential}
Consider a quantum particle moving in a periodic potential $W(x)=W(x+a)$. We assume that the lattice $W(x)$ has a deep minimum at $x_j$ within every unit cell $[j a, (j+1) a]$, where the particle can be localized. This is the case for example if $W(x) = \sum_j \tilde{W}(x-x_j)$ corresponds to a sum of many atomic potentials $\tilde{W}(x-x_j)$ created by nuclei located at positions $x_j$. Similarly, the NLST potential is periodic on the Bethe lattice, although the geometry is much more complicated in that case.

The idea behind tight-binding theory is to solve the problem of a single atomic potential $\tilde{W}(x-x_0)$ first. This yields a solution $\tilde{w}(x-x_0)$ localized around $x_0$. Then one assumes that the orbital $\tilde{w}(x-x_0)$ is a good approximation for the correct Wannier function $w(x-x_0)$ defined for the full potential $W(x)$, see Fig.~\ref{fig:TrugmanLoops} (c). The tight-binding orbital $\tilde{w}(x-x_0)$ defined by the potential $\tilde{W}(x-x_0)$ around $x_0$ is similar to the holon state defined by the LST Hamiltonian $\H_{\rm LST}$ around a given spinon position $\vec{x}_{\rm s}$.

The potential energy mismatch $W(x)-\tilde{W}(x)$ can now be treated as a perturbation. Most importantly, it induces transitions between neighboring orbitals $\tilde{w}(x-x_j)$ and $\tilde{w}(x-x_{j\pm1})$. This leads to a nearest neighbor tight-binding hopping element
\begin{equation}
t_{\rm tb} = \bra{\tilde{w}(x-x_j)} W(x) - \tilde{W}(x) \ket{\tilde{w}(x-x_{j \pm 1})}.
\end{equation}
For this perturbative treatment to be valid, it is sufficient to have a small spatial overlap between the two neighboring orbitals, 
\begin{equation}
\nu=| \bra{\tilde{w}(x-x_j)} \tilde{w}(x-x_{j\pm 1}) \rangle | \ll 1.
\end{equation}
The energy difference $W(x-x_j) - \tilde{W}(x-x_{j\pm1})$ on the other hand can be sizable. 

In practice, the most common reason for a small wavefunction overlap $\nu \ll 1$ is a high potential barrier which has to be overcome by the particle in order to tunnel between two lattice sites. In this case, once the barrier becomes too shallow, $\nu$ becomes sizable and the tight-binding approach breaks down. As another example, consider two superconducting quantum dots separated by a dirty metal. Even without a large energy barrier, the wavefunction of the superconducting order parameter decays exponentially outside the quantum dot due to disorder \cite{Larkin1983,Spivak1991}. This leads to an exponentially small wavefunction overlap $\nu \ll 1$ and justifies a tight-binding treatment.

Similar to the case of the string theory of magnetic polarons, the effective Hilbert space used in conventional tight-binding calculations is over-complete. To see this, note that the  tight-binding Wannier function $\tilde{w}(x-x_j)$ corresponding to lattice site $j$ is defined on the space $\mathscr{H}_j$ of complex functions mapping $\mathbb{R} \to \mathbb{C}$ with site $j$ in the center. Assuming that every such Wannier function is defined in its own copy of this Hilbertspace $\mathscr{H}_j$, we obtain -- by definition -- that the resulting tight-binding wavefunctions are mutually orthogonal. But in reality, the physical Hilbert space $\mathscr{H}_{\rm phys}$ consists of just one copy of the space of all functions mapping $\mathbb{R} \to \mathbb{C}$, and after identifying all states in $\mathscr{H}_j$ with a physical state in $\mathscr{H}_{\rm phys}$, the resulting tight-binding Wannier functions in $\mathscr{H}_{\rm phys}$ are no longer orthogonal in general: i.e. $\nu \neq 0$. As long as $\nu \ll 1$, they still qualify as good approximations for the true, mutually orthogonal Wannier functions $w(x-x_j)$, which justifies the tight-binding approximation.

\subsubsection{Tight-binding theory in the spinon-holon picture}
We can now draw an analogy for a single hole in the $t-J_z$ model. The true physical Hilbert space corresponds to all spin configurations and holon positions, see Eq.~\eqref{eq:BasisStatesDistortionField}. The over-complete Hilbert space is defined by copies of the Bethe lattice, each centered around a spinon positioned on the square lattice. As a result of Trugman loops, certain string configurations can be associated with \emph{two} spinon positions, see Fig.~\ref{fig:TrugmanLoops} (c).

We start by describing the tight-binding formalism for the Trugman loop hopping elements. When $t \ll J_z$, large energy barriers strongly suppress spinon hopping, see Fig.~\ref{fig:TrugmanLoops} (b). As a result, the overlap of the holon state corresponding to a spinon at $\vec{x}_{\rm s}$ with the holon state bound to a spinon at $\vec{x}_{\rm s}+\vec{r}$ is small,
\begin{equation}
 \nu_n(\vec{r}) = |\bra{\psi_n(\vec{x}_s+\vec{r})} \psi_n(\vec{x}_s) \rangle | \ll 1,
 \label{eq:overlapsTrugmanLoops}
\end{equation}
and the tight-binding approximation is valid. Here $\vec{r}$ denotes a vector connecting sites from the same sublattice and $n$ is the vibrational quantum number of the holon state. Note that in the definition of $\nu$ in Eq.~\eqref{eq:overlapsTrugmanLoops}, two states from the overcomplete basis are assumed to have unit overlap if they correspond to the same physical state. The holon wave functions on the Bethe lattice for different spinon positions $\vec{x}_{\rm s}$ are the equivalent of the tight-binding orbitals $\tilde{w}(x-x_j)$ corresponding to lattice sites $x_j$ discussed above. We will show below that the condition Eq.~\eqref{eq:overlapsTrugmanLoops} remains true even when the barrier becomes shallow, $J_z \ll t$, and that even in this regime the following tight-binding calculation is accurate.

Transitions between states associated with different spinon positions $\vec{x}_{\rm s}$ and $\vec{x}_{\rm s} + \vec{r}$ are induced by the potential energy mismatch $\H_{\rm NLST} - \H_{\rm LST}$, see Fig.~\ref{fig:TrugmanLoops} (b). This takes the role of $W(x) - \tilde{W}(x)$ from the conventional tight-binding calculation. Accordingly, the tight-binding spinon hopping element is given by
\begin{equation}
t_{\rm T}(n) =  \bra{\psi_n(\vec{r}_s+\vec{e}_x+\vec{e}_y)}  \H_{\rm NLST} - \H_{\rm LST}  \ket{\psi_n(\vec{r}_s)}.
\label{eq:tTrugTightBinding}
\end{equation}
Here we have set $\vec{r} = \vec{e}_x+\vec{e}_y$ which corresponds to the simplest Trugman loop process inducing spinon hopping diagonally across the plaquette. A more precise expression is obtained by noting that for long strings $\H_{\rm NLST} - \H_{\rm LST}$ also leads to longer-range spinon hopping, which should be counted separately. Below we will discuss in detail which contributions we include to obtain an accurate expression, see Eq.~\eqref{eqTrugmanTBfromLST}, for the diagonal Trugman loop spinon hopping $t_{\rm T}$ from Eq.~\eqref{eq:Htrugman}.

To understand the meaning of Eq.~\eqref{eq:tTrugTightBinding}, consider the simplest configuration $\ket{\vec{x}_{\rm s}, \Sigma}$ for which $( \H_{\rm NLST} - \H_{\rm LST} ) \ket{\vec{x}_{\rm s}, \Sigma} \neq 0$. It has a string length of $\ell=3$ and can be reached by performing three quarters of a loop around a plaquette, starting with either a spinon in the lower left corner and going counter-clockwise, or a spinon in the top right corner and going clockwise. This is illustrated in Fig.~\ref{fig:TrugmanLoops} (a) and (c). 

Now we consider the case of a shallow potential barrier $J_z \ll t$ and show that the overlap $\nu$ in Eq.~\eqref{eq:overlapsTrugmanLoops} is exponentially suppressed, even for vibrationally highly excited states with a large average string length. The overlap $\nu$ can be written as a sum over terms of the form $\psi_\ell^* \psi_{6-\ell}$, where the wavefunction $\psi_\ell$ defined on the Bethe lattice is related to the effective one-dimensional wavefunction $\phi_\ell$ by Eq.~\eqref{eq:reParametrize} with a typical extent given by $L_0$. For simplicity we approximate $|\phi_\ell |^2 \approx 1/L_0$ for $\ell < L_0$ and $\phi_\ell=0$ otherwise. Following Bulaevskii et al.~\cite{Bulaevskii1968} we equate the average kinetic and potential energies in the effective Schr\"odinger equation \eqref{eq:ContiSchrEq}, $1/2m^*L_0^2 = 4 J_z S^2 L_0$, to obtain the following estimate
\begin{equation}
L_0 \approx (2S)^{-2/3} \l \frac{t}{J_z} \r^{1/3} (z-1)^{1/6} \approx 1.2  \l \frac{t}{J_z} \r^{1/3} 
\label{eq:EstimateL0}
\end{equation}
for the typical length of the string connecting the spinon to the holon. In the second step we used $S=1/2$ and $z=4$ for the 2D anti-ferromagnet.

Because the Bethe lattice has a fractal structure with $z-1$ new branches emerging from each node, the amplitude $|\psi_\ell |^2$ is reduced by the additional exponential factor $\lambda_\ell^2$ as compared to $|\phi_\ell |^2$, see Eq.~\eqref{eq:reParametrize}. It holds $|\psi_\ell |^2 \approx (z-1)^{1-\ell} / (4 L_0)$, from which we obtain 
\begin{equation}
|\psi_\ell |^2 \approx 0.6 \l \frac{J_z}{t} \r^{1/3} \times 3^{- \ell}
\label{eq:EstimatePsiEll}
\end{equation}
using the estimate for $L_0$ from above. In terms of the original $t-J_z$ model, the exponential suppression of $|\psi_\ell |^2$ with $\ell$ can be understood as a consequence of the exponentially many spin configurations that can be realized by the motion of the hole. From Eq.~\eqref{eq:EstimatePsiEll} we see that already for $\ell=3$ we obtain $|\psi_3 |^2 \approx 0.023 ~ (J_z/t)^{1/3} \ll 1$ and thus $\nu \ll 1$. Even for excited string states the exponential suppression of the amplitude $| \psi_\ell |^2$ by $\lambda_\ell^2$ on the Bethe lattice leads to $|\psi_3|^2 \approx 0.028 |\phi_3 |^2 \ll 1$. Therefore the perturbative treatment of Trugman loops is justified for the \emph{entire} range of parameters $t/J_z$. We demonstrate this by an exact numerical calculation for a closely related toy model which is presented in Appendix \ref{apdx:TBtrugmanLoops}.

Next we provide additional physical intuition why the tight-binding approach works even in the regime when the potential barrier is shallow, $J_z \ll t$. In this limit, the kinetic energy of the holon is minimized by delocalizing the holon symmetrically over all possible $z-1$ directions at each node on the Bethe lattice. This leads to the exponential decay of the holon wavefunction $\psi_\ell$ with $\ell$, which is equal for all directions on the Bethe lattice. It also yields a zero-point energy of $- 2 \sqrt{z-1} t$ when $J_z=0$. In order to minimize its energy further, the holon could make use of the non-linearity of the string potential described by $\H_{\rm NLST}$: By occupying preferentially the directions on the Bethe lattice corresponding to Trugman loops, the average potential energy is lowered. This mechanism is very ineffective, because it requires localizing the holon on the particular set of states corresponding to the Trugman loops. This costs kinetic energy of order $t$: The average kinetic zero-point energy of a holon moving along one fixed direction is given by $-2 t$ instead of $-2 \sqrt{z-1} t$. Therefore, in the limit $t \gg J_z$, the kinetic term in the Hamiltonian dictates a symmetric distribution of the holon over all possible directions, which allows to treat perturbatively the effect of the reduced string potential for the very few specific directions defined by the Trugman loop string configurations.

\subsection{Application: Spinon hopping elements\\ in the $t-J_z$ model}
\label{subsec:SpinonHoppingtJz}
Next we apply the tight-binding theory of Trugman loops introduced above and derive a closed expression for the effective spinon hopping elements using Eq.~\eqref{eq:tTrugTightBinding}. 

\subsubsection{Contributing string configurations}
For every plaquette there exist two Trugman loops contributing to $t_{\rm T}$: a clockwise and a counter-clockwise one. In Fig.~\ref{fig:TrugmanLoops} (b) we show the energies $V^{\rm NLST}_\ell$ for states along one of these loops. For strings of length $\ell_1 \geq 3$, measured from the first spinon position $\vec{x}_s$, they differ from the expression $V_\ell = 4 J_z S^2 \ell$ assumed in LST by
\begin{equation}
\delta V_{\ell_1} = V^{\rm NLST}_{\ell_1} - V_{\ell_1} = - J_z S^2 \times (2, 8, 16, 26),
\end{equation}
for $\ell_1 = (3,4,5,6)$. These configurations overlap with strings of length $\ell_2 = 6 - \ell_1$, measured from the second spinon position $\vec{x}_s \pm \vec{e}_x \pm \vec{e}_y$ on the opposite side of the plaquette in the square lattice. Using Eq.~\eqref{eq:tTrugTightBinding} these states contribute an amount $2 \sum_{\ell_1=3}^6 \delta V_{\ell_1} \lambda_{6-\ell_1} \phi_{6-\ell_1}^* \lambda_{\ell_1} \phi_{\ell_1}$ to the Trugman loop hopping element $t_{\rm T}$. 

\begin{figure}[t!]
\centering
\epsfig{file=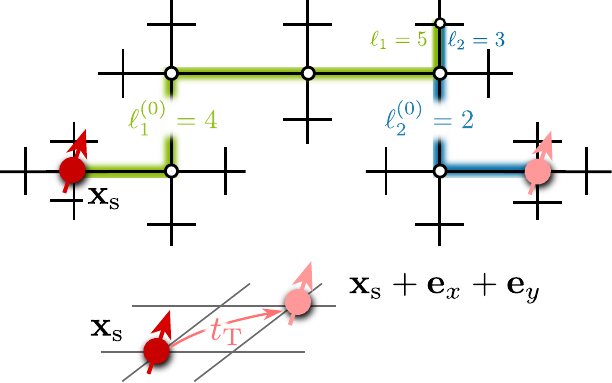, width=0.39\textwidth} $\qquad$
\caption{Contributions to the tight-binding Trugman loop spinon hopping element $t_{\rm T}$ from site $\vec{x}_s$ to $\vec{x}_s + \vec{e}_x + \vec{e}_y$. Strings of length $\ell_1$ measured from the first spinon position $\vec{x}_s$ overlap with strings of length $\ell_2$ measured from the second spinon position. The parts of the strings along the Trugman loop trajectory have lengths $\ell_{1,2}^{(0)}$ respectively, with $\ell_{1}^{(0)} + \ell_{2}^{(0)} = 6$.}
\label{fig:TBhoppingElements}
\end{figure}

In addition there are overlaps for longer strings with $\ell_1+\ell_2 > 6$. Let us start from a configuration with strings of lengths $\ell_1^{(0)}=4$ from spinon one, and $\ell_2^{(0)}=2$ from spinon two, such that combining them yields the Trugman loop process. In particular, $\ell_1^{(0)} + \ell_2^{(0)}=6$. Now a set of $(z-2) (z-1)^{n-1}$ string configurations of lengths $\ell_1 = \ell_1^{(0)} + n$ and $\ell_2=\ell_2^{(0)}+n$ (measured from the two spinon positions respectively) can be constructed, for which the last $n > 0$ segments can be identified on the Bethe lattice, see Fig.~\ref{fig:TBhoppingElements}. For all these states, the mismatch of the corresponding potential energies defined relative to the two spinon positions (LST versus NLST) is given by $\delta V_{\ell_1^{(0)}}$. Together all these configurations with $\ell_1^{(0)} = 3,4,5$ contribute $2 \sum_{\ell_1^{(0)}=3}^5 \delta V_{\ell_1^{(0)}} \sum_{n=1}^\infty (z-2) (z-1)^{n-1} \lambda_{\ell_2^{(0)}+n} \phi_{\ell_2^{(0)}+n}^* \lambda_{\ell_1^{(0)}+n} \phi_{\ell_1^{(0)}+n}$ to the tight-binding Trugman loop hopping element $t_{\rm T}$, where $\ell_2^{(0)} = 6 - \ell_1^{(0)}$ is assumed in the sum. 

The only missing set of strings with $\ell_1+\ell_2 > 6$ consists of cases where the first string (with length $\ell_1$ measured from the first spinon) includes the position of the second spinon on the Bethe lattice. This situation corresponds to $\ell_1^{(0)} = 6$ and $\ell_2^{(0)}=0$ using the notation from the previous paragraph. In this case there exist $(z-1)^{n}$ configurations of this type, for most of which one obtains a mismatch of potential energies of $\delta V_7 = - 24 J_z S^2$. Along certain directions the magnitude of the energy mismatch between LST and NLST is larger, and this corresponds to a case where a second Trugman loop follows the first one. Because it gives rise to longer-range spinon hopping, we will neglect these contributions in the following and take into account the constant energy mismatch of $\delta V_7 = \delta V_6 + 2 J_z S^2$ in the tight-binding calculation of $t_{\rm T}$. This class of strings thus contributes $2 \delta V_7 \sum_{n=1}^\infty (z-1)^n \lambda_{n} \phi_{n}^* \lambda_{6+n} \phi_{6+n}$ to $t_{\rm T}$. 

\subsubsection{Trugman loop hopping elements}
Summarizing, we obtain the following expression for the tight-binding Trugman loop hopping element
\begin{multline}
t_{\rm T} = 2 \sum_{\ell_1^{(0)}=3}^6 \sum_{n=0}^\infty  ~ \lambda_{6-\ell_1^{(0)}+n}  \lambda_{\ell_1^{(0)}+n} \phi_{6-\ell_1^{(0)}+n}^* \phi_{\ell_1^{(0)}+n} \\
\times  \bigg\{  \delta_{n,0} ~ \delta V_{\ell_1^{(0)}} + (1 - \delta_{n,0}) \bigg[ (1 - \delta_{\ell_1^{(0)},6})  \frac{z-2}{z-1} \delta V_{\ell_1^{(0)}}  \\
+ \delta_{\ell_1^{(0)},6} \l \delta V_{\ell_1^{(0)}} + 2 J_z S^2 \r \bigg] \bigg\} \times  (z-1)^{n}
\label{eqTrugmanTBfromLST}
\end{multline}
within LST. Note that the exponential factor $(z-1)^n$ is canceled by powers of $(z-1)$ appearing in the coefficients $\lambda_\ell$ relating the wavefunction on the Bethe lattice to the effective one-dimensional wavefunction $\phi_\ell$.

\begin{figure}[t!]
\centering
\epsfig{file=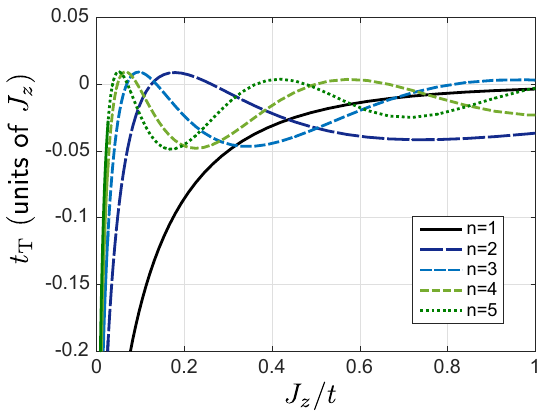, width=0.42\textwidth} $\quad$
\caption{Spinon hopping $t_{\rm T}$ in the $t-J_z$ model for $S=1/2$. We applied the tight-binding theory presented in the main text to the rotational ground state of the magnetic polaron in LST, for the vibrational ground state ($n=1$) and the first four vibrationally excited states ($n=2,...,5$). The maximum string length was $\ell_{\rm max}=200$.}
\label{fig:SpinonFromString}
\end{figure}

In Fig.~\ref{fig:SpinonFromString} we used Eq.~\eqref{eqTrugmanTBfromLST} to calculate the diagonal spinon hopping elements $t_{\rm T}$ for the lowest vibrational string states $n=1,...,5$. For small $J_z/t \ll 1$ we find the largest hoppings in units of $J_z$. For the ro-vibrational ground state of the string, $n=1$, the Trugman loop hopping $t_{\rm T}<0$ is always negative. This is due to the fact that we can choose $\phi_\ell > 0$ in the ground state, such that all overlaps in Eq.~\eqref{eqTrugmanTBfromLST} contribute a negative amount to $t_{\rm T}$. For vibrationally excited states $\phi_\ell$ has nodes where the wave function changes sign, thus reducing the Trugman loop hopping element. This effect explains the oscillatory behavior of $t_{\rm T}(J_z)$ observed as a function of $J_z$ for vibrationally excited states.

\subsubsection{Comparison to exact diagonalization}
To test the strong coupling description of magnetic polarons in the $t-J_z$ model, we compare our prediction for the dispersion relation to exact numerical simulations in small systems. First we study the shape of the dispersion in a $4$-by-$4$ lattice, followed by a discussion of the magnetic polaron bandwidth $W$ which has been calculated in larger systems in Refs.~\cite{Poilblanc1992,Chernyshev1999}. 

In the parton theory of magnetic polarons, we use a product ansatz for the strong coupling wave function where the holon is bound to the spinon, see Eq.~\eqref{eqStrongCplgWvfct}. As a result the momentum $\vec{k}$ of the magnetic polaron is entirely carried by the spinon; Its dispersion relation is determined from the effective spinon dispersion which is obtained from Eq.~\eqref{eq:Htrugman}:
\begin{equation}
\epsilon_{\rm s}(\vec{k}) = 2 t_{\rm T} \left[  \cos \l k_x + k_y \r +  \cos \l k_x - k_y \r \right].
\label{eq:spinonDispersiontJz}
\end{equation}

\begin{figure}[t!]
\centering
\epsfig{file=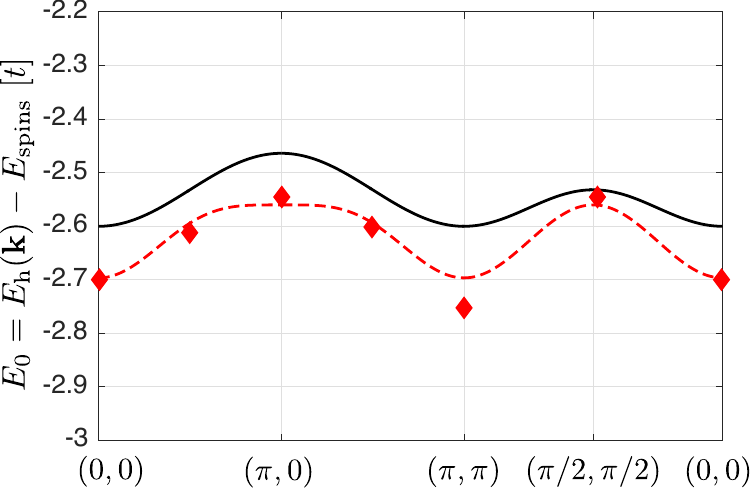, width=0.45\textwidth} $\qquad$
\caption{The magnetic polaron dispersion $E_0(\vec{k})$ is shown for the $t-J_z$ model at $J_z=0.2 t$ and $S=1/2$, relative to the energy $E_{\rm spins}$ of the ground state without the hole. The predictions from LST are compared to exact diagonalization (ED) for a small system with periodic boundary conditions. The solid black line is the LST prediction for an infinite system with a single hole. The dashed line is obtained by using the spinon dispersions for a $4 \times 4$ system, where the spinon hopping elements $t_{\rm T}$ are calculated in the infinite system. Note that an overall energy shift was added to the dashed line to simplify a direct comparison of the predicted bandwidths.}
\label{fig:DispersionComparisonEDIsing}
\end{figure}

In Fig.~\ref{fig:DispersionComparisonEDIsing} our results for the magnetic polaron dispersion relation are shown for $t/J_z=0.2$. We note that the ground state energy of the magnetic polaron on the $4 \times 4$ torus is rather well reproduced by LST. The minimum of the dispersion relation is predicted at $(0,0)$ and $(\pi,\pi)$ by LST. These two states are exactly degenerate in an infinite system because the N\'eel state breaks the sublattice symmetry. In the exact numerical calculation, the eigenstates with the two lowest energies are also located at momenta $(0,0)$ and $(\pi,\pi)$. We note that the numerical results show an energy difference $\sim 0.25 J_z$ between $(0,0)$ and $(\pi,\pi)$. This is a finite-size effect, indicating that the $4 \times 4$ lattice is not large enough to break the discrete translational symmetry of the $t-J_z$ model and form a N\'eel state. 

The shape of the dispersion in the $4 \times 4$ system deviates somewhat from the expectation in the infinite system. In particular, the energies at $(\pi,0)$ and $(\pm \pi/2, \pm \pi/2)$ are equal in this case. It is understood that this is a consequence of a particular symmetry of the $4 \times 4$ system \cite{Martinez1991}. We can also easily understand this from the tight-binding theory for spinon hopping: On a $4 \times 4$ cylinder there exists an additional Trugman loop leading to next nearest neighbor hopping along the lattice direction. In this case the holon moves around the torus one-and-a-half times in a straight manner. As a result we have to add an additional term $t_{\rm T} \left[ \cos(2k_x) + \cos(2 k_y) \right]$ to the spinon dispersion relation in Eq.~\eqref{eq:spinonDispersiontJz}, which leads to an exact degeneracy of $(\pi,0)$ and $(\pm \pi/2, \pm \pi/2)$. This dispersion relation is plotted as a dashed line on top of the data in Fig.~\ref{fig:DispersionComparisonEDIsing}. Note that we added a small overall energy shift to the dashed line to simplify a comparison with the shape of the exact dispersion relation.

\begin{figure}[t!]
\centering
\epsfig{file=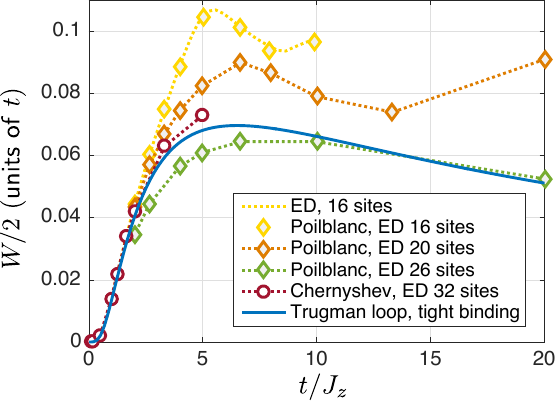, width=0.43\textwidth} $\quad$
\caption{Half the bandwidth $W/2$ of the lowest magnetic polaron band in the $t-J_z$ model: We compare exact numerical calculations for different system sizes by Poilblanc et al.~ \cite{Poilblanc1992} and Chernyshev and Leung~\cite{Chernyshev1999} to our tight-binding theory of Trugman loop processes.}
\label{fig:CompBandwithstJz}
\end{figure}

Comparison of the exact bandwidth with the LST calculation shows sizable quantitative deviations for the small $4 \times 4$ lattice. We expect that this is mostly due to finite size effects, as indicated for example by the absence of a degeneracy between momenta $(0,0)$ and $(\pi,\pi)$ as discussed above. We found that this effect becomes even more dramatic for smaller values of $J_z/t$. For example, at $J_z/t=0.1$ the energy difference between $(0,0)$ and $(\pi,\pi)$ is comparable to the entire bandwidth.

To study finite-size effects more systematically, we compare the bandwidth $W$ obtained from our tight-binding calculation to exact numerical results in larger systems obtained by Poilblanc et al.~ \cite{Poilblanc1992} and Chernyshev and Leung~\cite{Chernyshev1999}. From Eq.~\eqref{eq:spinonDispersiontJz} we see that the bandwidth can be directly related to the Trugman loop hopping, $W = 8 |t_{\rm T}|$. In Fig.~\ref{fig:CompBandwithstJz} we find excellent agreement between our tight-binding calculation and exact numerics in the regime where finite-size effects are small, for $t \lesssim 2 J_z$. For larger values of $t/J_z$, the numerical results begin to depend more sensitively on system size $N$. For the largest systems the behavior is even non-monotonic with $N$. The tight-binding result is closest to the data obtained for the largest system of $32$ sites solved by Chernyshev and Leung~\cite{Chernyshev1999}. 

As a function of $t/J_z$, we also observe non-monotonic behavior. For small $t / J_z$ the Trugman loop hopping is exponentially suppressed due to the presence of a large energy barrier. The amplitude $|t_{\rm T}|$ (in units of $t$) reaches a maximum around $t/J_z \approx 6$. For larger $t/J_z$ the strings become very long, leading to a saturation of $t_{\rm T}$ at a fraction of $J_z$ and thus a slow decay of $|t_{\rm T}| / t$. The numerics for finite-size systems shows a second increase of the bandwidth beyond some critical hopping $t > t_c$. For the $4 \times 4$ system this value corresponds to the point where we observed that the bandwidth becomes comparable to the energy splitting between momenta $(0,0)$ and $(\pi,\pi)$. This signals that the translational symmetry is not broken, which is a strong indication for a finite-size effect. Indeed, the critical value $t_c/J_z$ quickly increases for larger system sizes. This is expected from the dependence of the average string length on $t/J_z$.

\subsubsection{Contributions from longer loops}
Longer strings can generate higher-order loops, which gives rise to further-neighbor spinon hopping processes preserving the sublattice index. To a good approximation, they can be neglected because they are exponentially suppressed by the string length. Consider for example the simplest Trugman loop which renormalizes the next-neighbor hopping of the spinon linearly along the lattice direction. It involves a string of length $\ell=10$, four units longer than for the simplest Trugman loop. This already reduces the corresponding tight-binding element by a factor of $3^{-4} \approx 0.012$. Similarly, there are four loops involving two plaquettes which renormalize the diagonal spinon hopping. They can change the value of $t_{\rm T}$ on the level of $4 \times 3^{-4} \approx 5 \%$.

\section{Dynamical properties of magnetic polarons}
\label{sec:DynProp}

In this section we apply the strong coupling parton theory of magnetic polarons introduced in the previous section. We consider ultracold atom setups and calculate the dynamics of a single hole in the AFM for various experimentally relevant situations. We discuss how such experiments allow to test the parton theory of magnetic polarons. In the following, it will be assumed that the temperature $T \ll J$ is well below $J$ where corrections by thermal fluctuations are negligible.

\subsection{Effective Hamiltonian}
\label{subsec:SpinonDynamicsEffHam}
The basic assumption in the strong coupling parton theory is that the holon can adiabatically follow the spinon dynamics. Together they form a magnetic polaron, which can be described by an operator $\hat{f}^\dagger_{j,n}$ creating a magnetic polaron at site $j$. Formally we can write
\begin{equation}
\hat{f}^\dagger_{j,n} \ket{0} = \sd_j \ket{0} \otimes \ket{\psi^{(n)}}_{\rm BL},
\end{equation}
where $\sd_j$ creates a spinon on site $j$ of the square lattice; $\ket{\psi^{(n)}}_{\rm BL}$ denotes the wave function of the spinon-holon bound state on the Bethe lattice with ro-vibrational quantum number $n$.

The effective Hamiltonian of the magnetic polaron in the $t-J_z$ model reads,
\begin{equation}
\H_{\rm eff} = \sum_{n,j}  E(n) \hat{f}^\dagger_{j,n} \hat{f}_{j,n} +\sum_{n, \langle \ij \rangle_d}  t_{\rm T}(n)  \l \hat{f}^\dagger_{j,n} \hat{f}_{i,n} + \hc \r.
\end{equation}
The energy $E(n)$ of the spinon-holon bound state, as well as the Trugman loop hopping element $t_{\rm T}(n)$ (Eq.~\eqref{eq:tTrugTightBinding}) depend on the quantum number $n$, see Figs.~\ref{fig:StringExcitations} and \ref{fig:SpinonFromString}. Inter-band transitions, where $n$ is changed by a hopping process of the spinon, can be safely neglected in the strong coupling limit.

\subsection{Ballistic propagation: spinon dynamics}
\label{subsec:SpinonDynamics}
We begin by studying coherent spinon dynamics. As an initial state we consider a spinon localized in the origin of the square lattice, and the holon in its ro-vibrational ground state,
\begin{equation}
\ket{\Psi_0} = \hat{f}^\dagger_{0,1} \ket{0} =  \sd_0 \ket{0} \otimes \ket{\psi^{(n=1)}}_{\rm BL}.
\label{eqPsiIniLocGS}
\end{equation}
A protocol how this state can be prepared experimentally is presented below.

\begin{figure}[t!]
\centering
\epsfig{file=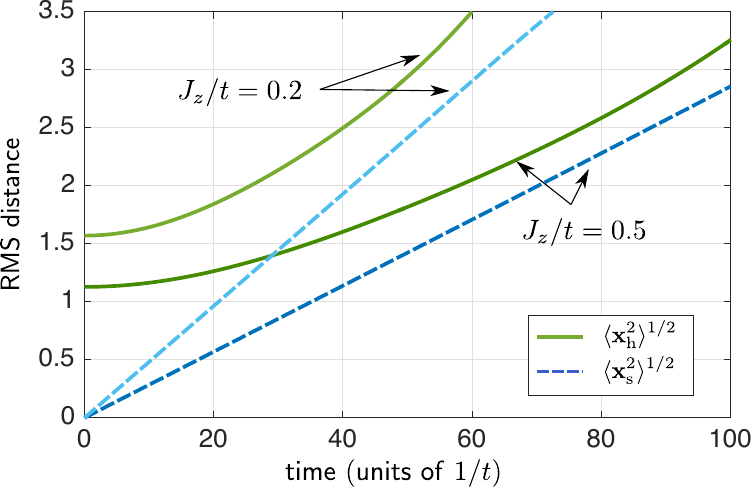, width=0.48\textwidth}
\caption{The root-mean square (RMS) radii $\langle \hat{\vec{x}}_{\rm s, h}^2 \rangle^{1/2}$ of the spinon and hole density distributions $\langle \hat{n}_{\rm s, h}(\vec{x}) \rangle$ are calculated as function of time, for $S=1/2$. We start from a localized spinon and the holon is initially in its ro-vibrational ground state. The dynamics of the hole distribution is dominated by the ballistic expansion of the spinon at long times.}
\label{fig:SpinonRelease}
\end{figure}

\subsubsection{Spinon expansion}
The dynamics of the magnetic polaron can be characterized by the root-mean square (rms) radii of the density distributions $\hat{n}_{\rm s,h}(\vec{x})$ of the spinon and the holon respectively,
\begin{equation}
\langle \hat{\vec{x}}_{\rm s,h}^2 \rangle^{1/2} = \l \sum_{\vec{x}} \vec{x}^2 \langle \hat{n}_{\rm s,h}(\vec{x}) \rangle \r^{1/2}.
\label{eq:defRMS}
\end{equation}
Both quantities can be measured using quantum gas microscopes. In such experiments the holon density is obtained directly by averaging over sufficiently many measurements of the hole position. By imaging the spin configuration \cite{Boll2016} the position of the spinon as well as the string configuration can also be estimated, see discussion at the end of Sec.~\ref{subsec:compNLST}.

In Fig.~\ref{fig:SpinonRelease} the rms radii are shown as a function of the evolution time. Because the spinon is fully localized initially, it expands ballistically and its rms radius grows linearly in time. The initial holon extent, in contrast, is determined by the spinon-holon wave function on the Bethe lattice. When the radius of the spinon density distribution $\hat{n}_{\rm s}(\vec{x})$ becomes comparable to that of the holon distribution, the hole density $\hat{n}_{\rm h}(\vec{x})$ also starts to expand ballistically with the same velocity $v_s$ as the spinon distribution. 

\begin{figure}[t!]
\centering
\epsfig{file=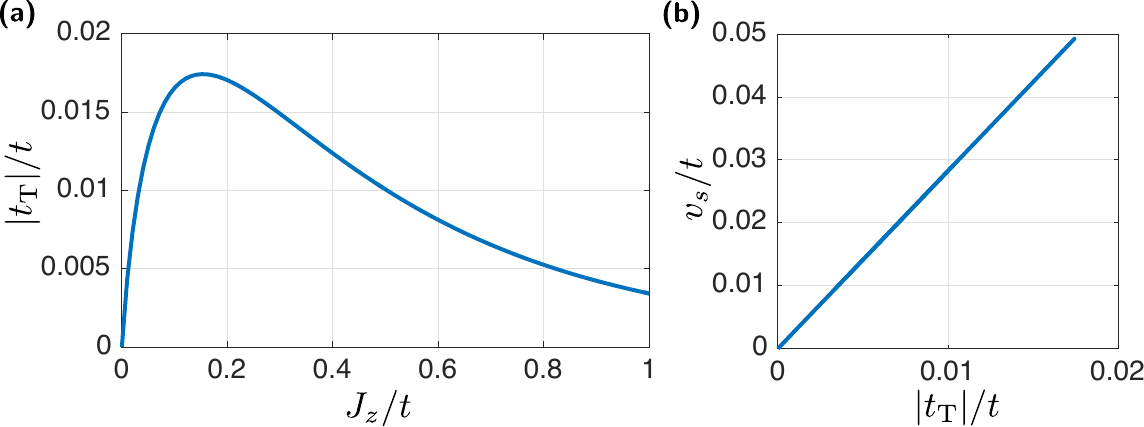, width=0.48\textwidth}
\caption{In the $t-J_z$ model the spinon velocity $v_s$ is determined by the Trugman loop hopping element $t_{\rm T}$ of the spinon, shown for various $J_z/t$ and $S=1/2$ in (a). The spinon velocity $v_s$, defined by the rms radius of the spinon density distribution, is directly proportional to $t_{\rm T}$ (b).}
\label{fig:SpinonReleaseVelocities}
\end{figure}

In the $t-J_z$ model the velocity $v_s$ is directly proportional to the Trugman loop hopping element $t_{\rm T}$ of the spinon, $v_s \approx 3 |t_{\rm T}|$, see Fig.~\ref{fig:SpinonReleaseVelocities}. Experimentally this relation allows a direct measurement of the Trugman loop hopping element.

\subsubsection{Preparation of the initial state}
\label{subsubsec:PrepSchemeMagPol}
Experimentally the initial state \eqref{eqPsiIniLocGS} can be prepared by first pinning the hole on the central site of the lattice using a localized potential of strength $g$. When $g$ is decreased slowly compared to the energy gap to the first vibrationally excited state, the holon adiabatically follows its ground state on the Bethe lattice. 

When the potential is lowered quickly compared to the spinon hopping, during a time $\tau \gg 1 / |t_{\rm T}|$, we can assume that the spinon remains localized on the central site during the quench. Because of the symmetry around the initial spinon position, the rotational quantum numbers of the holon eigenstate on the Bethe lattice do not change. They remain trivial as in the fully localized initial state. Thus, the central trapping potential can only couple different vibrationally excited states of the holon on the Bethe lattice during the adiabatic dynamics. In the limit when $g=0$ these eigenstates are separated by an energy gap $\Delta E \sim t^{1/3} J_z^{2/3} \gg J_z$. Hence one can adiabatically prepare the ro-vibrational ground state by choosing $1 / |t_{\rm T}| \ll \tau \ll 1/\Delta E$. During this time $\tau$ required for the preparation scheme, the spinon essentially remains localized.

\begin{figure}[t!]
\centering
\epsfig{file=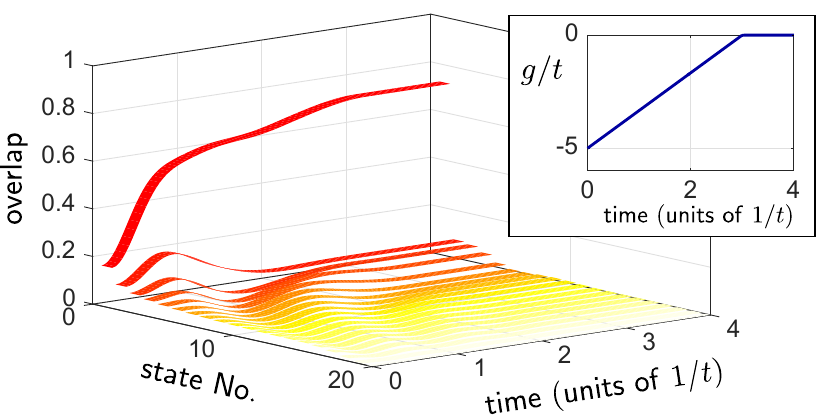, width=0.48\textwidth}
\caption{Adiabatic preparation of the ro-vibrational ground state of the holon. We start from a hole which is localized in the center by a pinning potential of strength $g$. The latter is decreased over a time $\tau = 5/t$ (inset) and the overlap of the holon state with the vibrational eigenstates of the magnetic polaron are calculated. We used LST for $S=1/2$ and neglected spinon dynamics. This is justified for the considered value $J_z/t=0.2$ in the strong coupling regime.}
\label{fig:MagPolPrep}
\end{figure}

\begin{figure*}[t!]
\centering
\epsfig{file=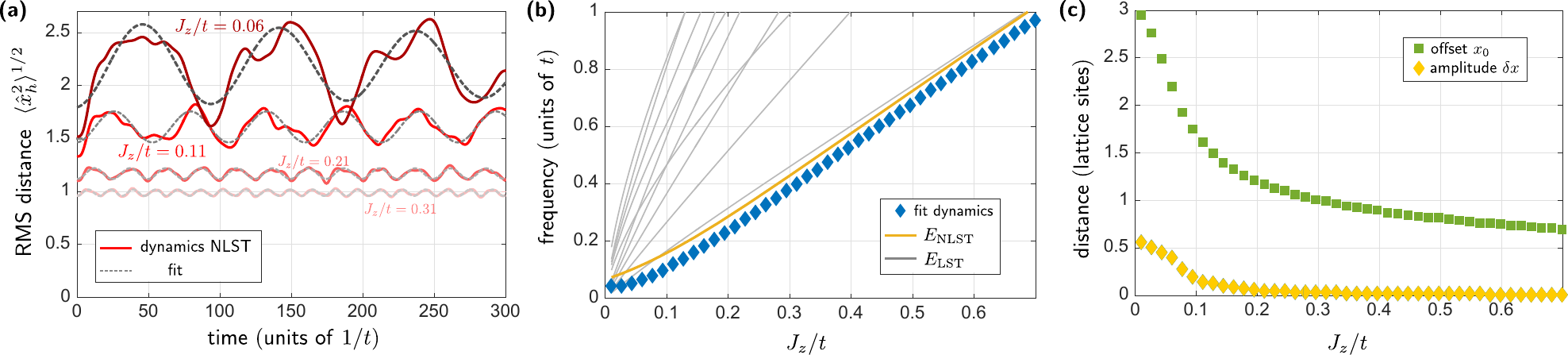, width=0.99\textwidth}
\caption{Bloch oscillation spectroscopy of magnetic polarons, at $S=1/2$. We start from a magnetic polaron in its ro-vibrational ground state and assume that the spinon is localized in the center of the lattice. Then a weak force $F=0.05 t/a$ is applied along the $x$-direction at time $t_0=0$. (a) Subsequently, the rms radius of the hole distribution along the force, $\langle \hat{x}_{\rm h}^2 \rangle^{1/2}$, begins to oscillate at different frequencies depending on the ratio $J_z/t$. It was calculated in the spinon frame. The strongest oscillations come from rotational excitations at the frequency $\omega$, but admixtures from vibrational states are also visible. (b) We extract the frequency $\omega$ from fits to data like shown in (a) and plot it as a function of $J_z/t$. It is approximately equal to the excitation energy $\Delta E$ of the first rotational eigenstate of the holon on the Bethe lattice. (c) The offset $x_0$ and the amplitude $\delta x$ of the oscillations observed in (a) are plotted as a function of $J_z/t$. In these calculations we used NLST in the spinon frame, assuming a maximum string length $\ell_{\rm max}=9$.}
\label{fig:ForceSpectroscopy}
\end{figure*}

In Fig.~\ref{fig:MagPolPrep} we calculate the overlap of the initially localized holon state with the vibrational eigenstates of the magnetic polaron in the absence of a pinning potential ($g=0$) and using LST. As shown in the inset of the figure, we start from $g=-5 t$ and decrease the magnitude of the potential linearly over a time $\tau = 3 / t$. This is sufficiently fast to neglect spinon dynamics and, as explained above, this also justifies to ignore rotationally excited eigenstates. 

Fig.~\ref{fig:MagPolPrep} demonstrates that even without optimizing the adiabatic preparation scheme, large overlaps with the ro-vibrational ground state of the holon (around $80 \%$ in this case) can be readily achieved using this method.

\subsection{Bloch oscillation spectroscopy of the magnetic polaron spectrum}
\label{subsec:BreathingOscillations}
In the last section we explained how the slow dynamics of the hole density distribution allows to experimentally measure spinon properties (specifically $t_{\rm T}$ of the vibrational ground state). Now we present a scheme allowing to study the rotational excitations of the holon experimentally and measure the corresponding excitation energies discussed in Sec.~\ref{secStringExcitations}. 

As in the previous section, our starting point is a magnetic polaron in its ro-vibrational ground state. For simplicity we assume that the spinon is localized initially. As shown in Sec.~\ref{subsubsec:PrepSchemeMagPol} this state can be adiabatically prepared in experiments with a quantum gas microscope. 

To populate the first excited state, which has a non-trivial rotational quantum number, we apply a force to the fermions,
\begin{equation}
\H_F =  \sum_{j,\sigma} \vec{F} \cdot \vec{x}_j  ~ \cd_{j, \sigma}  \c_{j, \sigma}.
\label{eq:Hforce}
\end{equation}
This breaks the rotational symmetry around the initial spinon position. In a system at half filling with a single hole, the force corresponds to a potential $\H_F = - \vec{F} \cdot \hat{\vec{x}}_{\rm h}$ acting on the hole at position $\hat{\vec{x}}_{\rm h}$. A similar situation has been considered before by Mierzejewski et al. in Ref.~\cite{Mierzejewski2011}, where interesting transport properties of the magnetic polaron have been predicted in the regime of a strong force. The term Eq.~\eqref{eq:Hforce} in the Hamiltonian can be easily realized for ultracold atoms using a magnetic field or an optical potential gradient. 

Here we study the opposite limit of a weak force, $F \lesssim J_z$. The frequency of Bloch oscillations is given by $\omega_{\rm B} = a F$, where $a$ is the lattice constant. We assume that $\omega_{\rm B}$ is small compared to the gap $\Delta E$ to the first ro-vibrational excited state of the holon,
\begin{equation}
\omega_{\rm B} \lesssim \Delta E. 
\label{eq:conditionSmallPopulation}
\end{equation}

We consider a situation where the force is suddenly switched on at time $t_0$. In the subsequent time evolution, the population of the first excited state begins to oscillate with a high frequency given by $\omega = \Delta E$ (we set $\hbar = 1$). This also manifests in oscillations of the density distribution of the holon, with the same frequency $\omega$. Because the density distribution of the hole can be directly measured, this allows to extract the excitation energy of the first rotationally excited state of the magnetic polaron experimentally. 

In Fig.~\ref{fig:ForceSpectroscopy} (a) we calculate the rms radius $\langle \hat{x}_{\rm h}^2 \rangle^{1/2}$ along the direction $x$ of the applied force $\vec{F} = F \vec{e}_x$, as a function of time. For various values of $J_z/t$ we observe clear, although not perfectly harmonic, oscillations. To extract their frequency and amplitude we performed the following fit to our numerical results,
\begin{equation}
\langle \hat{x}_{\rm h}^2 \rangle^{1/2} = x_0 + \delta x \cos (\omega t + \phi_0) e^{- \gamma t}.
\end{equation}
The fit parameters are the offset $x_0$, the amplitude $\delta x$, the frequency $\omega$ and phase $\phi_0$, and the decay rate $\gamma$ of the oscillations. 

In Fig.~\ref{fig:ForceSpectroscopy} (b) we plot the frequencies $\omega$ extracted from fits to the holon distribution for times $t_0,...,t_0+300/t$. These data points strongly depend on $J_z/t$ and are in excellent agreement with the energy gap to the first excited state of the magnetic polaron. We checked that the deviations from the exact value of the excitation energy $\Delta E$ are of the order of the Bloch oscillation frequency, $\Delta E - \omega \approx \omega_{\rm B}$. The decay rates $\gamma \ll \omega_{\rm B}$ were negligible for the case $Fa=0.05$ considered in Fig.~\ref{fig:ForceSpectroscopy}.

In Fig.~\ref{fig:ForceSpectroscopy} (c) we show the amplitude $\delta x$ and the offset $x_0$ as a function of $J_z/t$. We observe that the experimentally most interesting regime corresponds to cases where $J_z$ is not much larger than the Bloch oscillation frequency $\omega_{\rm B}$. Here the amplitude is sizable and the signal-to-noise ratio required in an experiment is not too small. In Fig.~\ref{fig:ForceSpectroscopy} we have kept the force $F$ constant. In order to obtain larger amplitudes of the oscillations it can also be tuned to larger values when $J_z/t$ is large.

\subsubsection{Effect of spinon dynamics}
Our calculations so far were performed using NLST in the spinon frame, see Fig.~\ref{fig:ForceSpectroscopy}. Now we discuss the effects of spinon dynamics on the density distribution of the hole. First we note that the force also acts on the spinon: When the spinon moves from site $i$ at position $\vec{x}_i$ to site $j$ at $\vec{x}_j$, the holon can follow adiabatically. The center of mass of the holon distribution is always given by the spinon position. When it changes from $\vec{x}_i$ to $\vec{x}_j$, this leads to an overall energy shift $\vec{F} \cdot (\vec{x}_i - \vec{x}_j)$ which corresponds to a force $\vec{F}$ acting on the spinon. 

In the $t-J_z$ model the spinon dynamics are slow, with a velocity set by the Trugman loop hopping element $t_{\rm T} \ll J_z$. This allows to realize a regime where the force $F \gg t_{\rm T}/a$ is large compared to $t_{\rm T}$. In this case the spinon is localized in a Wannier Stark state and it cannot move along the direction $\vec{e}_x$ of the force. At the same time the force can be chosen to be small compared to the excitation energy $\sim J_z$ of the first rotational state. This is necessary in order to observe coherent oscillations, see Eq.~\eqref{eq:conditionSmallPopulation}. Finally, the motion of the spinon in the direction $\vec{e}_y$ orthogonal to the force has no effect on the rms radius of the hole distribution in $x$-direction which is calculated in Fig.~\ref{fig:ForceSpectroscopy} (a). 

If the spinon hopping is comparable to the force, as expected for isotropic Heisenberg interactions between the spins, the spinon will undergo Bloch oscillations. When Eq.~\eqref{eq:conditionSmallPopulation} is satisfied, their frequency $\omega_{\rm B}$ is slow compared to the oscillations $\omega$ of the holon distribution. The resulting density of the hole in the laboratory frame, which can be directly measured experimentally, is a convolution of the spinon density and the holon distribution function. Thus by performing a Fourier analysis of the hole distribution, both frequency components $\omega_{\rm B}$ and $\omega$ can be extracted and analyzed separately.

\subsection{Far-from equilibrium dynamics: releasing localized holes}
\label{subsec:FarFromEqul}
So far our analysis was restricted to situations close to equilibrium, where mostly the ro-vibrational ground state of the magnetic polaron was populated. Now we consider the opposite limit where the hole is initially localized on the central site and the deep pinning potential is suddenly switched off at time $t_0$. This corresponds to the initial state
\begin{equation}
\ket{\Psi_0} = \sd_0 \hd(0) \ket{0} = \sum_n \psi_n \hat{f}^\dagger_{0,n} \ket{0}.
\end{equation}
In this case many different vibrational states $n$ of the magnetic polaron are populated (with amplitudes $\psi_n$). Recall that $\hd(0)$ creates a holon in the center of the Bethe lattice defined in the spinon frame, see Sec.~\ref{subsubsec:StringTheory}. From the symmetry of the initial state under $\hat{C}_4$ rotations of the holon around the spinon, it follows that no rotational excitations are created. While we focus on the zero temperature case here, the same problem was discussed in the opposite limit of infinite temperature and for $J_z=0$ in Refs.~\cite{Carlstrom2016PRL,Nagy2017PRB}.

\begin{figure}[t!]
\centering
\epsfig{file=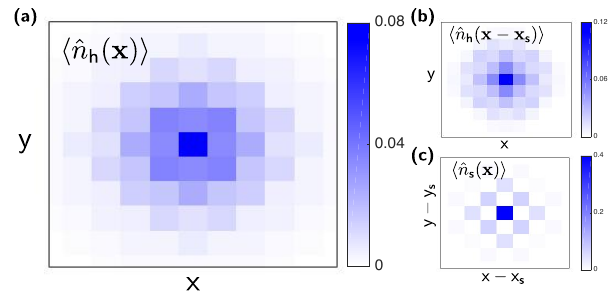, width=0.48\textwidth}
\caption{The hole distribution in the laboratory frame $\langle \hat{n}_{\rm h}(\vec{x}) \rangle$ (a) is obtained as a convolution of the spinon and holon wavefunctions. We consider far-from equilibrium dynamics of a single hole in the $t-J_z$ model which is initially localized in the center. The results are shown at a time of $50 /t$ and $J_z/t = 0.5$ was chosen. Our calculations are performed within LST for $S=1/2$ and assuming a maximum string length of $\ell_{\rm max}=30$. In (b) the holon distribution $\langle \hat{n}_{\rm}(\vec{x}-\vec{x}_{\rm s}) \rangle$ in the spinon frame is shown, and in (c) the spinon density in the laboratory frame is shown at the same final time $50 / t$.}
\label{fig:Convolution}
\end{figure}

\begin{figure}[t!]
\centering
\epsfig{file=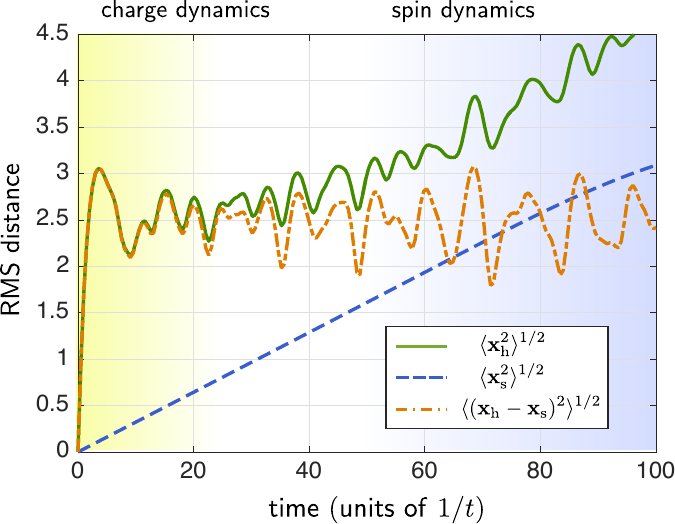, width=0.47\textwidth}
\caption{Far-from equilibrium dynamics of an initially localized hole in the $t-J_z$ model at $J_z=0.5 t$, as in Fig.~\ref{fig:Convolution}. The root-mean square (RMS) distance $\langle \hat{\vec{x}}_{\rm h}^2 \rangle^{1/2}$ of the hole density distribution $\langle \hat{n}_{\rm h}(\vec{x}) \rangle$ is dominated by fast charge dynamics at short times, and by slow spin dynamics at long times.}
\label{fig:HolonRelease}
\end{figure}

\emph{Theoretical technique.--}
To describe the far-from equilibrium dynamics after the quench at time $t_0$ we use LST and solve the multi-band problem defined in Sec.~\ref{subsec:SpinonDynamicsEffHam}. Here the vibrational quantum number $n$ acts like an effective band index. Note that the rotational symmetry of the Bethe lattice around the spinon position is unbroken. In particular, we calculate the density of the hole $\langle \hat{n}_{\rm h} (\vec{x}) \rangle$ in the laboratory frame which is given by
\begin{equation}
\langle \hat{n}_{\rm h} (\vec{x}_j) \rangle = \bra{\Psi} \hd_j \h_j \ket{\Psi}.
\end{equation}
Here $\hd_j$ is the holon operator in the laboratory frame corresponding to site $j$ on the square lattice. To evaluate the last expression we need the matrix elements
\begin{equation}
\mathcal{M}_{n',n}^{i',i}(j) = \bra{0} \hat{f}_{i',n'} \hd_j \h_j \hat{f}_{i,n}^\dagger \ket{0} \propto \delta_{i,i'}
\label{eqMatrixElementsNh}
\end{equation}
which we calculated using Monte Carlo sampling over states on the Bethe lattice.

\emph{Numerical results.--}
In Fig.~\ref{fig:Convolution} (a) we show the density distribution of the hole after the quench. It can be understood as a convolution of the spinon and the holon wavefunctions, whose density distributions are calculated in Fig.~\ref{fig:Convolution} (b) and (c). Note however that interference terms between terms from different bands $n \neq n'$, included in Eq.~\eqref{eqMatrixElementsNh}, are taken into account in Fig.~\ref{fig:Convolution} (a).

In Fig.~\ref{fig:HolonRelease}, we analyze the rms radius of the hole density distribution in the laboratory frame, for the same situation as in Fig.~\ref{fig:Convolution}. We consider a large value of $J_z/t=0.5$ but checked that the qualitative behavior is identical at smaller values of $J_z/t$. Our theoretical approach was benchmarked in Fig.~\ref{fig:QMC}. There we compared our results to time-dependent quantum Monte Carlo calculations at short-to-intermediate times and found excellent agreement with predictions by the LST. 

As a main result of Fig.~\ref{fig:HolonRelease}, we observe a clear separation of spinon and holon dynamics. At short times, the hole distribution is entirely determined by fast holon dynamics on the Bethe lattice, with a characteristic time scale $\sim 1/t$, see also Fig.~\ref{fig:QMC}. Additionally, for short times up to $\sim 20/t$ the rms radii $\langle \hat{\vec{x}}_{\rm h}^2 \rangle^{1/2}$ and $\langle (\hat{\vec{x}}_{\rm h} - \hat{\vec{x}}_{\rm s})^2 \rangle^{1/2}$ calculated in the lab and spinon frames respectively almost coincide in Fig.~\ref{fig:HolonRelease}.

At long times, we observe a ballistic expansion of the hole distribution. This behavior is similar to the coherent spinon dynamics discussed in Sec.~\ref{subsec:SpinonDynamics}, see Fig.~\ref{fig:SpinonRelease}. Similar to that case, we observe that the rms radius of the hole is equal, up to a constant offset, to the rms radius of the spinon distribution $\langle \hat{\vec{x}}_{\rm s}^2 \rangle^{1/2}$ at long times. The characteristic velocity is determined by the Trugman loop hopping element $t_{\rm T}$ of the spinon. It is much smaller than the bare hole hopping $t$, explaining the large separation of time scales observed in Fig.~\ref{fig:HolonRelease}. 

The separation of characteristic spinon and holon timescales is a hallmark of the parton theory of magnetic polarons. It can be understood as a precursor of true spin-charges separation: on short length scales (short times), the holon behaves almost as if it was free, resembling the physics of asymptotic freedom known from quarks in high-energy physics. Only on long length scales (long times) the dynamics of magnetic polarons become dominated by the slow spinon, and it becomes apparent that the spinon and the holon are truly confined.

Another key indicator for the string theory of magnetic polarons is the scaling of the energy spacings between low-energy vibrational excitations with the non-trivial power-law $t^{1/3} J_z^{2/3}$, see Ref.~\cite{Bulaevskii1968} and Sec.~\ref{secStringExcitations}. It is tempting to search for the same universal power-law in the far-from equilibrium dynamics considered here. To this end, we analyzed the position $t_1$ in time of the first pronounced maximum of the rms radius of the hole. For example, in Fig.~\ref{fig:HolonRelease} it is located around $t_1 \approx 3.5/t$. By repeating this procedure for various ratios $J_z/t$ we found a trivial power-law $t_1 \propto J_z$. 

To understand why we did not obtain the non-trivial power-law $t^{1/3} J_z^{2/3}$ we note that in the initial state the hole was localized on a single lattice site. The continuum approximation leading to the Schr\"odinger equation \eqref{eq:ContiSchrEq} is thus not justified and lattice effects become important. In particular this violates the scale invariance of the Schr\"odinger equation which gives rise to the characteristic $J_z^{2/3}$ power-law.

If on the other hand one starts from a low-energy state where the holon is distributed over several lattice sites, universal charge dynamics can be observed. Such situations can be realized experimentally by adiabatically releasing the hole from a pinning potential, see Sec.~\ref{subsubsec:PrepSchemeMagPol}. A similar situation has been studied by Golez et al. in Ref.~\cite{Golez2014} where a different quench was considered starting from the ground state of the magnetic polaron. By rescaling the time axes by a factor of $t^{1/3} J_z^{2/3}$ the authors of Ref.~\cite{Golez2014} demonstrated a collapse of their data at various values of $J_z/t$ to a single universal curve for sufficiently short times.

\subsection{Spectroscopy of magnetic polarons}
\label{subsec:Spectroscopy}
The most direct way to confirm the $t^{1/3} J_z^{2/3}$ power-law describing the energy of vibrationally excited states is to measure the spectral function $A(\omega,k)$ of the magnetic polaron. Self-consistent Greens function calculations by Liu and Manousakis \cite{Liu1991} as well as diagrammatic Monte Carlo calculations by Mishchenko et al.~\cite{Mishchenko2001} assuming isotropic Heisenberg couplings between the spins have been performed within the $t-J$ model. They showed strong evidence that the spectrum consists of several broad peaks above the magnetic polaron ground state, which have an energy spacing scaling like $t^{1/3} J^{2/3}$. 

The spectral function can be measured in solid state systems using ARPES. Similarly, for ultracold fermions radio-frequency  \cite{Chin2004,Stewart2008}, Bragg \cite{Veeravalli2008} and lattice modulation \cite{Jordens2008} spectroscopy measurements have been carried out. In quantum gas microscopes the spectrum can also be measured with full momentum and energy resolution \cite{Bohrdt2017spec} by modulating the tunneling rate of ultracold atoms into an empty probe system.

\begin{figure}[b!]
\centering
\epsfig{file=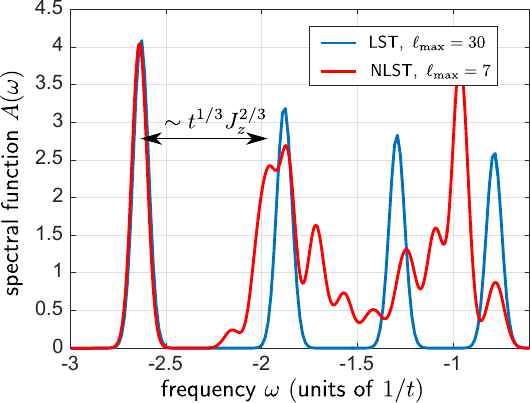, width=0.43\textwidth} $\quad$
\caption{The spectral function $A(\omega)$ is calculated in the strong coupling regime, here for $J_z/t=0.2$ and $S=1/2$. We have calculated the spectral weights $Z_n$ and energies $E_n$ of all eigenstates $n$ and broadened them by hand for clearer presentation. LST predicts a series of approximately equally spaced peaks, which evolve into a broad continuum if string-interactions are included in the NLST. The gap from the magnetic polaron peak at the lowest energy to the first excited states with appreciable spectral weight scales with the non-trivial power-law $t^{1/3} J_z^{2/3}$ characteristic of the string theory.}
\label{fig:LocalSpectrum}
\end{figure}

Here we briefly discuss the properties of the magnetic polaron spectrum in the strong coupling regime where $t \gg J_z$. Because the dispersion of the magnetic polaron is determined by the small Trugman loop hopping $t_{\rm T}$, the momentum dependence of the spectrum can be neglected, i.e. $A(\omega,k) \approx A(\omega)$. This allows for a purely local measurement of the spectral function, which significantly simplifies implementations in a quantum gas microscope \cite{Kollath2007,Bohrdt2017spec}.

In general, a series of peaks is expected which correspond to vibrationally excited states of the string. As shown in Fig.~\ref{fig:LocalSpectrum} and previously in Ref.~\cite{Kane1989} this is indeed what LST predicts. In the NLST substantial corrections are present at high energies which lead to a broadening of the spectral lines. At low energies we observe a gap both within LST and NLST which scales like $t^{1/3} J_z^{2/3}$.

In a measurement of the spectral function a fermion is removed from the system and the spinon and the holon are always created at the same site. As a consequence, the initial state of the holon on the Bethe lattice is rotationally invariant around the spinon position. This gives rise to a selection rule and leads to vanishing spectral weight of string states with rotational excitations. Thus the latter cannot be observed in $A(\omega)$. This explains why the spectral gap in Fig.~\ref{fig:LocalSpectrum} corresponds to the gap to vibrational excitations, in contrast to the rotational states observed in the Bloch oscillation spectroscopy of Sec.~\ref{subsec:BreathingOscillations}. In that case the rotational symmetry was explicitly broken by the applied force.

\section{Extensions to the $t-J$ model}
\label{sectJoutlook}

So far we have formulated the strong coupling parton description of magnetic polarons specifically for the $t-J_z$ model. As an important outcome of our work, the approach can be generalized to models including quantum fluctuations of the surrounding spin system. Most importantly, we can apply it to describe the dynamics of holes in the $t-J$ model, which is believed to play a central role in the understanding of high-temperature superconductivity. In the following we summarize how such generalizations work and present first results of this method relevant for current experiments with quantum gas microscopes. A more detailed discussion will be provided in forthcoming papers.

Our work on the $t-J_z$ model enables two possible extensions to the $t-J$ model. The first uses the generalized $1/S$ expansion, see Sec.~\ref{subsec:LargeSholehopping} and Appendix \ref{apdxQuantFlucGenS}. It is based on the idea to start from a classical N\'eel state which is distorted by the holon motion as in the case of the $t-J_z$ model. Transverse couplings between the spins are then included by applying linear spin wave theory around the distorted classical state. Even in the presence of quantum fluctuations this approach allows to introduce spinons and strings in the theory. The properties of the resulting mesons, for example the string tension, are renormalized in this case due to polaronic dressing by magnon fluctuations. 

Here we discuss a second extension, which is based on an analogy with the squeezed space picture of the one-dimensional $t-J$ model \cite{Ogata1990,Kruis2004a,Hilker2017}. Instead of labeling the basis states by the eigenvalues of the spin operators $\hat{S}^z_j$ on lattice sites $j$, we keep track of the holon trajectory $\Sigma$ defined on the Bethe lattice. The motion of the hole changes the original positions of the spins, and their quantum state is defined in a pure spin system without doping on the original square lattice; we identify the latter with the two-dimensional analogue of squeezed space. Similar to Eq.~\eqref{eqHpartontJz} in the case of the $t-J_z$ model, the Hilbert space used in the parton theory of a single hole in the $t-J$ model becomes
\begin{equation}
\mathscr{H}_{\rm p} = \mathscr{H}_{\rm s} \otimes \mathscr{H}_{\Sigma} \otimes \mathscr{H}_{\rm mag}.
\label{eqHpartontJ}
\end{equation}
The first two terms $\mathscr{H}_{\rm s}$ and $\mathscr{H}_{\Sigma}$ describe the geometric string connecting the spinon at one end with the holon at the other end, i.e. the meson degrees of freedom. The last term $\mathscr{H}_{\rm mag}$ includes quantum fluctuations in the surrounding spin system -- e.g. magnons in the case of a quantum N\'eel state.

At strong couplings, when $t \gg J$, we can make a product ansatz to describe the magnetic polaron in the $t-J$ model,
\begin{equation}
\ket{\psi_{\rm mag. pol.}} = \ket{\psi_{\rm spinon}} \otimes \ket{\psi_{\rm holon}} \otimes \ket{\psi_{\rm spins}}_{\rm sq}.
\label{eqStrongCplgWvfctTJ}
\end{equation}
The last term describes the underlying spins in squeezed space. For a single hole in the $t-J$ model we choose it to be the ground state of a 2D Heisenberg AFM on a square lattice. The strong coupling ansatz is a direct generalization of Eq.~\eqref{eqStrongCplgWvfct} for the case of the $t-J_z$ model, where $\ket{\psi_{\rm spins}}_{\rm sq}$ is given by a classical N\'eel state. 

As a first step, we confirm that spinons and holons are confined in the $t-J$ model. We use the strong coupling wavefunction Eq.~\eqref{eqStrongCplgWvfctTJ} as a variational ansatz and note that geometric strings in general correspond to highly excited states of the surrounding spin system. Two spins which are nearest neighbors (NN) in squeezed space can become next nearest neighbors (NNN) in real space after the holon has moved through the system, and vice-versa. 

To estimate the energy $E_\Sigma$ of the resulting state, we consider straight holon trajectories for simplicity, for which $E_\Sigma \propto \ell$ is proportional to the length $\ell$ of the geometric string. The string tension obtained within this LST approximation is given by
\begin{equation}
\frac{dE_\Sigma}{d\ell} = 2 J (C_2 - C_1). 
\label{eqdESigmadEll}
\end{equation}
Here $C_{1}$ and $C_2$ denote the NN and NNN spin-spin correlation functions in squeezed space. In the case of the $t-J_z$ model $C_1 = - C_2 = -S^2$, and by setting $J = J_z$ we obtain the string tension $4 J_z S^2$ which we previously used for the LST, see Eq.~\eqref{eq:HLST}. 

Because the ground state expectation values $C_{1,2}$ can be determined numerically for the 2D Heisenberg AFM, we can calculate the LST string tension for the $t-J$ model: $dE_\Sigma / d\ell  \approx 1.1 J$. For $S=1/2$ the string tension for the $t-J$ model in units of $J$ is about $10 \%$ larger than for the $t-J_z$ model in units of $J_z$. This can be understood by noting that the isotropic term $J \hat{\vec{S}}_i \cdot \hat{\vec{S}}_j$ in the $t-J$ model includes couplings in $x$ and $y$-direction in spin space, in addition to the Ising coupling relevant to the $t-J_z$ model. As a result, we expect that spinons and holons are confined in the $t-J$ model with typical string lengths of comparable size in both models. 

Next we check our assumption in Eq.~\eqref{eqStrongCplgWvfctTJ} that the motion of the hole only has a negligible effect on the spins in squeezed space. To this end we compare the typical time scales for spin-exchange processes, $\tau_{\rm fluc} = 1/J$, with the typical time $\tau_{\rm h} = \ell_0 / t$ it takes the holon to cover a distance given by the average string length $\ell_0$. If $\tau_{\rm h} < \tau_{\rm fluc}$, quantum fluctuations do not have sufficient time to adapt to the new configuration of the geometric string and we expect that spin correlations in squeezed space are only weakly modified by the presence of the hole. Indeed, in the entire strong coupling regime $t \gg J$ we find that $\tau_{\rm h} \ll \tau_{\rm fluc}$. When $t/J = 3$, as in the case relevant to high-temperature superconductors \cite{Lee2006}, we obtain $\tau_{\rm h} / \tau_{\rm fluc} \approx 0.4$. This suggests that the strong coupling wavefunction \eqref{eqStrongCplgWvfctTJ} provides a good description of magnetic polarons in the $t-J$ model.

\begin{figure}[t!]
\centering
\epsfig{file=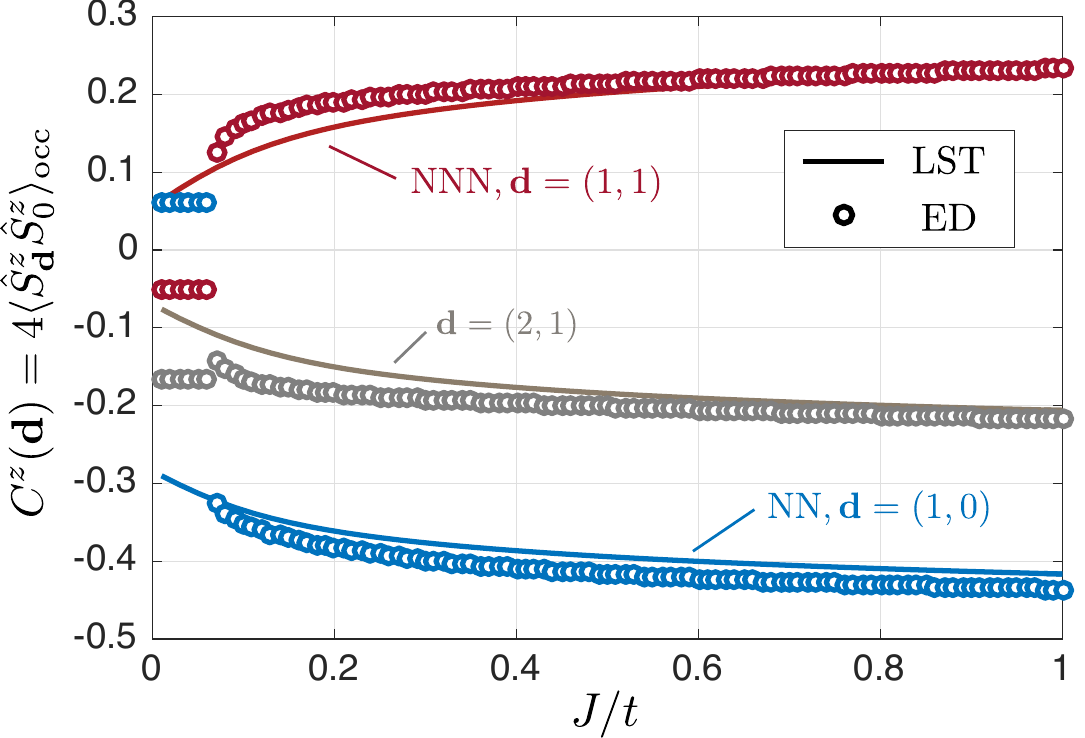, width=0.48\textwidth}
\caption{Spin-spin correlation functions $C^z(\vec{d}) = 4 \langle \hat{S}^z_{\vec{d}} \hat{S}^z_0 \rangle_{\rm occ}$ in a weakly doped $t-J$ model. We compare results by exact diagonalization (ED) of one hole in a $4 \times 4$ system with periodic boundary conditions to predictions from LST based on the strong-coupling wavefunction Eq.~\eqref{eqStrongCplgWvfctTJ}. The correlators are evaluated only for states where both sites $\vec{0}$ and $\vec{d}$ are occupied by spins, as indicated by the notation $\langle ... \rangle_{\rm occ}$. In the ED calculations we set $S^z_{\rm tot}=-1/2$ and simulated the sector with total conserved momentum of $\vec{k}=(\pi/2,\pi/2)$.}
\label{figCorrtJ}
\end{figure}

As an application of the strong coupling meson theory, we use it to calculate local spin-spin correlations in the presence of a hole. They are directly accessible using quantum gas microscopes \cite{Boll2016} and can be used to test our predictions experimentally. In Fig.~\ref{figCorrtJ} we compare exact numerical simulations of the $t-J$ model for a single hole with results from LST. Remarkably, the strong coupling meson theory can explain quantitatively the dependence of the spin correlations on the ratio $J/t$, caused entirely by doping. It merely combines the holon motion on the Bethe lattice with a set of spin correlators in a pure spin system without doping: Essentially NN spin correlations in real space acquire contributions from NNN, or even more distant, correlators in squeezed space when the holon motion is taken into account. This leads to the observed decrease (increase) of NNN (NN) correlations. 

We find good agreement of LST with ED simulations in Fig.~\ref{figCorrtJ} for NN correlators in the regime $J \gtrsim 0.1 t$. For correlators between sites which are further apart, there is still perfect qualitative agreement, although some quantitative differences are visible. Only at very small values of $J/t \lesssim 0.1$ we find strong deviations of LST from ED results, which can be associated with a transition to a state with strong ferromagnetic NN correlations reminiscent of the Nagaoka effect \cite{Nagaoka1966}. We expect that this transition is strongly influenced by the finite system size.

In a forthcoming work, we show how spinon dynamics can be included and more properties of magnetic polarons in the $t-J$ model can be understood from a strong-coupling meson description. The most dramatic difference as compared to the $t-J_z$ model is that spinon motion is possible not only via Trugman loops, but also by direct spin-exchange processes. This readily explains why some properties of holes in the $t-J$ and $t-J_z$ models -- including for example the linear dependence of their energy on $t^{1/3} J^{2/3}$ \cite{Mishchenko2001}-- are identical, whereas the dispersion relations are entirely different.

Finally, we note that the strong coupling ansatz in Eq.~\eqref{eqStrongCplgWvfctTJ} can be generalized to finite temperatures by using a product of density matrices, as described for one dimension in Ref.~\cite{Hilker2017}. In this case the correlators $C_1$ and $C_2$ determining the string tension in Eq.~\eqref{eqdESigmadEll} are calculated at finite temperature. The resulting weaker correlations lead to a reduced string tension, and thus to longer strings. As shown in Ref.~\cite{Nagy2017PRB}, even at infinite temperature and for $J \ll t$, the Hilbert space $\mathscr{H}_\Sigma$ defined by geometric strings provides a qualitatively accurate description of the single-hole $t-J$ model.

\section{Summary and Outlook}
\label{sec:Summary}
When holes are doped into an anti-ferromagnet at low concentration, they form quasiparticles called magnetic polarons. In this paper we have introduced a strong coupling parton theory for magnetic polarons in the $t-J_z$ model. Starting from first principles, we introduced a parton construction where the magnetic polaron is described as a bound state of a spinon and a holon. The spinon carries its fractional spin $S=1/2$ and the holon its charge $Q=-1$ quantum numbers. The two partons are connected by a string of displaced spins. This description combines earlier theoretical approaches to magnetic polarons, see in particular Refs.~\cite{Bulaevskii1968,Trugman1988,Kane1989,Manousakis2007}. Our formalism can be generalized to Hamiltonians including quantum fluctuations, e.g. the isotropic $t-J$ model. 

Ultracold atoms provide a new opportunity to investigate individual magnetic polarons experimentally, on a microscopic level. In this paper, we have discussed several realistic setups to measure their properties and study their dynamics far from equilibrium. We start from a single hole which can be pinned by a tightly focused laser beam. To investigate magnetic polarons in equilibrium using quantum gas microscopes, we proposed an adiabatic preparation scheme where the pinning potential is adiabatically released. If the laser potential localizing the hole is suddenly switched off, on the other hand, interesting far-from equilibrium dynamics can be studied.

Using the capabilities of quantum gas microscopy, direct observations of the constituent partons forming the magnetic polaron are possible. We analyzed this problem and showed that the full distribution function $p_\ell$ of string lengths $\ell$ connecting spinons and holons can be accurately measured.

As a hallmark of the parton theory, we point out the existence of rotational as well as vibrational excited states of magnetic polarons. Similar to mesons in high-energy physics, which are understood as bound states of two confined quarks, these excited states give rise to resonances with a well-defined energy. Creating the first vibrational excitation costs an energy $\sim t^{1/3} J_z^{2/3}$, whereas the first rotational excitation only costs an energy $\sim J_z$. While vibrational excitations of magnetic polarons have been discussed previously in the context of angle-resolved photoemission spectroscopy (ARPES), rotational resonances are invisible in ARPES spectra due to selection rules. To observe rotational states of magnetic polarons we propose to use Bloch oscillation spectroscopy: In this method a weak force is applied to the hole, which drives transitions to the first rotationally excited state at energy $\Delta E$. This causes oscillations of the hole density distribution at frequency $\Delta E$, which can be measured in experiments with ultracold atoms.  

We have applied the strong coupling parton theory to study dynamical properties of holes inside an anti-ferromagnet. Our calculations are based on a strong-coupling product ansatz for describing the spinon and holon parts of the wavefunctions. In contrast to models with transverse couplings between the spins, the dynamics of spinons in the $t-J_z$ model is generated entirely by Trugman loop trajectories of the holon which restore the surrounding N\'eel order. To calculate the resulting hopping strength of the spinon we developed a tight-binding formalism which is valid for arbitrary values of $J_z/t$. Our predictions are in excellent agreement with exact calculations for the largest system sizes which are numerically feasible \cite{Poilblanc1992,Chernyshev1999}. 

We have considered a far-from equilibrium situation, where a single hole is created in a N\'eel state by removing the central spin. As a benchmark, we have performed time-dependent quantum Monte Carlo calculations for the $t-J_z$ model at short-to-intermediate times. We compared them to our strong coupling parton theory and obtained very good quantitative agreement for the accessible times. At short times, we predict universal holon expansion. At intermediate times $\sim 1/J_z$, we observe a saturation of the string length connecting the spinon and the holon. At much longer times, the ballistic expansion of the spinon becomes dominant. 

The strong coupling parton description of magnetic polarons can be extended to study a larger class of problems. In a first approach, we will include magnon corrections describing quantum fluctuations around the classical N\'eel state. This allows us to study, for example, the dispersion relation of magnetic polarons in the usual $t-J$ model with Heisenberg interactions between the spins. In addition, spin-hole correlation functions can be calculated, which are directly accessible in experiments with quantum gas microscopes \cite{Boll2016}. The dynamics of the magnon cloud forming around the spinon-holon bound state can also be studied. In a second approach, we generalize the concept of squeezed space known from one dimension \cite{Ogata1990} and calculate properties of the magnetic polaron in the $t-J$ model using a variational wavefunction. In this paper we have presented first results of this approach, which can be tested using current experiments with fermionic quantum gas microscopes.

The couplings to phonons present in real solids can also be included in our theory using the Landau-Pekar wavefunction \cite{Landau1948}. Far-from equilibrium dynamics can then be studied, where the spinon becomes correlated with the magnons and the phonons. This can be of particular interest to describe recent experiments where cuprates have been driven far-from equilibrium by a short laser pulse \cite{Fausti2011}.

\section*{Acknowledgements}
The authors acknowledge fruitful discussions with Immanuel Bloch, Debanjan Chowdhury, Ignacio Cirac, Sebastian Eggert, Christian Gross, Timon Hilker, Sebastian Huber, Michael Knap, Corinna Kollath, Zala Lenarcic, Izabella Lovas, Salvatore Manmana, Efstratios Manousakis, Anton Mazurenko, Giovanna Morigi, Matthias Punk, Subir Sachdev, Guillaume Salomon, Richard Schmidt, Todadri Senthil, Tao Shi, Lev Vidmar, Gergely Zarand, Johannes Zeiher and Zheng Zhu. Support from Harvard-MIT CUA, NSF Grant No. DMR-1308435 and from AFOSR Quantum Simulation MURI, AFOSR grant number FA9550-16-1-0323 is gratefully acknowledged. FG acknowledges support by the Gordon and Betty Moore foundation under the EPIQS program. AB acknowledges support from the Technical University of Munich - Institute for Advanced Study, funded by the German Excellence Initiative and the European Union FP7 under grant agreement 291763, the DFG grant No. KN 1254/1-1 and from the Studienstiftung des deutschen Volkes. DG acknowledges support from the Harvard Quantum Optics Center and the Swiss National Science Foundation.

\newpage

\appendix

\section{Time-dependent quantum Monte Carlo calculations}
\label{apdx:TdepQMC}

We determine the hole's propagation in the $t-J_z$ model, Eq.~\eqref{eq:model} with $J_\perp=0$, at short and intermediate times using a real-time quantum Monte Carlo procedure. This allows us to take into account the interactions between the spins and the hopping of the hole in a numerically exact way. Our results are shown in Fig.~\ref{fig:QMC}. In the case of a single hole in the system, the Hamiltonian $\hat{\mathcal{H}} = \hat{\mathcal{H}}_t + \hat{\mathcal{H}}_J$ in Eq.~\eqref{eq:model} can be rewritten as
\begin{eqnarray}
\hat{\mathcal{H}}_t &=& t \sum_{\langle i, j \rangle} \hat{h}_i^\dagger \, \hat{h}_j \, \tilde{\mathcal{F}}^\dagger_{ij} + \hc \\
\hat{\mathcal{H}}_J &=& J_z\sum_{\langle i,j \rangle} \hat{S}_i^z \, \hat{S}_j^z,
\end{eqnarray}
where the operator $\h_i$ annihilates a holon at site $i$. As the hole hops from site $i$ to $j$, the operator $\tilde{\mathcal{F}}^\dagger_{ij} = \sum_{\sigma} \hat{c}^\dagger_{j,\sigma} \, \hat{c}_{i,\sigma}$ moves the spin at site $i$ to site $j$. Note that the effect of $\tilde{\mathcal{F}}_{ij}$ is equivalent to that of $\hat{\mathcal{F}}_{ij}$, but we expressed it in terms of the original fermion operators now.

In the following we determine the dynamics of a hole created by removing the central spin from the classical N\'eel state, as discussed also in Sec.~\ref{subsec:FarFromEqul}. This initial state will be denoted by $| \Psi_0 \rangle$. We consider a system at zero spin temperature, but our method can easily be generalized to a spin environment at finite temperature. We determine the probability $p_i(\tau)$ of finding the hole at site $i$ after a propagation time $\tau$. It is given by $p_i (\tau) = \langle \Psi (\tau)|  \h_i^\dagger \, \h_i  | \Psi (\tau) \rangle$, where $|\Psi(\tau)\rangle = e^{-i \hat{\mathcal{H}} \tau} \, | \Psi_0\rangle$ denotes the quantum state of the system at time $\tau$.

Since both the hopping and the Ising interaction conserve the $z$ component of the  spin, we work in the interaction picture representation to solve the system's dynamics. 
The quantum state of the system at time $\tau$ is defined  in this representation as 
\begin{equation}
| \Psi_I(\tau) \rangle = e^{i \hat{\mathcal{H}}_J  \tau} \, | \Psi (\tau) \rangle,
\label{eq:Psi_I}
\end{equation}
and its time evolution is governed by
\begin{equation}
i \partial_\tau |\Psi_I(\tau) \rangle = \hat{\mathcal{H}}_{t, I}(\tau) \, | \Psi_I(\tau) \rangle,
\label{eq:interaction_picture}
\end{equation}
with $\hat{\mathcal{H}}_{t, I} (\tau) = e^{i \hat{\mathcal{H}}_J \tau} \, \hat{\mathcal{H}}_t \, e^{-i \hat{\mathcal{H}}_J \tau}.$
We solve Eq.~\eqref{eq:interaction_picture} using the time-ordered exponential \cite{Negele2008},
\begin{equation}
|\Psi_I(\tau) \rangle = T_{\tau^\prime} e^{-i \int_0^\tau d\tau^\prime \, \H_J(\tau^\prime)} \, |\Psi\rangle,
\label{eq:Psi_I_time_evolution}
\end{equation}
where $T_\tau$ denotes the time ordering operator. The last equation also determines the state of the system $| \Psi(\tau) \rangle$ in the Schr\"odinger picture through Eq.~\eqref{eq:Psi_I}.

\begin{figure*}[t!]
\centering
\epsfig{file=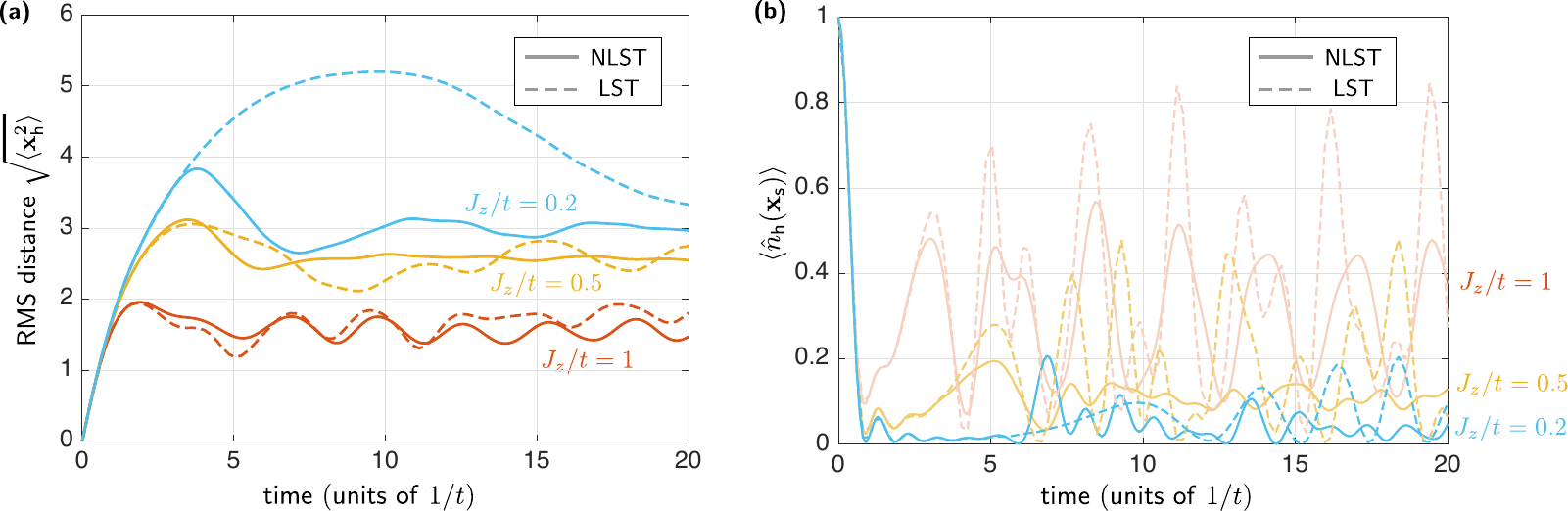, width=0.96\textwidth}
\caption{Comparison of LST and NLST for $S=1/2$. The starting point is a single hole created in the lattice by removing the central spin. (a) We calculate the rms radius of the hole density distribution as a function of time. Note that LST includes slow spinon dynamics, whereas NLST assumes a localized spinon for simplicity. In (b) we calculate the probability to find the spinon and the holon on the same lattice site. The maximum string lengths used are $\ell_{\rm max}=10$ (NLST) and $\ell_{\rm max}=30$ (LST).}
\label{fig:LSTvsNLST}
\end{figure*}

As the hole propagates through the lattice, it explores all possible paths simultaneously. Every such random walk path can be associated with a phase. This becomes apparent by expanding the time evolution in Eq.~\eqref{eq:Psi_I_time_evolution} in powers of the hopping Hamiltonian 
\begin{equation}
e^{-i \, \hat{\mathcal{H}} \tau } = e^{-i \H_J \tau} \, T_{\tau^\prime} \, \sum_{n=0}^\infty \frac{(-i \, \tau)^n}{n!}\left( \frac{1}{\tau}\int_0^\tau \, d\tau^\prime \, \hat{\mathcal{H}}_{t,I} (\tau^\prime) \right)^n.
\label{eq:SSE}
\end{equation}
Each power of the hopping $\hat{\mathcal{H}}_{t, I}(\tau)$ generates a step of the hole onto its $z=4$ neighbors. Therefore, the $n$th order expansion of the time evolution corresponds to the sum over all random walk paths of length $n$. In the following we use Eq.~\eqref{eq:SSE} to determine the motion of the hole. 

The dynamics of the spin background is governed by the time time evolution operator $e^{-i \hat{\mathcal{H}}_J \tau}$. As the hopping of the hole does not create spin flips, the spin environment remains in an eigenstate of the Ising spin Hamiltonian. After each step of the hole, the energy of the system can change, and the hole picks up a phase shift associated with the change in the Ising energy. 
Since in Eq.~\eqref{eq:SSE}, we integrate over the times when the hole hops from one site to the next, the phase factors are averaged out, which leads to the dephasing of the hole's wave function as induced by the spin environment. This suppresses the coherence of the hole's dynamics and leads to a slower propagation, as compared to the $J_z = 0$ non-interacting case, as we demonstrate in Fig.~\ref{fig:QMC}.

We determine the hole's dynamics numerically by sampling the paths using a real-time quantum Monte Carlo algorithm \cite{Sandvik1991,Carlstrom2016PRL,Nagy2017PRB}. In order to account for the $\frac{\tau^n}{n!}$ prefactor in the expansion as well as for the $z^n$ phase space, we sample random walk paths of length $n$ from a Poisson distribution with probability $\mathbb{P}_n \propto \frac{(z t)^n}{n!}$. Moving along each of the sampled paths, the hole modifies the spin environment by permuting the spins. We store the final spin state $|\Gamma_\alpha \rangle$ of the system for each path $\alpha$. The amplitudes of these states are determined by the $(-i)^n$ prefactors as well as the amplitudes arising from the time integrals over the phase factors picked up by the hole in Eq.~\eqref{eq:SSE}. We evaluate these $n$ dimensional integrals using Monte Carlo sampling. The resulting pairs $\{ \lambda_\alpha(\tau), |\Gamma_\alpha \rangle\}$ of amplitudes $\lambda_\alpha(\tau)$ and spin states $|\Gamma_\alpha \rangle$ determine the quantum state of the system,
\begin{equation}
|\Psi(\tau) \rangle = \sum_\alpha \lambda_\alpha(\tau) \, |\Gamma_\alpha \rangle,
\end{equation} 
where the summation runs over all possible random walk paths $\alpha$.

During the Monte Carlo sampling of the paths, we store the final quantum states of the system and add up all amplitudes corresponding to the same final state. This allows us to determine the transition probabilities $p_i(\tau)$, while we make use of the normalization condition $\sum_i p_i(\tau) = 1$ \cite{Carlstrom2016PRL}. As was discussed in Ref.~\cite{Nagy2017PRB}, the paths describing the time evolution of the states $|\Psi(\tau) \rangle$ and $\langle \Psi(\tau) |$ in the expression $p_i (\tau) = \langle \Psi (\tau)|  \h_i^\dagger \, \h_i  | \Psi (\tau) \rangle$ are sampled independently in order to avoid systematic numerical errors at intermediate times.

\section{Magnetic polaron dynamics -\\ NLST versus LST}
\label{apdx:NLSTvsLST}
In this appendix, we compare results for the hole dynamics obtained from linear and non-linear string theories. We consider the problem of a hole created in the N\'eel state by removing the central spin, as in Sec.~\ref{subsec:FarFromEqul}. For short times we obtained excellent agreement between LST and NLST, which we benchmarked by comparing to time-dependent Monte Carlo calculations in Fig.~\ref{fig:QMC}. 

In Fig.~\ref{fig:LSTvsNLST} (a) we show the rms radius of the hole density distribution for longer times. Note that LST calculations include the slow spinon dynamics, whereas the NLST results are performed without taking into account the motion of the spinon. For $J_z=t$ we find good agreement between NLST and LST. At long times the ballistic spinon expansion can be observed in the LST result, which is not included in NLST. The length of the string in the long-time limit is the same in LST and NLST in this case. 

For $J_z/t=0.5$ we observe quantitative deviations between our results from NLST and LST in Fig.~\ref{fig:LSTvsNLST} (a). In particular, the LST predicts stronger oscillations at long times. In the NLST many quantum states with slightly different energies are included. This is expected to cause dephasing, which explains the observed suppression of coherent string dynamics. 

For $J_z/t=0.2$ we observe large deviations between LST and NLST. The rms radius predicted by NLST suddenly saturates at intermediate times, while it continues to grow in the LST. This is an artifact of the NLST result, where the maximum string length used in the calculations was $\ell_{\rm max}=10$. For $J_z/t=0.2$ longer strings need to be included. Our LST calculations use strings up to a length $\ell_{\rm max}=30$, which is sufficient to obtain fully converged results. Note that LST can easily go to even longer values of the string length. 

In Fig.~\ref{fig:LSTvsNLST} (b) we also compare predictions for the probability of finding the spinon and the holon on the same site. As before we obtain excellent agreement for short times. For longer times, the LST predicts more oscillations, but the qualitative behavior is the same in both theories.

\section{The generalized $1/S$ expansion\\ and inclusion of quantum fluctuations}
\label{apdxQuantFlucGenS}

In this appendix we introduce the formalism for the generalized $1/S$ expansion and explain how it can be used to include quantum fluctuations in the strong coupling parton description of magnetic polarons. 

\subsection{Shortcoming of the conventional $1/S$ expansion}
\label{subsubsecShortComingConventionalS}
In the conventional $1/S$ expansion of magnetic polaron problems \cite{SchmittRink1988,Kane1989,Liu1991}, $S$ only enters in the corresponding Schwinger-boson constraint, $\sum_\sigma \bd_{j \sigma} \b_{j \sigma} + \hd_j \h_j = 2 S$. In this case the holon hopping is described by Eq.~\eqref{eq:HtSB} with the operator $\hat{\mathcal{F}}_{ij}(S) \equiv \hat{\mathcal{F}}_{ij}(1/2)$ from Eq.~\eqref{eq:Fij12}, for all values $S$. Note that the conventional Schwinger-boson constraint is respected by the resulting operator $\H_t$. 

Let us consider the effect of such holon motion on the local N\'eel order parameter $\hat{\Omega}_{j}$, see Eq.~\eqref{eqDefOmgaj}. Within the conventional extension of the $t-J_z$ model to large values of $S$, the motion of the holon from site $i$ to $j$ is accompanied by changes of the spins $S^z_{i}$ and $S^z_{j}$ by $\pm 1/2$, see Fig.~\ref{fig:largeSholeHopping} (b). As a result, the sign of the local N\'eel order parameter $\Omega_j$ cannot change when $S \gg 1/2$ is large, unless the holon performs multiple loops. 

We conclude that the conventional $1/S$ expansion cannot capture correctly the destruction of the N\'eel order parameter by the holon motion, at least when $S \gg 1/2$ is large. This effect is particularly pronounced for a hole inside a one dimensional spin chain with $S=1/2$. In this case a single hole introduces a domain wall into the system, where the local N\'eel order parameter changes sign -- a direct manifestation of spin-charge separation \cite{Kruis2004a,Hilker2017}. The conventional $1/S$ expansion cannot capture this effect because it only allows for small local changes of the underlying N\'eel order parameter. 

The shortcoming of the conventional $1/S$ expansion explained above is also reflected by the fact that the new Schwinger-boson constraint in Eq.~\eqref{eq:constraintTJ} is not satisfied for $S > 1/2$. This means that the holon can move without fully distorting the underlying N\'eel order in this case. When the generalized constraint from Eq.~\eqref{eq:constraintTJ} is enforced, the local N\'eel order can be destroyed by a single hole hopping event. This is an important feature of the $t-J_z$ model at $S=1/2$, and it is fully captured  -- even for large values of $S$ -- by the generalized model which we introduce now.

\subsection{Generalized holon hopping for large $S$}
\label{subsubsec:LargeSholonHopping}
In order to ensure that the generalized Schwinger-boson constraint Eq.~\eqref{eq:constraintTJ} is satisfied by $\H_t$ from Eq.~\eqref{eq:HtSB}, the operator $\hat{\mathcal{F}}_{ij}(S)$ in Eq.~\eqref{eq:HtSB} has to depend explicitly on the value of $S$. We will refer to the new term $\hat{\mathcal{F}}_{ij}(S)$ as the \emph{generalized holon-hopping} operator. 

To construct $\hat{\mathcal{F}}_{ij}(S)$ we recall the physical origin of nearest-neighbor hole hopping in the original model Eq.~\eqref{eq:model}: The fermion initially occupying site $j$ moves to site $i$ without changing its spin state, see Fig.~\ref{fig:largeSholeHopping} (c). For $S=1/2$ this process is accurately described within the Schwinger-boson language by the operator $\hat{\mathcal{F}}_{ij}(1/2)$ from Eq.~\eqref{eq:Fij12}, which changes the $z$-component of the spins on sites $i$ and $j$ by $\pm 1/2$ respectively. 

For larger $S>1/2$ the entire spin state is transferred from site $j$ to $i$ by the operator $\hat{\mathcal{F}}_{ij}(S)$. Because expressing this process in terms of Schwinger-bosons leads to a highly non-linear and cumbersome expression, we define it using the eigenstates $\ket{n_\uparrow^l, n_\downarrow^l}$ of $\bd_{l,\uparrow} \b_{l,\uparrow}$ (eigenvalue $n_\uparrow^l$) and $\bd_{l,\downarrow} \b_{l,\downarrow}$ (eigenvalue $n_\downarrow^l$) on site $l=i,j$ now: 
\begin{equation}
\hat{\mathcal{F}}_{ij}(S) = \sum_{n_\uparrow + n_\downarrow = 2S} \ket{0^j,0^j,n_\uparrow^i, n^i_\downarrow} \bra{n_\uparrow^j, n_\downarrow^j,0^i,0^i}.
\label{eq:HijSdef}
\end{equation}
Note that $\hd_j \h_i \hat{\mathcal{F}}_{ij}(S)$ respects the constraint in Eq.~\eqref{eq:constraintTJ}, because the operator $\hat{\mathcal{F}}_{ij}(S)$ only acts on the subspace with zero Schwinger-bosons on site $i$ and $2S$ Schwinger-bosons on site $j$.

\subsection{Inclusion of quantum fluctuations}
\label{subsubsecInlcusionQuantFluc}

The generalized $1/S$ expansion presented in Sec.~\ref{subsec:LargeSholehopping} can be extended to include the effects of quantum fluctuations. On the one hand this allows to describe excitations in the $t-J_z$ models with $S > 1/2$ which do not change the sign of the N\'eel order parameter. In such cases the magnitude of the $z$ component $\hat{S}^z_j$ changes but its sign is unmodified. Excitations of this type are present for example at finite temperatures. On the other hand, more general models including flip-flop terms $\hat{S}^+_i \hat{S}^-_j$ in the spin Hamiltonian can be treated this way. Most importantly this includes the SU(2) invariant $t-J$ model, which we discuss in more detail using the present formalism in a forthcoming work. 

Quantum fluctuations around the classical N\'eel state are included in the usual way \cite{Huse1988,Sachdev1989,Kane1989} using the Holstein-Primakoff approximation \cite{Holstein1940}. Here we also allow for distortions of the classical N\'eel state which are introduced by the motion of the holon. They are described by the Ising variables $\tilde{\tau}^z_j$ which represent the orientation of the spins, see Sec.~\ref{subsubsecFormalismDistortion}. By performing a Holstein-Primakoff approximation around the distorted state we obtain the following representation of spin operators,
\begin{flalign}
\hat{S}_j^z = & \tilde{\tau}_j^z \l S - \ad_j \a_j \r, 
\label{eq:HolsteinPrimakoff1} \\
\hat{S}_j^{\tilde{\tau}_j^z} =\sqrt{2S}  &~  \a_j, \quad \hat{S}_j^{\tilde{\tau}_j^z} = \sqrt{2S} ~ \ad_j.
\label{eq:HolsteinPrimakoff2}
\end{flalign}
These expressions are correct up to leading order in $\a_j$, but higher orders can also be retained \cite{Holstein1940}.

The operators $\a_j$ in Eqs.~\eqref{eq:HolsteinPrimakoff1} and \eqref{eq:HolsteinPrimakoff2} can also be expressed in terms of the Schwinger-bosons introduced in Sec.~\ref{subsubsec:SchwingerBosonRep}. For $\tilde{\tau}_j^z=1$ one obtains $\a_j = \b_{j, \downarrow}$, whereas $\a_j = \b_{j, \uparrow}$ when $\tilde{\tau}_j^z=-1$. Fluctuations of the respective second component $\b_{j, \uparrow}$ ($\b_{j, \downarrow}$) which is condensed when $\tilde{\tau}_j^z=1$ (if $\tilde{\tau}_j^z=-1$) adds higher-order corrections \cite{Kane1989}.

Using the representation from Eqs.~\eqref{eq:HolsteinPrimakoff1}, \eqref{eq:HolsteinPrimakoff2} we can rewrite the spin-$S$ Ising Hamiltonian $\H_J$ from Eq.~\eqref{eq:model} as
\begin{multline}
\H_{J} =   - 2 N S^2 J_z + \frac{S}{2} \sum_\ij  \biggl\{ 2 S J_z \l 1 - \hat{\sigma}_\ij^z \r  \\
+  \l 1 + \hat{\sigma}_\ij^z \r \left[  J_z \l \ad_i \a_i + \ad_j \a_j \r \right] \\
-   \l 1 - \hat{\sigma}_\ij^z \r \left[ J_z \l  \ad_i \a_i + \ad_j \a_j \r  \right] \biggr\}.
\label{eq:HheisenbergFluc}
\end{multline}
Here $N$ denotes the number of lattice sites and distortions of the N\'eel order are described by the field $\hat{\sigma}_\ij^z$ defined in Eq.~\eqref{eqDefSigmaij}. The expression in Eq.~\eqref{eq:HheisenbergFluc} can be easily generalized to include transverse couplings in spin space.

\subsection{Broken strings: magnon corrections}
As an example for applications of quantum fluctuations in the $1/S$ expansion, we sketch how defects of the strings can be treated perturbatively. A detailed discussion will be provided in a forthcoming publication. Consider a holon performing a Trugman loop starting from a state with non-zero string length. This allows the holon to tunnel diagonally across a plaquette in the square lattice, without moving the spinon. The same effect is obtained when next-nearest neighbor hopping processes with amplitude $t'$ are included in the microscopic $t-J_z$ Hamiltonian. In both cases the string is ripped apart, and the resulting state cannot be described within the Hilbert space of the string theory. 

To include the effect of broken strings for $S=1/2$ into our formalism, we suggest to use the conventional $1/S$ expansion for describing next-nearest neighbor hopping processes. I.e. we treat the string as unbroken, but create magnon excitations $\a_i$ when the holon is hopping to a next-nearest neighbor site. This reflects the distortion of the physical spin state: On the one hand the spin configuration for $S=1/2$ is correctly described within the Holstein-Primakoff approximation, see Eq.~\eqref{eq:HolsteinPrimakoff1}. On the other hand, the energy gain from the new spin configuration is taken into account.

\begin{figure*}[t!]
\centering
\epsfig{file=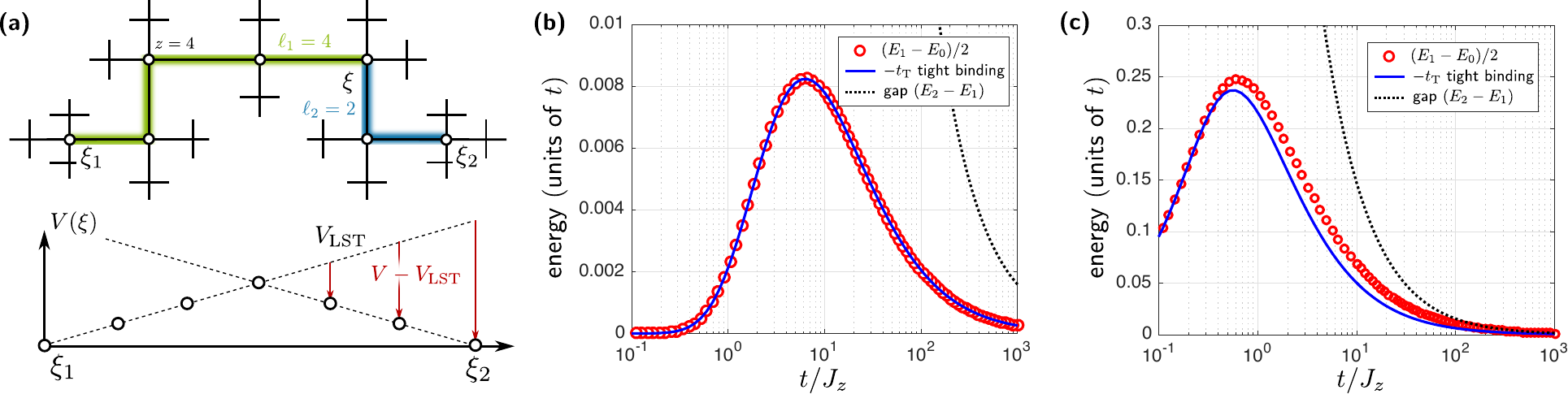, width=0.95\textwidth}
\caption{Toy model for Trugman loop tunneling. (a) We consider a single particle hopping between the sites $\xi$ of a Bethe lattice with coordination number $z$. The potential energy $V(\xi) = J_z {\rm min} (\ell_1, \ell_2)$ is determined by the smaller of the two distances to the two spinon sites $\xi_1$ and $\xi_2$. The distance $\ell_{\rm T}$ between $\xi_1$ and $\xi_2$ corresponds the length of the Trugman loop. In (b) and (c) we compare the energy splitting $(E_1-E_0)/2$ of the lowest two eigenstates of the toy model to the tight-binding Trugman loop hopping element $t_{\rm T}$ and the energy gap $\Delta = E_2-E_1$ to higher excited states. Parameters are $z=4$ and $\ell_{\rm T}=6$ in (b) and $z=3$ and $\ell_{\rm T} = 2$ in (c).}
\label{fig:BLtoy}
\end{figure*}

\section{Tight-binding description of Trugman loops on the Bethe lattice}
\label{apdx:TBtrugmanLoops}
In this appendix, we benchmark the tight-binding description of Trugman loop processes introduced in the main text. To this end we study a toy-model and show that a tight-binding description of Trugman loop processes is justified even when the hopping rate $t \gg J_z$ is large compared to the potential energy $\sim J_z$.

Like before, we assume that the Bethe lattice provides a valid description of the many-body Hilbert space, where every site corresponds to a holon trajectory. As explained in Sec.~\ref{subsubsec:StringTheory} this representation of the Hilbert space is over-complete due to the presence of Trugman loops \cite{Trugman1988}, i.e. holon trajectories which leave the N\'eel order intact but move the holon between two sites corresponding to the same sublattice. On the Bethe lattice such trajectories correspond to sites with the same minimal potential energy $V_{\ell=0} = - 2 J_z S^2$, see Eq.~\eqref{eq:HLST}. 

When $t \ll J_z$, the typical string length $\ell_0 \ll 1$ is very short and the holon is tightly localized around a site on the Bethe lattice with the minimal energy $V_{\ell=0}$. This picture of self-trapped holes in an anti-ferromagnet has been put forward by Bulaevskii et al.~\cite{Bulaevskii1968}. For longer string lengths, the holon wavefunction $\phi_\ell$ decays exponentially because it cannot penetrate the potential barrier $V_\ell$. The picture of fully localized holes is not correct, however: due to quantum tunneling through the barrier the probability of the holon performing a Trugman loop and effectively moving the spinon is finite. In the regime $t \ll J_z$ where the barrier is large, the induced Trugman loop hopping $t_{\rm T}$ introduced in the main text can be calculated using standard tight-binding theory. 

When $t \gtrsim J_z$ the potential barrier $V_\ell$ between two sites on the Bethe lattice corresponding to different spinon positions becomes shallow. In this case, the effective one-dimensional wavefunction $\phi_\ell$ does not decay strongly for the relevant lengths $\ell \leq 6$ and a different mechanism is required to justify the tight-binding description explained in the main text. We will show that the exponential decay with $\ell$ of the wavefunction $\psi_{\ell,s}$, which is defined for a string of length $\ell$ and with directions $s$ on the Bethe lattice (see Sec.~\ref{secStringExcitations}), is sufficient to make the tight-binding calculation reliable.

\subsection{Toy model}
\label{subsec:TrugmanLoopTBtoy}
Consider the Hilbert space defined by a single particle hopping between the sites of a Bethe lattice (tunneling amplitude $t$) with coordination number $z$. We choose two sites $\xi_{1,2}$ in the Bethe lattice, corresponding to two possible spinon positions, where the potential energy $V(\xi_{1,2})=0$ vanishes. Away from these two points the potential energy increases linearly with the distance $\ell$ on the Bethe lattice: I.e., for any given site $\xi$ of the Bethe lattice we define the potential energy as 
\begin{equation}
V(\xi) = J_z {\rm min} (\ell_1,\ell_2)
\label{eqVxiToy}
\end{equation}
where $\ell_{1,2}$ are the distances from $\xi$ to $\xi_{1,2}$ respectively. This situation is illustrated in Fig.~\ref{fig:BLtoy} (a) for a distance $\ell_T=6$ between $\xi_1$ and $\xi_2$. It closely resembles the situation considered in the original $t-J_z$ model, where infinitely many spinon positions $\xi_n$ with the property $V(\xi_n) = {\rm min}_\xi V(\xi)$ exist.

The spectrum of the toy model can be easily calculated numerically to any desired precision. When $t \ll J_z$, we expect two almost degenerate ground states with an energy splitting $E_1-E_0 = 2 t_{\rm T}^{\rm eff}$ and an energy gap $\Delta = E_2-E_1$ to vibrationally excited states. Here $t_{\rm T}^{\rm eff}$ is the effective Trugman loop tunneling element. In Fig.~\ref{fig:BLtoy} (b) and (c) we compare our results for the energy splitting to a tight-binding calculation based on orbitals obtained by assuming a completely linear string potential $V_\ell = J_z \ell$. For $t \ll J_z$ we find excellent agreement of the tight-binding calculation with the exact result. We also calculate the gap $\Delta$ to higher excited states and find that it is much larger than the energy splitting due to Trugman loop tunneling on the Bethe lattice. 

In Fig.~\ref{fig:BLtoy} (b) and (c) we observe a similar behavior even when $t \gg J_z$. For toy-model parameters $z=4$ and the length of the Trugman loop $\ell_{\rm T}=6$, see Fig.~\ref{fig:BLtoy} (b), we find that the tight-binding calculation predicts the tunneling correctly for arbitrary values of $t / J_z$ to high precision (to within $\sim 0.5 \%$). When the Trugman loop is shorter, $\ell_{\rm T}=2$, and the coordination number $z=3$ in the toy model, see Fig.~\ref{fig:BLtoy} (c), we observe sizable quantitative deviations of the tight-binding theory from the exact result for $t \gtrsim J_z$. In this case the gap $\Delta$ to higher excited states becomes comparable to the energy spacing between the lowest two states for the largest values of $t/J_z$.

The reason why the tight-binding theory works so well, even when the potential barrier along the Trugman loop trajectory shallow, is the fractal structure of the Bethe lattice. When $t \gg J_z$, the amplitude of a low-energy wavefunction $\psi_{\ell,s}$ is distributed symmetrically over all possible $z-1$ directions at every node on the Bethe lattice, see Eq.~\eqref{eq:reParametrize}. This minimizes the kinetic energy, yielding an energy shift of $- 2 \sqrt{z-1} t$, and leads to the exponential decay of $\psi_{\ell,s} \sim (z-1)^{\ell/2} \phi_\ell$ compared to the effective one-dimensional string wavefunction $\phi_\ell$. In order to gain energy from the non-linearity $V(\xi) - J_z \ell_1$ or $V(\xi) - J_z \ell_2$ of the string potential from Eq.~\eqref{eqVxiToy}, the wavefunction would need to be localized in one particular direction of the Bethe lattice, along the Trugman loop trajectory. This is energetically unfavorable because it only leads to a potential energy gain $\sim J_z$ but costs localization energy $\sim t$: Localizing the particle in one particular direction only leads to a zero-point energy $- 2 t$.

\bibliographystyle{unsrt}

\end{document}